\documentclass[opre,nonblindrev]{informs3} % current default for manuscript submission

\DoubleSpacedXI % Made default 4/4/2014 at request
\usepackage{endnotes}
\usepackage{epsf}
\usepackage{graphicx}
\usepackage{amsmath} 
\usepackage{bm}
\usepackage{amssymb}
\usepackage{amsfonts}
\usepackage[ruled,vlined]{algorithm2e}
\usepackage{graphicx} 
\usepackage{caption}
\usepackage{subcaption}
\usepackage{enumerate}
\usepackage[shortlabels]{enumitem}
\usepackage{epstopdf}
\usepackage{multirow}
\usepackage{ctable}
\usepackage{epstopdf}
\usepackage{url}
\usepackage[english]{babel}
\usepackage[compact]{titlesec}
\newcounter{claimcounter}

\newcounter{assumpcounter}

\newenvironment{Proof}[1]{\par\noindent\textit{Proof:}\space#1}{\hfill $\blacksquare$}

\usepackage{times}
\usepackage{helvet}
\usepackage{courier}
\usepackage{wrapfig}

\let\footnote=\endnote
\let\enotesize=\normalsize
\def\notesname{Endnotes}%
\def\makeenmark{$^{\theenmark}$}
\def\enoteformat{\rightskip0pt\leftskip0pt\parindent=1.75em
  \leavevmode\llap{\theenmark.\enskip}}

\TheoremsNumberedThrough     % Preferred (Theorem 1, Lemma 1, Theorem 2)
\ECRepeatTheorems

\EquationsNumberedThrough    % Default: (1), (2), ...
\begin{document}
%%%%%%%%%%%%%%%%

%\setcounter{page}{1}
%\DOI{10.1287/xxxx.0000.0000}%

\RUNAUTHOR{Joseph and Bhatnagar}

\RUNTITLE{A Cross Entropy based Stochastic Approximation Algorithm for Reinforcement Learning with Linear Function Approximation}

% \TITLE{Bundling Information Goods of Decreasing Value}
\TITLE{A Cross Entropy based Stochastic Approximation Algorithm for Reinforcement Learning with Linear Function Approximation}

\ARTICLEAUTHORS{%
\AUTHOR{Ajin George Joseph}
\AFF{Department of Computer Science and Automation, Indian Institute of Science, Bangalore, India, \EMAIL{ajin@csa.iisc.ernet.in}} %, \URL{}}
\AUTHOR{Shalabh Bhatnagar}
\AFF{Department of Computer Science and Automation, Indian Institute of Science, Bangalore, India, \EMAIL{shalabh@csa.iisc.ernet.in}} %, \URL{}}
} % end of the block

\ABSTRACT{%
In this paper, we provide a new algorithm for the problem of prediction in Reinforcement Learning, \emph{i.e.}, estimating the Value Function of a Markov Reward Process (MRP) using the linear function approximation architecture, with memory and computation costs scaling quadratically in the size of the feature set. The algorithm is a multi-timescale variant of the very popular Cross Entropy (CE) method which is a model based search method to find the global optimum of a real-valued function. This is the first time a model based search method is used for the prediction problem. The application of CE to a stochastic setting is a completely unexplored domain. A proof of convergence using the ODE method is provided. The theoretical results are supplemented with experimental comparisons. The algorithm achieves good performance fairly consistently on many RL benchmark problems. This demonstrates the competitiveness of our algorithm against least squares and other state-of-the-art algorithms in terms of computational efficiency, accuracy and stability.
% Enter your abstract
}%

%\KEYWORDS{deterministic inventory theory; infinite linear programming duality;
%  existence of optimal policies; semi-Markov decision process; cyclic schedule}


\maketitle
%%%%%%%%%%%%%%%%%%%%%%%%%%%%%%%%%%%%%%%%%%%%%%%%%%%%%%%%%%%%%%%%%%%%%%

% Samples of sectioning (and labeling) in OPRE
% NOTE: (1) \section and \subsection do NOT end with a period
%       (2) \subsubsection and lower need end punctuation
%       (3) capitalization is as shown (title style).
%
%\section{Introduction.}\label{intro} %%1.
%\subsection{Duality and the Classical EOQ Problem.}\label{class-EOQ} %% 1.1.
%\subsection{Outline.}\label{outline1} %% 1.2.
%\subsubsection{Cyclic Schedules for the General Deterministic SMDP.}
%  \label{cyclic-schedules} %% 1.2.1
%\section{Problem Description.}\label{problemdescription} %% 2.

% Text of your paper here

\section{Introduction and Preliminaries}
In this paper, we follow the Reinforcement Learning (RL) framework as described in \cite{sutton1998introduction, white1993survey, bertsekas1995dynamic}. The basic structure in this setting is the discrete time Markov Decision Process (MDP) which is a 4-tuple ($\mathbb{S}$, $\mathbb{A}$, $\mathrm{R}$, $\mathrm{P}$), where $\mathbb{S}$ denotes the set of \textit{states} and $\mathbb{A}$ is the set of \textit{actions}. $\mathrm{R}: \mathbb{S} \times \mathbb{A} \times \mathbb{S} \rightarrow \mathbb{R}$ is the \textit{reward function} where $\mathrm{R}(s, a, s^{\prime})$ represents the reward obtained in state $s$ after taking action $a$ and transitioning to $s^{\prime}$. Without loss of generality, we assume that the reward function is bounded, \emph{i.e.}, $|\mathrm{R}(.,.,.)| \leq \mathrm{R}_{\mathrm{max}} < \infty$. $\mathrm{P}:\mathbb{S} \times \mathbb{A} \times \mathbb{S} \rightarrow [0,1]$ is the \textit{transition probability kernel}, where $\mathrm{P}(s, a, s^{\prime}) = \mathbb{P}(s^{\prime} | s, a)$ is the probability of next state being $s^{\prime}$ conditioned on the fact that the current state is $s$ and action taken is $a$. We assume that the state and action spaces are finite with $|\mathbb{S}|=n$ and $|\mathbb{A}|=b$. A \textit{stationary policy} $\pi:\mathbb{S} \rightarrow \mathbb{A}$ is a function from states to actions, where $\pi(s)$ is the action taken in state $s$. A given policy $\pi$ along with the transition kernel $\mathrm{P}$ determines the state dynamics of the system. For a given policy $\pi$, the system behaves as a Markov Reward Process (MRP) with transition matrix $\mathrm{P}^{\pi}(s,s^{\prime})$ = $\mathrm{P}(s,\pi(s),s^{\prime})$. The policy can also be stochastic in order to incorporate  exploration. In that case, for a given $s \in \mathbb{S}$, $\pi(.\vert s)$ is a probability distribution over the action space $\mathbb{A}$.

For a given policy $\pi$, the system evolves at each discrete time step and this process can be captured as a sequence of triplets $(\mathbf{s}_{t}, \mathbf{r}_{t}, \mathbf{s}^{\prime}_{t}), t \geq 0$, where $\mathbf{s}_{t}$ is the random variable which represents the current state at time $t$, $\mathbf{s}^{\prime}_{t}$ is the transitioned state from $\mathbf{s}_{t}$ and $\mathbf{r}_{t} = \mathrm{R}(\mathbf{s}_t, \pi(\mathbf{s}_t), \mathbf{s}^{\prime}_{t})$ is the reward associated with the transition. In this paper, we are concerned with the problem of \textit{prediction}, {\em i.e.}, estimating the long run $\gamma$-discounted cost $V^{\pi} \in \mathbb{R}^{\mathbb{S}}$ (also referred to as the \textit{Value function}) corresponding to the given policy $\pi$. Here, given $s \in \mathbb{S}$, we let 
\begin{equation}
V^{\pi}(s) \triangleq \mathbb{E}\left[\sum_{t=0}^{\infty}\gamma^{t}\mathbf{r}_{t} \big \vert \mathbf{s}_{0} = s\right],
\end{equation}
where $\gamma \in [0,1)$ is a constant called  the \textit{discount factor} and $\mathbb{E}[\cdot]$ is the expectation over sample trajectories of states obtained in turn from  $\mathrm{P}^{\pi}$ when starting from the initial state $s$. $V^{\pi}$ is represented as a vector in $\mathbb{R}^{\vert\mathbb{S}\vert}$. $V^{\pi}$ satisfies the well known \textit{Bellman equation} under policy $\pi$, given by
\begin{equation}
V^{\pi} = \mathrm{R}^{\pi} + \gamma\mathrm{P}^{\pi}V^{\pi} \triangleq T^{\pi}V^{\pi},
\end{equation}
where $\mathrm{R}^\pi \triangleq (\mathrm{R}^\pi(s),s\in\mathbb{S})^\top$ with $\mathrm{R}^{\pi}(s)$ = $\mathbb{E}\left[\mathbf{r}_{t} \vert \mathbf{s}_t = s\right]$, $V^\pi \triangleq (V^\pi(s),s\in\mathbb{S})^\top$ and $T^\pi V^\pi \triangleq ((T^\pi V^\pi)(s),$
$ s \in \mathbb{S})^\top$, respectively. Here $T^\pi$ is called the \textit{Bellman operator}. If the model information, \emph{i.e.}, $\mathrm{P}^{\pi}$ and $\mathrm{R}^{\pi}$  are available, then we can obtain the value function $V^{\pi}$ by solving analytically the linear system $V^{\pi} = (I-\gamma \mathrm{P}^{\pi})^{-1}\mathrm{R}^{\pi}$.

However, in this paper, we follow  the usual RL framework, where we assume that the model, is inaccessible; only a sample trajectory $\{(\mathbf{s}_t, \mathbf{r}_{t}, \mathbf{s}^{\prime}_{t})\}_{t=1}^{\infty}$ is available where at each instant $t$, state $\mathbf{s}_t$ of the triplet $(\mathbf{s}_t, \mathbf{r}_{t}, \mathbf{s}^{\prime}_{t})$ is sampled using an arbitrary distribution $\nu$ over $\mathbb{S}$, while the next state $\mathbf{s}^{\prime}_{t}$ is sampled using $\mathrm{P}^{\pi}(\mathbf{s}_t,.)$ and $\mathbf{r}_{t}$ is the immediate reward for the transition. The value function $V^{\pi}$ has to be estimated here from the given sample trajectory. 

To further make the problem more arduous, the number of states $n$ may be large in many practical applications, for example, in games such as chess and backgammon. Such combinatorial blow-ups exemplify the underlying problem with the value function estimation, commonly referred to as the \textit{curse of dimensionality}. In this case, the value function is unrealizable due to both storage and computational limitations. Apparently one has to resort to  approximate solution methods where we sacrifice precision for computational tractability. A common approach in this context is the function approximation method \cite{sutton1998introduction}, where we approximate the value function of unobserved states using available training data. 

In the \textit{linear function approximation technique}, a linear architecture consisting of a set of $k$, $n$-dimensional feature vectors, $1 \leq k \ll n$, $\{\phi_{i} \in \mathbb{R}^{\mathbb{S}}\}$, $1 \leq i \leq k$,  is chosen \emph{a priori}.  For a state $s \in \mathbb{S}$, we define
\begin{equation}\label{eqn:phieq}
\phi(s) \triangleq \begin{bmatrix}
         \phi_{1}(s) \\
          \phi_{2}(s) \\
          \vdots \\
          \phi_{k}(s) \\          
\end{bmatrix}_{k \times 1}, \hspace{2mm}      
\Phi \triangleq  \begin{bmatrix}
\phi(s_{1})^{\top} \\
\phi(s_{2})^{\top} \\
\vdots \\
\phi(s_{n})^{\top}
\end{bmatrix}_{n \times k},
\end{equation}
where the vector $\phi(\cdot)$ is called the \textit{feature vector}, while $\Phi$ is called the \textit{feature matrix}.

Primarily, the task in linear function approximation is to find a weight vector $z \in \mathbb{R}^{k}$ such that the predicted value function $\Phi z \approx V^{\pi}$. Given $\Phi$, the best approximation of $V^{\pi}$ is its projection on to the  subspace $\{\Phi z |  z \in \mathbb{R}^{k}\}$ (column space of $\Phi$) with respect to an arbitrary norm. Typically, one uses the weighted norm $\Vert.\Vert_{\nu}$ where $\nu(\cdot)$ is an arbitrary distribution over $\mathbb{S}$.  The norm  $\Vert.\Vert_{\nu}$ and its associated linear projection operator $\Pi^{\nu}$ are defined as
\begin{eqnarray}\label{eq:norm}
\Vert V \Vert_{\nu}^{2} = \sum_{i=1}^{n}V(i)^{2}\nu(i), \hspace{4mm}
\Pi^{\nu} = \Phi(\Phi^{\top}D^{\nu}\Phi)^{-1}\Phi^{\top}D^{\nu},
\end{eqnarray}
\normalsize
where $D^{\nu}$ is the diagonal matrix with $D^{\nu}_{ii} = \nu(i), i=1, \dots ,n$.
So a familiar objective in most approximation algorithms is to find a vector $z^{*} \in \mathbb{R}^{k}$ such that $\Phi z^{*} \approx \Pi^{\nu}V^{\pi}$. 

Also it is important to note that the efficacy of the learning method depends on both the features $\phi_{i}$  and the parameter $z$ \cite{lagoudakis2003least}. Most commonly used features include Radial Basis Functions (RBF), Polynomials, Fourier Basis Functions \cite{konidaris2011value}, Cerebellar Model Articulation Controller (CMAC) \cite{eldracher1994function} \emph{etc}. In this paper, we assume that a carefully chosen set of features is available \emph{a priori}.

The existing algorithms can be broadly classified as $(i)$ Linear methods which include Temporal Difference (TD) \cite{sutton1988learning}, Gradient Temporal Difference (GTD \cite{sutton2009convergent}, GTD2 \cite{sutton2009fast}, TDC \cite{sutton2009fast}) and Residual Gradient (RG) \cite{baird1995residual} schemes,  whose computational complexities are linear in $k$ and hence are good for large values of $k$ and $(ii)$ Second order methods which include Least Squares Temporal Difference (LSTD) \cite{bradtke1996linear, boyan2002technical} and Least Squares Policy Evaluation (LSPE) \cite{nedic2003least} whose computational complexities are quadratic in $k$ and are useful for moderate values of $k$. Second order methods, albeit computationally expensive, are seen to be more data efficient than others except in the case when trajectories are very small \cite{dann2014policy}.

Eligibility traces \cite{sutton1988learning} can be integrated into most of these algorithms to improve the convergence rate. Eligibility trace is a mechanism to accelerate learning by blending temporal difference methods with Monte Carlo simulation (averaging the values) and weighted using a geometric distribution with parameter $\lambda \in [0,1)$. The algorithms with eligibility traces are named with $(\lambda)$ appended, for example TD$(\lambda)$, LSTD$(\lambda)$ \emph{etc}. In this paper, we do not consider the treatment of eligibility traces.

Sutton's \textit{TD($\lambda$) algorithm with function approximation} \cite{sutton1988learning}  is one of the fundamental algorithms in RL. TD($\lambda$) is an online, incremental algorithm, where at each discrete time $t$, the weight vectors are adjusted to better approximate the target value function. The simplest case of the one-step TD learning, \emph{i.e.} $\lambda = 0$, starts with an initial vector $\mathbf{z}_{0}$ and the learning continues at each discrete time instant $t$ where a new prediction vector $\mathbf{z}_{t+1}$ is obtained using the recursion,
\[\mathbf{z}_{t+1} = \mathbf{z}_{t} + \alpha_{t+1}\delta_{t}(\mathbf{z}_t)\phi(\mathbf{s}_t).\]
In the above, $\alpha_t$ is the learning rate which satisfies $\sum_{t}\alpha_{t} = \infty$, $\sum_{t}\alpha_{t}^{2} < \infty$ and $\delta_t(z) \triangleq \mathbf{r}_{t} + \gamma z^{\top}\phi(\mathbf{s}^{\prime}_{t}) - z^{\top}\phi(\mathbf{s}_{t})$  is called the \textit{Temporal Difference (TD)-error}.
In on-policy cases where Markov Chain is ergodic and the sampling distribution $\nu$ is the stationary distribution of the Markov Chain, then with $\alpha_t$ satisfying the above conditions and with $\Phi$ being a full rank matrix, the convergence of TD(0) is guaranteed \cite{tsitsiklis1997analysis}. But in off-policy cases, \emph{i.e.}, where the sampling distribution $\nu$ is not the stationary distribution of the chain, TD(0) is shown to diverge \cite{baird1995residual}. 

By applying stochastic approximation theory, the limit point $z^{*}_{\texttt{TD}}$ of TD$(0)$ is seen to satisfy
\begin{equation}
0 = \mathbb{E}\left[\delta_{t}(z)\phi(\mathbf{s}_t)\right] 
  = Az - b, 
\end{equation}
where $A = \mathbb{E}\left[\phi(\mathbf{s}_{t})(\phi(\mathbf{s}_{t}) - \gamma \phi(\mathbf{s}^{\prime}_{t}))^{\top}\right]$ and $b = \mathbb{E}\left[\mathbf{r}_{t}\phi(\mathbf{s}_{t})\right].$
This gives rise to the \textit{Least Squares Temporal Difference (LSTD)} algorithm \cite{bradtke1996linear, boyan2002technical}, which at each iteration $t$, provides estimates $A_t$ of matrix $A$ and $\mathbf{b}_t$ of vector $b$, and upon termination of the algorithm at time $T$, the approximation vector $\mathbf{z}_{T}$ is evaluated by
$\mathbf{z}_{T} = (A_T)^{-1}\mathbf{b}_T$.

\textit{Least Squares Policy Evaluation (LSPE)} \cite{nedic2003least} is a multi-stage algorithm where in the first stage, it obtains $\mathbf{u}_{t+1} = \argmin_{u}\Vert \Phi \mathbf{z}_{t} - T^{\pi}\Phi u \Vert^{2}_{\nu} \hspace{1mm}$ using the least squares method.
In the subsequent stage, it minimizes the fix-point error using the recursion $\mathbf{z}_{t+1} = \mathbf{z}_{t} + \alpha_{t+1}\left(\mathbf{u}_{t+1} - \mathbf{z}_{t}\right)$. 

Van Roy and Tsitsiklis \cite{tsitsiklis1997analysis} gave a different characterization for the limit point $z^{*}_{\texttt{TD}}$ of TD(0) as the fixed point of the \textit{projected Bellman operator} $\Pi^{\nu}T^{\pi}$,
\begin{equation}
\Phi z = \Pi^{\nu}T^{\pi} \Phi z.
\end{equation}
This characterization yields a new error function, the \textit{Mean Squared Projected Bellman Error (MSPBE)} defined as
\begin{equation}
\textrm{MSPBE}(z) \triangleq \Vert \Phi z - \Pi^{\nu}T^{\pi} \Phi z \Vert_{\nu}^{2}\hspace*{1mm}, \hspace*{4mm} z \in \mathbb{R}^{k}. \end{equation}
In \cite{sutton2009fast, sutton2009convergent}, this objective function is maneuvered to derive  novel $\Theta(k)$ algorithms like \textit{GTD}, \textit{TDC} and \textit{GTD2}. GTD2 is a multi-timescale algorithm given by the following recursions:
\begin{eqnarray}
\mathbf{z}_{t+1} = \mathbf{z}_{t} + \alpha_{t+1}\left(\phi(\mathbf{s}_{t}) - \gamma \phi(\mathbf{s}^{\prime}_{t})\right)(\phi(\mathbf{s}_t)^{\top}\mathbf{v}_{t}),\\
\mathbf{v}_{t+1} = \mathbf{v}_{t} + \beta_{t+1}(\delta_{t}(\mathbf{z}_t) - \phi(\mathbf{s}_t)^{\top}\mathbf{v}_{t})\phi(\mathbf{s}_t). \hspace{7mm}
\end{eqnarray}
The learning rates $\alpha_{t}$ and $\beta_{t}$ satisfy $\sum_{t}\alpha_{t} = \infty$, $\sum_{t}\alpha_{t}^{2} < \infty$ and $\beta_{t} = \eta \alpha_{t}$, where $\eta > 0$.

Another pertinent error function is the \textit{Mean Square Bellman Residue ($\mathrm{MSBR}$)} which is defined as 
\begin{equation}
\mathrm{MSBR}(z) \triangleq \mathbb{E}\left[(\mathbb{E}\left[\delta_t(z)| \mathbf{s}_t\right])^{2}\right], z \in \mathbb{R}^{k}. 
\end{equation}
\textrm{MSBR} is a measure of how closely the prediction vector represents the solution to the Bellman equation. 

\textit{Residual Gradient (RG)} algorithm \cite{baird1995residual}  minimizes the error function MSBR directly using stochastic gradient search. RG however requires double sampling, \emph{i.e.}, generating two independent samples $\mathbf{s}^{\prime}_{t}$ and $\mathbf{s}^{''}_{t}$ of the next state when in the current state $\mathbf{s}_{t}$. The recursion is given by
\begin{equation}
\mathbf{z}_{t+1} = \mathbf{z}_{t} + \alpha_{t+1}\left(\mathbf{r}_{t}+\gamma \mathbf{z}_{t}^{\top}\phi^{\prime}_{t}- \mathbf{z}_{t}^{\top}\phi_{t}\right)\left(\phi_t-\gamma\phi^{''}_{t}\right),
\end{equation}
where $\phi_t \triangleq \phi(\mathbf{s}_{t})$, $\phi^{\prime}_{t} \triangleq \phi(\mathbf{s}^{\prime}_{t+1})$ and $\phi^{''}_{t} \triangleq \phi(\mathbf{s}^{''}_{t})$. Even though RG algorithm guarantees convergence, due to large variance, the convergence rate is small.

\begin{figure}[!h]
	\centering
	{\includegraphics[height=60mm, width=100mm]{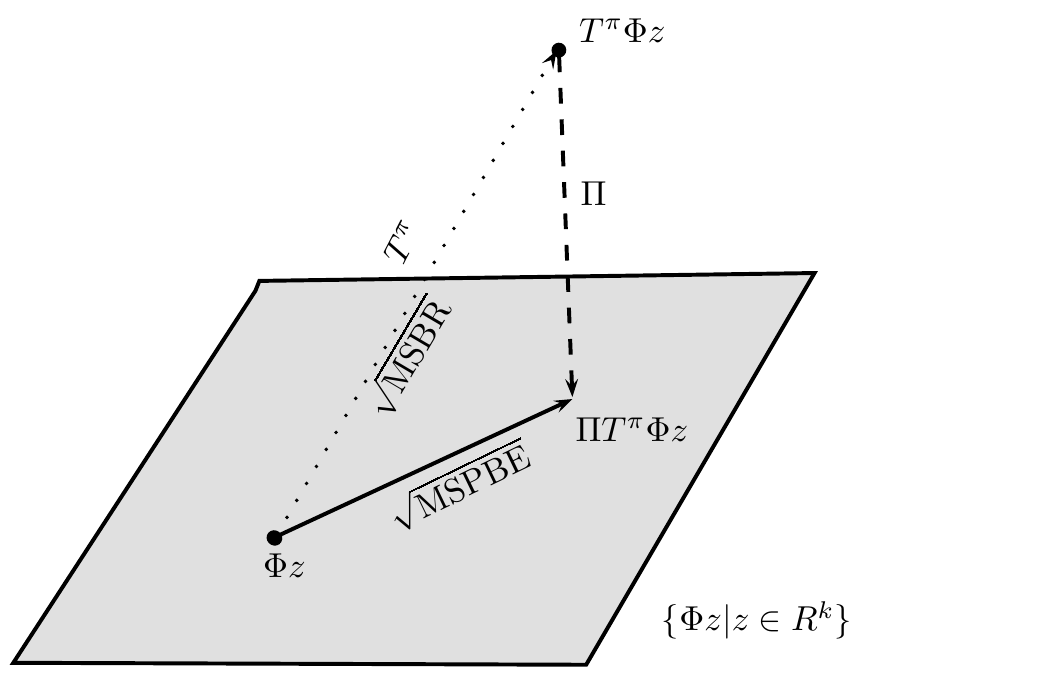}}
	\caption{Diagram depicting the relationship between the error functions \textrm{MSPBE} and \textrm{MSBR}.}\label{fig:errfnrel}
\end{figure}

If the feature set, \emph{i.e.}, the columns of the feature matrix $\Phi$ is linearly independent, then both the error functions MSBR and MSPBE are strongly convex.  However, their respective minima are related depending on whether the feature set is perfect or not. A feature set is \emph{perfect} if $V^{\pi} \in \{\Phi z \vert z \in \mathbb{R}^{k}\}$. If the feature set is perfect, then the respective minima of MSBR and MSPBE are the same. In the imperfect case, they differ. A relationship between MSBR and MSPBE  can be easily established as follows:
\begin{equation}
\mathrm{MSBR}(z) = \mathrm{MSPBE}(z) + \Vert T^{\pi}\Phi z - \Pi^{\nu} T^{\pi}\Phi z \Vert^{2}, \hspace*{2mm} z \in \mathbb{R}^{k}.
\end{equation}
A vivid depiction of the relationship is shown in Figure \ref{fig:errfnrel}.

\begin{table}
\begin{center}
\renewcommand{\arraystretch}{1.0}
\setlength\tabcolsep{1.1pt}
\begin{tabular}{llll}
\specialrule{.1em}{.02em}{.02em} 
\hline\noalign{\smallskip}
   $\vert$ Algorithm \hspace*{3cm}$\vert$ & Complexity \hspace*{15mm}$\vert$ & Error \hspace{18mm}$\vert$ &Eligibility Trace $\vert$\\
\noalign{\smallskip}
\hline
\specialrule{.1em}{.02em}{.02em} 
\noalign{\smallskip}
 	\hspace*{2mm}LSTD\hspace*{3cm} & $\Theta(k^{3})$ \hspace*{15mm}& MSPBE \hspace*{18mm}& Yes \\ 
	\hspace*{2mm}TD \hspace*{3cm} &$\Theta(k)$ \hspace*{15mm}& MSPBE \hspace*{18mm}& Yes \\ 
    \hspace*{2mm}LSPE \hspace*{3cm}& $\Theta(k^{3})$\hspace*{15mm} & MSPBE \hspace*{18mm}& Yes\\
	\hspace*{2mm}GTD \hspace*{3cm}& $\Theta(k)$\hspace*{15mm} & MSPBE \hspace*{18mm}& -\\
	\hspace*{2mm}GTD2 \hspace*{3cm}& $\Theta(k)$ \hspace*{15mm}& MSPBE \hspace*{18mm}& -\\
    \hspace*{2mm}RG \hspace*{3cm}& $\Theta(k)$ \hspace*{15mm}& MSBR \hspace*{18mm}& Yes\\	  
\hline
\specialrule{.1em}{.01em}{.01em} 
\end{tabular}
\end{center}
\caption{Comparison of the state-of-the-art function approximation RL algorithms}\label{tab:compalgtable}
\end{table}
Another relevant error objective is the \textit{Mean Square Error (MSE)} which is the square of the $\nu$-weighted distance from $V^{\pi}$ and is defined as
\begin{equation}
\mathrm{MSE}(z) \triangleq \Vert V^{\pi} - \Phi z \Vert^{2}_{\nu}\hspace*{1mm}, \hspace*{4mm} z \in \mathbb{R}^{k}.
\end{equation}
In \cite{williams1993tight} and \cite{scherrer2010should} the  relationship between MSE and MSBR is provided. It is found that, for a given $\nu$ with $\nu(s) > 0, \forall s \in \mathbb{S}$, 
\begin{equation}
\sqrt{\mathrm{MSE}(z)} \leq \frac{\sqrt{C(\nu)}}{1-\gamma}\sqrt{\mathrm{MSBR}(z)},
\end{equation}
where $C(\nu) = \max_{s,s^{\prime}}{\frac{\mathrm{P}(s, s^{\prime})}{\nu(s)}}$.
Another bound which is of considerable importance is the bound on the MSE of the limit point $z^{*}_{\mathrm{TD}}$ of the TD($0$) algorithm provided in \cite{tsitsiklis1997analysis}. It is found that
\begin{equation}
\sqrt{\mathrm{MSE}(z^{*}_{\mathrm{TD}})} \leq \frac{1}{\sqrt{1-\gamma^{2}}}\sqrt{\mathrm{MSE}(z^{\nu})},
\end{equation}
where $z^{\nu} \in \mathbb{R}^{k}$ satisfies $\Phi z^{\nu} = \Pi^{\nu}V^{\pi}$ and $\gamma$ is the discount factor. Table \ref{tab:compalgtable} provides a list of important TD based algorithms along with the associated error objectives. The algorithm complexities are also shown in the table.

Put succinctly, when linear function approximation is applied in an RL setting, the main task can be cast as an optimization problem whose objective function is one of the aforementioned error functions. Typically, almost all the state-of-the-art algorithms employ gradient search technique to solve the minimization problem. In this paper, we apply a gradient-free technique called the \textit{Cross Entropy (CE) method} instead to find the minimum. By `\textit{gradient-free}', we mean the algorithm does not incorporate information on the gradient of the objective function, rather uses the function values themselves. Cross Entropy method is  commonly subsumed within the general class of \textit{Model based search methods} \cite{zlochin2004model}. Other methods in this class are \textit{Model Reference Adaptive Search (MRAS)} \cite{hu2007model}, \textit{Gradient-based Adaptive Stochastic Search for Simulation Optimization (GASSO)} \cite{zhou2014simulation}, \textit{Ant Colony Optimization (ACO)} \cite{dorigo1997ant} and \textit{Estimation of Distribution Algorithms (EDAs)} \cite{muhlenbein1996recombination}. Model based search methods have been applied to the control problem\footnote{The problem here is to find the optimal basis of the MDP.} in \cite{hu2008model} and in basis adaptation\footnote{The basis adaptation problem is to find the best parameters of the basis functions for a given policy.} \cite{menache2005basis}, but this is the first time such a procedure has been applied to the prediction problem. However, due to certain limitations in the original CE method, it cannot be directly applied to the RL setting. In this paper, we have proposed a method to workaround these limitations of the CE method, thereby making it a good choice for the RL setting. Note that any of the aforementioned error functions can be employed, but in this paper, we attempt to minimize MSPBE as it offers the best approximation with less bias to the projection $\Pi^{\nu}V^{\pi}$  for a given policy $\pi$, using a single sample trajectory.
\subsubsection*{Our Contributions}
The \textit{Cross Entropy (CE) method} \cite{rubinstein2013cross, de2005tutorial} is a model based search algorithm to find the global maximum of a given real valued objective function. In this paper, we propose for the first time, an adaptation of this method to the problem of parameter tuning in order to find the best estimates of the value function $V^{\pi}$ for a given policy $\pi$ under the linear function approximation architecture. We propose a multi-timescale stochastic approximation algorithm which minimizes the MSPBE. The algorithm possesses the following attractive features:
\begin{enumerate}
\item
No restriction on the feature set.
\item
The computational complexity is quadratic in the number of features (this is a significant improvement compared to the cubic complexity of the least squares algorithms).
\item
It is competitive with least squares and other state-of-the-art algorithms in terms of accuracy.
\item
It is online with incremental updates.
\item
It gives guaranteed convergence to the global minimum of the MSPBE.
\end{enumerate} 
A noteworthy observation is that since MSPBE is a strongly convex function \cite{dann2014policy}, local and global minima overlap and the fact that CE method finds the global minima as opposed to local minima, unlike gradient search, is not really essential. Nonetheless, in the case of non-linear function approximators, the convexity property does not hold in general and so there may exist multiple local minima in the objective and the gradient search schemes would get stuck in local optima unlike CE based search. We have not explored the non-linear case in this paper. However, our approach can be viewed as a significant first step towards efficiently using model based search for policy evaluation in the RL setting.
\section{Proposed Algorithm: SCE-MSPBEM}
We present in this section our algorithm SCE-MSPBEM, acronym for \textit{Stochastic  Cross Entropy-Mean Squared Projected Bellman Error Minimization} that
minimizes the Mean Squared Projected Bellman Error (MSPBE) by incorporating a multi-timescale stochastic approximation variant of the Cross Entropy (CE) method.
\subsection{Summary of Notation:} Let $\mathbb{I}_{k \times k}$ and $0_{k \times k}$ be the identity matrix and the zero matrix with dimensions $k \times k$ respectively. Let $f_{\theta}(\cdot)$ be the probability density function (\emph{pdf}) parametrized by $\theta$ and $\mathbb{P}_{\theta}$ be its induced probability measure. Let $\mathbb{E}_{\theta}[\cdot]$ be the expectation \emph{w.r.t.} the probability distribution  $f_{\theta}(\cdot)$. We define the $(1-\rho)$-quantile of a real-valued function $\mathcal{H}(\cdot)$ \emph{w.r.t.} the probability distribution $f_{\theta}(\cdot)$ as follows:
\begin{equation}\label{eq:quantile}
\gamma^{\mathcal{H}}_{\rho}(\theta) \triangleq \sup\{l \in \mathbb{R} : \mathbb{P}_{\theta}(\mathcal{H}(\mathbf{x}) \geq l) \geq \rho \}, 
\end{equation}
Also $\lceil a \rceil$ denotes the smallest integer greater than $a$. For $A \subset \mathbb{R}^{m}$, let $I_{A}$ represent the indicator function, \emph{i.e.}, $I_{A}(x)$ is 1 when $x \in A$ and is 0 otherwise. We denote by $\mathbb{Z}_{+}$ the set of non-negative integers. Also we denote by $\mathbb{R}_{+}$ the set of non-negative real numbers. Thus, $0$ is an element of both $\mathbb{Z}_{+}$ and $\mathbb{R}_{+}$. In this section, $\mathbf{x}$ represents a random variable and $x$ a deterministic variable.
\subsection{Background: The CE Method}
To better understand our algorithm, we briefly explicate the original CE method first.
\subsubsection{Objective of CE}
The \textit{Cross Entropy (CE) method} \cite{rubinstein2013cross, hu2009performance, de2005tutorial} solves problems of the following form:
\[\textrm{ Find } \hspace*{4mm} x^{*} \in \argmax_{x \in \mathcal{X} \subset \mathbb{R}^{m}} \mathcal{H}(x),\]
where $\mathcal{H}(\cdot)$ is a multi-modal real-valued function and $\mathcal{X}$ is called the solution space.

The goal of the CE method is to find an optimal ``\textit{model}" or probability distribution over the solution space $\mathcal{X}$ which concentrates on the global maxima of $\mathcal{H}(\cdot)$. The CE method adopts an iterative procedure where at each iteration $t$, a search is conducted on a space of parametrized probability distributions $\{f_{\theta} \vert \theta \in \Theta\}$ on $\mathcal{X}$ over $\mathcal{X}$, where $\Theta$ is the parameter space, to find a distribution parameter $\theta_t$ which reduces the \emph{Kullback-Leibler (KL)} distance from the optimal model. The most commonly used class here is the \emph{exponential family of distributions}.\\
\textbf{Exponential Family of Distributions:} These are denoted as $\mathcal{C} \triangleq$ $\{f_{\theta}(x) = h(x)e^{\theta^{\top}\Gamma(x)-K(\theta)} \mid \theta\in \Theta \subset \mathbb{R}^d\}, \textrm{ where }$ $h:\mathbb{R}^{m} \longrightarrow \mathbb{R}$, $\Gamma:\mathbb{R}^{m} \longrightarrow \mathbb{R}^{d}$ and  $K:\mathbb{R}^{d} \longrightarrow \mathbb{R}$. By rearranging the parameters, we can show that the Gaussian distribution with mean vector $\mu$ and the covariance matrix $\Sigma$ belongs to $\mathcal{C}$. In this case,
\begin{equation} \label{eq:gdist}
f_{\theta}(x) = \frac{1}{\sqrt{(2\pi)^{m}|\Sigma|}}e^{-(x-\mu)^{\top}\Sigma^{-1}(x-\mu)/2},
\end{equation}
and so one may let
${\displaystyle h(x) = \frac{1}{\sqrt{(2\pi)^{m}}}}$, $\Gamma(x) = (x, xx^{\top})^{\top}$ \normalsize and  ${\displaystyle \theta = (\Sigma^{-1} \mu,\hspace*{1mm}-\frac{1}{2}\Sigma^{-1})^{\top}}$\normalsize.\vspace*{1mm}\\
\begin{itemize}
\item[$\circledast$]\textbf{Assumption (A1):} The parameter space $\Theta$ is compact.
\end{itemize}
\subsubsection{CE Method (Ideal Version)}
The CE method aims to find a sequence of model parameters ${\{\theta_t\}}_{t \in \mathbb{Z}^{+}}$, where $\theta_t \in \Theta$ and an increasing sequence of thresholds ${\{\gamma_{t}\}}_{t \in \mathbb{Z}^{+}}$ where $\gamma_t \in \mathbb{R}$, with the property that the support of the model identified using $\theta_{t}$, \emph{i.e.}, $\overline{\{x \vert f_{\theta_t}(x) \neq 0\}}$ is contained in the region $\{x \vert \mathcal{H}(x) \geq \gamma_{t}\}$. By assigning greater weight to higher values of $\mathcal{H}$ at each iteration, the expected behaviour of the probability distribution sequence should improve. The most common choice for $\gamma_{t+1}$ is $\gamma^{\mathcal{H}}_{\rho}(\theta_t)$, the $(1-\rho)$-quantile of $\mathcal{H}(\mathbf{x})$ \emph{w.r.t.} the probability distribution $f_{\theta_{t}}(\cdot)$,  where $\rho \in (0,1)$ is set \emph{a priori} for the algorithm. We take Gaussian distribution as the preferred choice for $f_{\theta}(\cdot)$ in this paper. In this case, the model parameter is $\theta = (\mu, \Sigma)^{\top}$ where $\mu \in \mathbb{R}^{m}$ is the mean vector and $\Sigma \in \mathbb{R}^{m \times m}$ is the covariance matrix.

The CE algorithm is an iterative procedure which starts with an initial value $\theta_0 = (\mu_{0}, \Sigma_0)^{\top}$ of the mean vector and the covariance matrix tuple and at each iteration $t$, a new parameter $\theta_{t+1} = (\mu_{t+1}, \Sigma_{t+1})^{\top}$ is derived from the previous value $\theta_t$ as follows:
%\begin{multline}\label{eq:opt1}
%\theta_{t+1} = \argmax_{\theta \in \Theta}\\\mathbb{E}_{\theta_{t}}\left\lbrace \mathcal{S}(\mathcal{H}(X))I_{\{\mathcal{H}(X) \geq \gamma_{t+1}\}}\left[\begin{aligned} -\frac{1}{2}\ln{((2\pi)^{m}\vert \Sigma \vert)}  + \\ -\frac{1}{2}(X-\mu)^{\top}\Sigma^{-1}(X-\mu)\end{aligned}\right]\right\rbrace
%\end{multline}
\begin{equation}\label{eq:opt1}
%\begin{aligned}
\theta_{t+1} = \argmax_{\theta \in \Theta}\mathbb{E}_{\theta_{t}}\left\lbrace S(\mathcal{H}(\mathbf{x}))I_{\{\mathcal{H}(\mathbf{x}) \geq \gamma_{t+1}\}}\log{f_\theta(\mathbf{x})}\right\rbrace,
%\end{aligned}
\end{equation}
$\mathrm{where} \hspace{1mm} S  \textrm{ is } \textrm{positive and strictly monontone.}\hspace{1mm}$\\
If the gradient \emph{w.r.t.} $\theta$ of the objective function in (\ref{eq:opt1}) is equated to 0 and using (\ref{eq:gdist}) for $f_\theta(\cdot)$, we obtain
\begin{gather} 
\mu_{t+1} = \frac{\mathbb{E}_{\theta_{t}}\left[\mathbf{g_{1}}\bm{(}\mathcal{H}(\mathbf{x}), \mathbf{x}, \gamma_{t+1}\bm{)}\right]}{\mathbb{E}_{\theta_{t}}\left[\mathbf{g_{0}}\bm{(}\mathcal{H}(\mathbf{x}), \gamma_{t+1}\bm{)}\right]} \triangleq \Upsilon_{1}(\mathcal{H}(\cdot), \theta_{t}, \gamma_{t+1}),\hspace{10mm} \label{eq:sigmaideal1}\\
\Sigma_{t+1} = \frac{\mathbb{E}_{\theta_{t}}\left[\mathbf{g_{2}}\bm{(}\mathcal{H}(\mathbf{x}), \mathbf{x}, \gamma_{t+1}, \mu_{t+1}\bm{)}\right]}{\mathbb{E}_{\theta_{t}}\left[\mathbf{g_{0}}\bm{(}\mathcal{H}(\mathbf{x}), \gamma_{t+1}\bm{)}\right]} \triangleq \Upsilon_{2}(\mathcal{H}(\cdot), \theta_{t}, \gamma_{t+1}, \mu_{t+1}).\hspace{1mm} \label{eq:sigmaideal2}
\end{gather}
\begin{eqnarray}
&\hspace*{-17mm}\textrm{ where }\hspace*{4mm}\mathbf{g_{0}}\bm{(}\mathcal{H}(x), \gamma\bm{)} \triangleq S(\mathcal{H}(x))I_{\bm{\{}\mathcal{H}(x) \geq \gamma\bm{\}}}, \hspace*{40mm}\\ 
&\mathbf{g_{1}}\bm{(}\mathcal{H}(x), x, \gamma\bm{)} \triangleq S(\mathcal{H}(x))I_{\bm{\{}\mathcal{H}(x) \geq \gamma\bm{\}}}x, \hspace*{31mm}\\
&\hspace*{0mm}\mathbf{g_{2}}\bm{(}\mathcal{H}(x), x, \gamma, \mu\bm{)} \triangleq S(\mathcal{H}(x))I_{\bm{\{}\mathcal{H}(x) \geq \gamma\bm{\}}}(x-\mu)(x-\mu)^{\top}.
\end{eqnarray}
\begin{remark}The function $S(\cdot)$ in (\ref{eq:opt1}) is positive and strictly monotone and is used to account for the cases when the objective function $\mathcal{H}(x)$ takes negative values for some $x$. One common choice is $S(x) = \exp(rx)$ where $r \in \mathbb{R}$ is chosen appropriately.
\end{remark}
\subsubsection{CE Method (Monte-Carlo Version)}
It is hard in general to evaluate $\mathbb{E}_{\theta_t}[\cdot]$ and $\gamma_t$, so  the stochastic counterparts of the equations (\ref{eq:sigmaideal1}) and (\ref{eq:sigmaideal2}) are used instead in the CE algorithm. This gives rise to the Monte-Carlo version of the CE method. In this stochastic version, the algorithm generates a sequence $\{\bar{\theta}_{t} = (\bar{\mu}_{t}, \bar{\Sigma}_{t})^{\top}\}_{t \in \mathbb{Z}_{+}}$, where at each iteration $t$, $N_{t}$ samples $\Lambda_{t} = \{\mathbf{x}_1, \mathbf{x}_2, \dots, \mathbf{x}_{N_t}\}$ are picked using the distribution $f_{\bar{\theta}_{t}}$ and the estimate of $\gamma_{t+1}$ is obtained as follows: $\bar{\gamma}_{t+1}=\mathcal{H}_{(\lceil(1-\rho)N_t\rceil)}$ where $\mathcal{H}_{(i)}$ is the $i$th-order statistic of $\{\mathcal{H}(\mathbf{x}_i)\}_{i=1}^{N_t}$. The estimate $\bar{\theta}_{t+1}=(\bar{\mu}_{t+1}, \bar{\Sigma}_{t+1})^{\top}$ of the model parameters $\theta_{t+1} = (\mu_{t+1}, \Sigma_{t+1})^{\top}$ is obtained as
\begin{eqnarray}
\bar{\mu}_{t+1} = \frac{\frac{1}{N_t}\sum_{i=1}^{N_t}\mathbf{g_{1}}\bm{(}\mathcal{H}(\mathbf{x}_{i}), \mathbf{x}_{i}, \bar{\gamma}_{t+1}\bm{)}}{\frac{1}{N_t}\sum_{i=1}^{N_t}\mathbf{g_{0}}\bm{(}\mathcal{H}(\mathbf{x}_{i}), \bar{\gamma}_{t+1}\bm{)}} \triangleq \bar{\Upsilon}_{1}(\mathcal{H}(\cdot), \bar{\theta}_{t}, \bar{\gamma}_{t+1}),\hspace{10mm} \label{eqn:mcvers1}\\
\bar{\Sigma}_{t+1} = \frac{\frac{1}{N_t}\sum_{i=1}^{N_t}\mathbf{g_{2}}\bm{(}\mathcal{H}(\mathbf{x}_{i}), \mathbf{x}_{i}, \bar{\gamma}_{t+1}, \bar{\mu}_{t+1}\bm{)}}{\frac{1}{N_t}\sum_{i=1}^{N_t}\mathbf{g_{0}}\bm{(}\mathcal{H}(\mathbf{x}_{i}), \bar{\gamma}_{t+1}\bm{)}} \triangleq \bar{\Upsilon}_{2}(\mathcal{H}(\cdot), \bar{\theta}_{t}, \bar{\gamma}_{t+1}, \bar{\mu}_{t+1}).\label{eqn:mcvers2}
\end{eqnarray}
An \textit{observation allocation rule} $\{N_{t} \in \mathbb{Z}_{+}\}_{t \in \mathbb{Z}_{+}}$ is used to determine the sample size. The Monte-Carlo version of the CE method is described in Algorithm \ref{algo:cemc}.\\
\begin{algorithm}[H]
%	\SetLine
	\textbf{Step 0:} Choose an initial \emph{p.d.f.} $f_{\bar{\theta}_0}(\cdot)$ on $\mathcal{X}$ where $\bar{\theta}_{0} = (\bar{\mu}_{0}, \bar{\Sigma}_{0})^{\top}$ and fix an $\epsilon > 0$\;
	\textbf{Step 1:} \textbf{[Sampling Candidate Solutions]} Randomly sample $N_{t}$  independent and identically distributed solutions $\Lambda_{t}=\{\mathbf{x}_1, \dots, \mathbf{x}_{N_t}\}$ using $f_{\bar{\theta}_t}(\cdot)$.\\
	\textbf{Step 2:} \textbf{[Threshold Evaluation]} Calculate the sample $(1-\rho)$-quantile $\bar{\gamma}_{t+1}=\mathcal{H}_{(\lceil (1-\rho)N_{t}\rceil)}$, where $\mathcal{H}_{(i)}$ is the $i$th-order statistic of the sequence $\{\mathcal{H}(\mathbf{x}_i)\}_{i=1}^{N_t}$\;
	\textbf{Step 3:} \textbf{[Threshold Comparison]}\\
	\eIf{$\bar{\gamma}_{t+1} \geq \bar{\gamma}^{*}_t+\epsilon$}{
		$\bar{\gamma}^{*}_{t+1} = \bar{\gamma}_{t+1}$,
	}{
		$\bar{\gamma}^{*}_{t+1} = \bar{\gamma}^{*}_{t}$.
	}
	\textbf{Step 3:}  \textbf{[Model Parameter Update]}\\
	$\bar{\theta}_{t+1}  =  (\bar{\mu}_{t+1}, \bar{\Sigma}_{t+1})^{\top} = \Big(\bar{\Upsilon}_{1}(\mathcal{H}(\cdot), \bar{\theta}_{t}, \bar{\gamma}^{*}_{t+1}), \bar{\Upsilon}_{2}(\mathcal{H}(\cdot), \bar{\theta}_{t}, \bar{\gamma}^{*}_{t+1}, \bar{\mu}_{t+1})\Big)^{\top}$.\\
	\textbf{Step 4:}  If the stopping rule is satisfied, then return $\bar{\theta}_{t+1}$, else set $t:=t+1$ and go to Step 1.
	\caption{The Monte-Carlo CE Algorithm\label{algo:cemc}}
\end{algorithm}
\begin{remark}The CE method is also applied in stochastic settings for which the objective function is given by $\mathcal{H}(x) = \mathbb{E}_{\mathbf{y}}\left[G(x, \mathbf{y})\right]$, where $\mathbf{y} \in \mathcal{Y}$ and $\mathbb{E}_{\mathbf{y}}[\cdot]$ is the expectation w.r.t. a probability distribution on $\mathcal{Y}$. Since the objective function is expressed in terms of expectation, it might be hard in some scenarios to obtain the true values of the objective function. In such cases, estimates of the objective function are used instead. The CE method is shown  to have global convergence properties in such cases too.
\end{remark}
\subsubsection{Limitations of the CE Method}
A significant limitation of the CE method is its dependence on the sample size $N_{t}$ used in Step 1 of Algorithm \ref{algo:cemc}. One does not know \textit{a priori the best value for the sample size $N_{t}$}. Higher values of $N_{t}$ while resulting in higher accuracy also require more computational resources. One often needs to apply brute force in order to obtain a good choice of $N_{t}$. Also as $m$, the dimension of the solution space, takes large values, more samples are required for better accuracy, making \textit{$N_{t}$ large} as well. This makes \textit{finding the $i$th-order statistic $\mathcal{H}_{(i)}$ in Step 2 harder}. Note that the order statistic $\mathcal{H}_{(i)}$ is obtained by sorting the list $\{\mathcal{H}(\mathbf{x}_1), \mathcal{H}(\mathbf{x}_2), \dots \mathcal{H}(\mathbf{x}_{N_{t}})\}$. The computational effort required in that case is $O(N_{t}\log{N_{t}})$ which in most cases is inadmissible. The other major bottleneck is the \textit{space required to store the samples $\Lambda_{t}$}. In situations when $m$ and $N_{t}$ are large, the storage requirement is a major concern. 

The CE method is also \textit{offline in nature.} This means that the function values $\{\mathcal{H}(\mathbf{x}_1), \dots, \mathcal{H}(\mathbf{x}_{N_{t}})\}$ of the sample set $\Lambda_{t} = \{\mathbf{x}_1, \dots, \mathbf{x}_{N_t}\}$ should be available before the model parameters can be updated in Step 4 of Algorithm \ref{algo:cemc}. So when applied in the prediction problem of approximating the value function $V^{\pi}$ for a given policy $\pi$ using the linear architecture defined in (\ref{eqn:phieq}) by minimizing the error function MSPBE$(\cdot)$, we require the estimates of $\{\mathrm{MSPBE}(\mathbf{x}_1), \dots, \mathrm{MSPBE}(\mathbf{x}_{N_{t}})\}$. This means that a sufficiently long traversal along the given sample trajectory has to be conducted to obtain the estimates before initiating the CE method. This does not make the CE method amenable to online implementations in RL, where the value function estimations are performed in real-time after each observation.

In this paper, we resolve all these shortcomings of the CE method by remodelling the same in the stochastic approximation framework and thus \textit{replacing the sample averaging operation in equations (\ref{eqn:mcvers1}) and (\ref{eqn:mcvers2}) with a bootstrapping approach} where we continuously improve the estimates based on past observations. Each successive estimate is obtained as a function of the previous estimate and a noise term. We replace the $(1-\rho)$-quantile estimation using the order statistic method in Step 2 of Algorithm \ref{algo:cemc} with a stochastic recursion which serves the same purpose, but more efficiently. The model parameter update in step 3 is also replaced with a stochastic recursion. We also bring in additional modifications to the CE method to adapt to a Markov Reward Process (MRP) framework and thus obtain an \textit{online version} of CE where the computational requirements are quadratic in the size of the feature set for each observation. To fit the online nature, we have developed an expression for the objective function MSPBE, where we are able to separate its deterministic and non-deterministic components. This separation is critical since the original expression of MSPBE is convoluted with the solution vector and the expectation terms and hence is unrealizable. The separation further helps to develop a stochastic recursion for estimating MSPBE. Finally, in this paper, we provide a proof of convergence of our algorithm using an ODE based analysis.

\subsection{Proposed Algorithm (SCE-MSPBEM) }
\textbf{Notation: } In this section, $\mathbf{z}$ represents a random variable and $z$ a deterministic variable.\\
SCE-MSPBEM is an \textit{ algorithm to approximate the value function $V^{\pi}$ (for a given policy $\pi$) with linear  function approximation, where the optimization is performed using a multi-timescale stochastic approximation variant of the CE algorithm}. Since the CE method is a maximization algorithm, the objective function in the optimization problem here is the negative of MSPBE. Thus,
\begin{eqnarray}\label{eqn:mspbeobj}
z^{*} = \argmin_{z \in \mathcal{Z} \subset \mathbb{R}^{k}} \mathrm{MSPBE}(z) = \argmax_{z \in \mathcal{Z} \subset \mathbb{R}^{k}}  \mathcal{J}(z),\\
\hspace{20mm} \mathrm{ where } \hspace{2mm} \mathcal{J} = -\mathrm{MSPBE}. \nonumber
\end{eqnarray}
Here $\mathcal{Z}$ is the solution space, \emph{i.e.}, the space of parameter values of the function approximator. We also define $\mathcal{J}^{*} \triangleq \mathcal{J}(z^{*})$.
\begin{remark}Since $\exists z  \in \mathcal{Z}$ such that $\Phi z = \Pi^{\nu}T^{\pi}\Phi z$, the value of $\mathcal{J}^{*}$ is  $0$.
\end{remark}
\begin{itemize}
\item[$\circledast$]\textbf{Assumption (A2):} The solution space $\mathcal{Z}$ is compact, \emph{i.e.}, it is closed and bounded.
\end{itemize}
In \cite{sutton2009fast} a compact expression for MSPBE is given as follows: 
\begin{equation}
\mathrm{MSPBE}(z) = (\Phi^{\top} D^{\nu} (T_{\pi} V_{z}-V_{z}))^{\top}(\Phi^{\top}D^{\nu}\Phi)^{-1}
(\Phi^{\top}D^{\nu}(T_{\pi} V_{z}-V_{z})).
\end{equation}
Using the fact that $V_{z} = \Phi z$, the expression $\Phi^{\top}D^{\nu} (T_{\pi}V_{z}-V_{z})$ can be rewritten as 
\vspace*{-2mm}
\begin{multline}\label{eqn:comp1}
\Phi^{\top}D^{\nu}(T_{\pi}V_{z}-V_{z}) = \mathbb{E}\left[\mathbb{E}\left[\phi_t(\mathbf{r}_{t}+
\gamma z^{\top}\phi^{\prime}_{t} - z^{\top}\phi_{t})| \mathbf{s}_t\right]\right]
\\ = \mathbb{E}\left[\mathbb{E}\left[\phi_t \mathbf{r}_{t} \vert \mathbf{s}_t\right]\right] +   \mathbb{E}\left[\mathbb{E}\left[\phi_t(\gamma\phi^{\prime}_{t} -\phi_t)^{\top} \vert \mathbf{s}_t\right]\right]z, \hspace*{3mm} \mathrm{where} \hspace{1mm} \phi_t \triangleq \phi(\mathbf{s}_t) \textrm{ and } \phi^{\prime}_t \triangleq \phi(\mathbf{s}^{\prime}_t).
\hspace*{5mm}
\end{multline}
\begin{equation}\label{eqn:comp2}
\hspace*{-45mm}\textrm{Also,} \hspace*{2cm} \Phi^{\top}D^{\nu}\Phi = \mathbb{E}\left[\phi_t \phi_{t}^{\top}\right].\hspace*{5cm}
\end{equation}
Putting all together we get,
\vspace*{-8mm}\\
\begin{multline}
\mathrm{MSPBE}(z) = \left(\mathbb{E}\left[\mathbb{E}\left[\phi_t \mathbf{r}_{t} \vert \mathbf{s}_t\right]\right] +   \mathbb{E}\left[\mathbb{E}\left[\phi_t(\gamma\phi^{\prime}_{t} -\phi_t)^{\top} \vert \mathbf{s}_t\right]\right]z\right)^{\top}(\mathbb{E}\left[\phi_t \phi_{t}^{\top}\right])^{-1} \\ \left(\mathbb{E}\left[\mathbb{E}\left[\phi_t \mathbf{r}_{t} \vert \mathbf{s}_t\right]\right] +   \mathbb{E}\left[\mathbb{E}\left[\phi_t(\gamma\phi^{\prime}_{t} -\phi_t)^{\top} \vert \mathbf{s}_t\right]\right]z\right) .
\end{multline}
\begin{equation}\label{eq:stdetobj}
= \left(\omega^{(0)}_{*} + \omega^{(1)}_{*} z\right)^{\top} \omega^{(2)}_{*} \left(\omega^{(0)}_{*} + \omega^{(1)}_{*} z\right),
\end{equation}
where $\omega^{(0)}_{*} \triangleq \mathbb{E}\left[\mathbb{E}\left[\phi_t \mathbf{r}_{t} \vert \mathbf{s}_t\right]\right]$, $\omega^{(1)}_{*} \triangleq  \mathbb{E}\left[\mathbb{E}\left[\phi_t(\gamma\phi^{\prime}_{t} -\phi_t)^{\top} \vert \mathbf{s}_t\right]\right]$ and $\omega^{(2)}_{*} \triangleq (\mathbb{E}\left[\phi_t \phi_{t}^{\top}\right])^{-1}$.

This is a quadratic function on $z$. Note that, in the above expression, the parameter vector $z$ and the stochastic component involving $\mathbb{E}[\cdot]$ are decoupled. Hence the stochastic component can be estimated independent of the parameter vector $z$. 

The goal of this paper is to adapt CE method into a MRP setting in an online fashion, where we solve the prediction problem which is a continuous stochastic optimization problem. The important tool we employ here to achieve this is the stochastic approximation framework. Here we take a slight digression to explain stochastic approximation algorithms.

\textbf{Stochastic approximation algorithms} \cite{borkar2008stochastic,kushner2012stochastic,robbins1951stochastic} are a natural way of utilizing prior information. It does so by discounted averaging of the prior information and are usually expressed as recursive equations of the following form:
\begin{equation}\label{eq:strec}
\mathbf{z}_{j+1} = \mathbf{z}_{j} + \alpha_{j+1}\Delta \mathbf{z}_{j+1},
\end{equation}
where $\Delta \mathbf{z}_{j+1} = q(\mathbf{z}_{j}) + b_{j} + \mathbb{M}_{j+1}$ is the \textit{increment term}, $q(\cdot)$ is a Lipschitz continuous function, $b_{j}$ is the \textit{bias term} with $b_{j} \rightarrow 0$ and $\{\mathbb{M}_{j}\}$ is a \textit{martingale difference noise sequence}, \emph{i.e.}, $\mathbb{M}_{j}$ is $\mathcal{F}_{j}$-measurable and integrable and $\mathbb{E}[\mathbb{M}_{j+1} \vert \mathcal{F}_{j}] = 0, \forall j$. Here $\{\mathcal{F}_{j}\}_{j \in \mathbb{Z}_{+}}$ is a filtration, where the $\sigma$-field $\mathcal{F}_{j} = \sigma(\mathbf{z}_i, \mathbb{M}_{i}, 1 \leq i \leq j, \mathbf{z}_{0})$. The learning rate $\alpha_{j}$ satisfies $\Sigma_{j} \alpha_{j} = \infty$, $\Sigma_{j} \alpha_{j}^{2} < \infty$.

We have the following well known result from \cite{borkar2008stochastic} regarding the limiting behaviour of the stochastic recursion (\ref{eq:strec}):
\begin{theorem}
Assume $\sup_{j} \Vert z_{j} \Vert < \infty$, $\mathbb{E}[\Vert \mathbb{M}_{j+1} \Vert^{2} \vert \mathcal{F}_{j}] \leq K(1+\Vert \mathbf{z}_{j} \Vert^{2}), \forall j$ and $q(\cdot)$ is Lipschitz continuous. Then the iterates
$\mathbf{z}_{j}$ converge almost surely to the compact connected internally chain transitive invariant set of the ODE:
\begin{equation}\label{eq:dtode}
\dot{z}(t) =  q(z(t)), t \in \mathbb{R}_{+}.
\end{equation}
\end{theorem}
Put succinctly, the above theorem establishes an equivalence between the asymptotic behaviour of the iterates $\{\mathbf{z}_{j}\}$ and the deterministic ODE (\ref{eq:dtode}). In most practical cases, the ODE have a unique globally asymptotically stable equilibrium at an arbitrary point $z^{*}$. It will then follow from the theorem that $\mathbf{z}_{j} \rightarrow z^{*}$ \emph{a.s}. However, in some cases, the ODE can have multiple isolated stable equilibria. In such cases, the convergence of $\{\mathbf{z}_{j}\}$ to one of these equilibria is guaranteed, however the limit point would depend on the noise and the initial value.

A relevant extension of the stochastic approximation algorithms is the multi-timescale variant. Here there will be multiple stochastic recursions of the kind (\ref{eq:strec}), each with possibly different learning rates. The learning rates defines the timescale of the particular recursion. So different learning rates imply different timescales. If the increment terms are well-behaved and the learning rates properly related (defined in Chapter 6 of \cite{borkar2008stochastic}), then the chain of recursions exhibit a well-defined asymptotic behaviour. See Chapter 6 \cite{borkar2008stochastic} for more details.

Now digression aside, note that the important stochastic variables of the ideal CE method are $\mathcal{H}, \gamma_k$, $\mu_k$, $\Sigma_k$ and $\theta_k$. Here, the objective function $\mathcal{H} = \mathcal{J}$. In our approach, we track these variables independently using stochastic recursions of the kind (\ref{eq:strec}). Thus we model our algorithm as a multi-timescale stochastic approximation algorithm  which tracks the ideal CE method. Note that the stochastic recursion is uniquely identified by their increment term, their initial value and the learning rate. We consider here these recursions in great detail.\vspace*{2mm}\\
\textbf{1. Tracking the Objective Function $\mathcal{J}(\cdot)$: } Recall that the goal of the paper is to develop an online and incremental prediction algorithm. This implies that algorithm has to learn from a given sample trajectory using an incremental approach  with a single traversal of the trajectory. The algorithm SCE-MSPBEM   operates online on a single sample trajectory $\{(\mathbf{s}_{t}, \mathbf{r}_{t}, \mathbf{s}^{\prime}_{t})\}_{t=0}^{\infty}$, where $\mathbf{s}_{t} \sim \nu(\cdot)$, $\mathbf{s}^{\prime}_{t} \sim \mathrm{P}^{\pi}(\mathbf{s}_{t}, \cdot)$ and $\mathbf{r}_{t} = \mathrm{R}(\mathbf{s}_{t}, \pi(\mathbf{s}_{t}), \mathbf{s}^{\prime}_{t})$.
\begin{itemize}
\item[$\circledast$] \textbf{Assumption (A3):}
 For the given trajectory $\{(\mathbf{s}_{t}, \mathbf{r}_{t}, \mathbf{s}^{\prime}_{t})\}_{t=0}^{\infty}$, let $\phi_t, \phi^{\prime}_{t}$, and $\mathbf{r}_{t}$ have uniformly bounded second moments. Also, $\mathbb{E}\left[\phi_{t}\phi_{t}^{\top}\right]$ is non-singular.
\end{itemize}
In the expression (\ref{eq:stdetobj}) for the objective function $\mathcal{J}(\cdot)$, we have isolated the stochastic and deterministic part. The stochastic part can be identified by the tuple $\omega_{*} \triangleq (\omega^{(0)}_{*}, \omega^{(1)}_{*}, \omega^{(2)}_{*})^{\top}$. So if we can find ways to track $\omega_{*}$, then it implies we could track $\mathcal{J}(\cdot)$. This is the line of thought we follow here. In our algorithm, we track $\omega_{*}$ using the time dependent variable $\omega_{t} \triangleq (\omega^{(0)}_{t}, \omega^{(1)}_{t}, \omega^{(2)}_{t})^{\top}$, where $\omega^{(0)}_{t} \in \mathbb{R}^{k}$, $\omega^{(1)}_{t} \in \mathbb{R}^{k \times k}$ and $\omega^{(2)}_{t} \in \mathbb{R}^{k \times k}$. Here $\omega^{(i)}_{t}$ independently tracks $\omega^{(i)}_{*}$, $1 \leq i \leq 3$. Note that tracking implies $\lim_{t \rightarrow \infty}\omega^{(i)}_{t} = \omega^{(i)}_{*}$, $1 \leq i \leq 3$. The increment term $\Delta \omega_{t} \triangleq (\omega^{(0)}_{t}, \omega^{(1)}_{t}, \omega^{(2)}_{t})^{\top}$ used for this recursion is defined as follows:
\begin{equation}\label{eq:omginc}
\hspace*{20mm}\left.
\begin{aligned}
\bigtriangleup{\omega}^{(0)}_{t}\hspace*{2mm} =& \hspace*{2mm}\mathbf{r}_{t}\phi_{t}-{\omega}^{(0)}_{t},\\
\bigtriangleup{\omega}^{(1)}_{t} \hspace*{2mm}=& \hspace*{2mm}\phi_{t}(\gamma\phi^{\prime}_{t}-\phi_{t})^{\top}-{\omega}^{(1)}_{t}, \\
\bigtriangleup{\omega}^{(2)}_{t}\hspace*{2mm} =& \hspace*{2mm}\mathbb{I}_{k \times k} -  \phi_{t} \phi_{t}^{\top}{\omega}^{(2)}_{t},
\end{aligned}
\hspace*{60mm}\right\}
\end{equation}
where $\phi_t \triangleq \phi(\mathbf{s}_{t})$ and $\phi^{\prime}_{t} \triangleq \phi(\mathbf{s}^{\prime}_{t})$. Now we define a new function $\bar{\mathcal{J}}(\omega_{t}, z) \triangleq -\left({\omega}^{(0)}_{t} + {\omega}^{(1)}_{t}z\right)^{\top}{\omega}^{(2)}_{t}\left({\omega}^{(0)}_{t} + {\omega}^{(1)}_{t}z\right)$. Note that this is the same expression as (\ref{eq:stdetobj}) except for $\omega_{t}$ replacing $\omega_{*}$. Since $\omega_{t}$ tracks $\omega_{*}$, it is easily verifiable that $\bar{\mathcal{J}}(\omega_{t}, z)$ indeed tracks $\mathcal{J}(z)$ for a given $z \in \mathcal{Z}$.

The stochastic recursions which track $\omega_{*}$ and the objective function $\mathcal{J}(\cdot)$ are defined in (\ref{eqn:alomgupd}) and (\ref{eqn:jval}) respectively. A rigorous analysis of the above stochastic recursion is provided in lemma \ref{lemma:lm1}. There we also find that the initial value $\omega_{0}$ is irrelevant.\vspace*{2mm}\\
\textbf{2. Tracking $\gamma_{\rho}(\mathcal{J}, \theta)$: } Here we are faced two difficult situations: $(i)$ the true objective function $\mathcal{J}$ is unavailable and $(ii)$ we have to find a stochastic recursion which tracks $\gamma_{\rho}(\mathcal{J}, \theta)$ for a given distribution parameter $\theta$. To solve $(i)$ we use whatever is available, \emph{i.e.} $\bar{\mathcal{J}}(\omega_{t}, \cdot)$ which is the best available estimate of the true function $\mathcal{J}(\cdot)$ at time $t$. In other words, we bootstrap. Now to address the second part we make use of the following lemma from \cite{homem2007study}. The lemma provides a characterization of the $(1-\rho)$-quantile of a given real-valued function $H$ \emph{w.r.t.} to a given probability distribution function $f_{\theta}$.
\begin{lemma}\label{lma:qnopt}
The $(1-\rho)$-quantile of a bounded real valued function $H(\cdot)$   $\Big(\textrm{with } H(x) \in [H_l, H_u]\Big)$ w.r.t the probability density function $f_{\theta}(\cdot)$ is reformulated as an optimization problem 
\begin{eqnarray}\label{eqn:optprob}
\gamma_{\rho}(H, \theta) = \argmin_{y \in [H_l, H_u]}\mathbb{E}_{\theta}\left[\psi(H(\mathbf{x}), y)\right], \end{eqnarray}
where   $\mathbf{x} \sim f_{\theta}(\cdot)$, $\psi(H(x), y) = (1-\rho)(H(x)-y)I_{\{H(x) \geq y\}}+\rho(y-H(x))I_{\{H(x) \leq y \}}$ and $\mathbb{E}_{\theta}[\cdot]$ is the expectation w.r.t. the p.d.f. $f_{\theta}(\cdot)$.
\end{lemma}
In this paper, we employ the time-dependent variable $\gamma_{t}$ to track $\gamma_{\rho}(\mathcal{J}, \cdot)$. The increment term in the recursion is the subdifferential $\nabla_{y} \psi$. This is because $\psi$ is non-differentiable as it follows from its definition from the above lemma. However subdifferential exists for $\psi$. Hence we utilize it to solve the optimization problem (\ref{eqn:optprob}) defined  in lemma \ref{lma:qnopt}. Here, we define an increment function (contrary to an increment term) and is defined as follows:
\begin{equation}
\begin{aligned}
\Delta \gamma_{t+1}(z) = -(1-\rho)I_{\{\bar{\mathcal{J}}(\omega_{t}, z) \geq \gamma_t\}}+\rho I_{\{\bar{\mathcal{J}}(\omega_{t}, z) \leq \gamma_t\}}
\end{aligned}
\end{equation}
The stochastic recursion which tracks $\gamma_{\rho}(\mathcal{J}, \cdot)$ is given in (\ref{eqn:algamma}). A deeper analysis of the recursion (\ref{eqn:algamma}) is provided in lemma \ref{lmn:gammacov}. In the analysis we also find that the initial value $\gamma_0$ is irrelevant.\vspace*{2mm}\\
\textbf{3. Tracking $\Upsilon_1$ and $\Upsilon_2$}: In the ideal CE method, for a given $\theta_t$, note that $\Upsilon_1(\theta_t, \dots)$ and $\Upsilon_2(\theta_t, \dots)$ form the subsequent model parameter $\theta_{t+1}$. In our algorithm, we completely avoid the sample averaging technique employed in the Monte-Carlo version. Instead, we follow the stochastic approximation recursion to track the above quantities. Two time-dependent variables $\xi^{(1)}_t$ and $\xi^{(2)}_t$ are employed to track $\Upsilon_1$ and $\Upsilon_2$ respectively. The increment functions used by their respective recursions are defined as follows:
\begin{eqnarray}
\Delta \xi^{(0)}_{t+1}(z)=&\mathbf{g_{1}}(\bar{\mathcal{J}}(\omega_{t}, z), z, \gamma_{t}) - \xi^{(0)}_{t}\mathbf{g_{0}}(\bar{\mathcal{J}}(\omega_t, z), \gamma_{t}), \label{eq:xi0inc}\\
\Delta \xi^{(1)}_{t+1}(z)=&\mathbf{g_{2}}(\bar{\mathcal{J}}(\omega_t, z), z, \gamma_t, \xi^{(0)}_{t}) - \xi^{(1)}_{t}\mathbf{g_{0}}(\bar{\mathcal{J}}(\omega_t, z), \gamma_{t}). \label{eq:xi1inc}
\end{eqnarray}
The recursive equations which track $\Upsilon_1$ and $\Upsilon_2$ are defined in (\ref{eqn:xi0}) and (\ref{eqn:xi1}) respectively. The analysis of these recursions is provided in lemma \ref{lmn:xiconv}. In this case  also, the initial values are irrelevant.\vspace*{2mm}\\
\textbf{4. Model Parameter Update: } In the ideal version of CE, note that given $\theta_t$, we have $\theta_{t+1} = (\Upsilon_1(\mathcal{H}(\cdot), \theta_t, \dots), \Upsilon_2(\mathcal{H}(\cdot), \theta_t, \dots))^{\top}$. This is a discrete change from $\theta_t$ to $\theta_{t+1}$. But in our algorithm, we adopt a smooth update of the model parameters. The recursion is defined in equation (\ref{eq:thupd}). We prove in theorem \ref{thm:main} that the above approach indeed provide an optimal solution to the optimization problem defined in (\ref{eqn:mspbeobj}).\vspace*{2mm}\\
\textbf{5. Learning Rates and Timescales: }The algorithm uses two learning rates $\alpha_{t}$ and $\beta_{t}$ which are deterministic, positive, nonincreasing and satisfy the following conditions:
\begin{equation}\label{eqn:learnrt}
\sum_{t=1}^{\infty}\alpha_{t} = \sum_{t=1}^{\infty}\beta_{t} = \infty, \hspace{8mm} 
\sum_{t=1}^{\infty}\left(\alpha_{t}^{2}+\beta_{t}^{2}\right) < \infty, \hspace{8mm}
\lim_{t \rightarrow \infty}\frac{\alpha_t}{\beta_t} = 0.
\end{equation}
In a multi-timescale stochastic approximation setting, it is important to understand the difference between timescale and learning rate. The timescale of a stochastic recursion is defined by its learning rate (also referred as step-size). Note that from the conditions imposed on the learning rates $\alpha_t$ and $\beta_t$ in (\ref{eqn:learnrt}), we have $\frac{\alpha_t}{\beta_t} \rightarrow 0$. So $\alpha_{t}$ decays to $0$  faster than $\beta_{t}$. Hence the timescale obtained from $\beta_t, t \geq 0$ is considered faster as compared to the other. So in a multi-timescale stochastic recursion scenario, the evolution of the recursion controlled by the faster step-sizes (converges faster to $0$)  is slower compared to the recursions controlled by the slower step-sizes. This is because the increments are weighted by their learning rates, \emph{i.e.}, the learning rates control the quantity of change that occurs to the variables when the update is executed. So the faster timescale recursions converge faster compared to its slower counterparts. Infact, when observed from a faster timescale recursion, one can consider the slower timescale recursion to be almost stationary. This attribute of the multi-timescale recursions are very important in the analysis of the algorithm. In the analysis, when studying the asymptotic behaviour of a particular stochastic recursion, we can consider the variables of other recursions which are on slower timescales to be constant. In our algorithm, the recursion of $\omega_t$ and $\theta_t$  proceed along the slowest timescale and so updates of $\omega_{t}$ appear to be quasi-static when viewed from the timescale on which the recursions governed by $\beta_{t}$ proceed. The  recursions of $\gamma_t, \xi^{(0)}_t$ and $\xi^{(1)}_t $ proceed along the faster timescale and hence have a faster convergence rate. The stable behaviour of the algorithm is attributed to the timescale differences obeyed by the various recursions.\vspace*{2mm}\\
\textbf{6. Sample Requirement:} The streamline nature inherent in the stochastic approximation algorithms demands only a single sample per iteration. Infact, we use two samples $\mathbf{z}_{t+1}$ (generated in (\ref{eq:dtmix})) and $\mathbf{z}^{p}_{t+1}$ (generated in (\ref{eqn:algammaold}) whose discussion is deferred for the time being).This is a remarkable improvement, apart from the fact that the algorithm is now online and incremental in the sense that whenever a new state transition $(\mathbf{s}_{t}, \mathbf{r}_{t}, \mathbf{s}^{\prime}_{t})$ is revealed, the algorithm learns from it by evolving the variables involved and directing the model parameter $\theta_t$ towards the degenerate distribution concentrated on the optimum point $z^{*}$.\vspace*{2mm}\\
\textbf{7. Mixture Distribution: }In the algorithm, we use a mixture distribution $\widehat{f}_{\theta_{t}}$  to generate the sample $\mathbf{x}_{t+1}$, where $\widehat{f}_{\theta_t} = (1-\lambda)f_{\theta_t} + \lambda f_{\theta_0}$ with $\lambda \in (0,1)$ the mixing weight. The initial distribution parameter $\theta_0$ is chosen \emph{s.t.} the density function $f_{\theta_0}$ is strictly positive on every point in the solution space $\mathcal{X}$, \emph{i.e.}, $f_{\theta_0}(x) > 0, \forall x \in \mathcal{X}$. {The mixture approach facilitates exploration of the solution space and prevents the iterates from getting stranded in suboptimal solutions.}\\
The SCE-MSPBEM algorithm is formally presented in Algorithm \ref{algo:sce-mspbem}.
\newpage
\scalebox{0.88}{
\begin{minipage}{1.1\linewidth}
\begin{algorithm}[H]
	\KwData{ $\alpha_t, \beta_t, c_t \in (0,1)$, $c_t \rightarrow 0$, $\epsilon_1, \lambda, \rho \in (0, 1)$, \hspace*{1mm} $S(\cdot): \mathbb{R} \rightarrow \mathbb{R}^{+}$;\\}
	\vspace*{1mm}
	\textbf{Initialization:} $\gamma_0 = 0$, $\gamma^{p}_0 = -\infty$, $\theta_0 = (\mu_0, \Sigma_0)^{\top}$, $T_0 = 0$, $\xi^{(0)}_{t} = 0_{k \times 1}$, $\xi^{(1)}_{t} = 0_{k \times k}$, \hspace*{20mm}$\omega^{(0)}_{0} = 0_{k \times 1}, \omega^{(1)}_{0} = 0_{k \times k}, \omega^{(2)}_{0} = 0_{k \times k}, \theta^{p}=NULL$;\\
	\vspace*{1mm}
	\ForEach{ $(\mathbf{s}_{t}, \mathbf{r}_{t}, \mathbf{s}^{\prime}_{t})$ of the trajectory }{
	\vspace*{2mm}
	\begin{equation}\label{eq:dtmix}
	\begin{aligned}
	\widehat{f}_{\theta_{t}} = (1-\lambda)f_{\theta_{t}} + \lambda f_{\theta_0};\\
	\mathbf{z}_{t+1}   \sim \widehat{f}_{\theta_{t}}(\cdot);	
	\end{aligned}
	\end{equation}
	$\bullet$ \textbf{[Objective Function Evaluation]}
	\vspace*{-3mm}
	\begin{eqnarray}
	&\omega_{t+1} = \omega_{t} + \alpha_{t+1}\Delta \omega_{t+1};\label{eqn:alomgupd}\\
	&\bar{\mathcal{J}}(\omega_{t}, \mathbf{z}_{t+1}) = -\left(\omega^{(0)}_{t} + \omega^{(1)}_{t}\mathbf{z}_{t+1}\right)^{\top}\omega^{(2)}_{t}\left(\omega^{(0)}_{t} + \omega^{(1)}_{t}\mathbf{z}_{t+1}\right);\label{eqn:jval}
	\end{eqnarray}
	$\blacktriangleright$ \textbf{[Threshold Evaluation]}
	\vspace*{-10mm}\\
	\begin{equation}\label{eqn:algamma}
	\begin{aligned}
	\gamma_{t+1} = \gamma_{t} -  \beta_{t+1} \Delta \gamma_{t+1}(\mathbf{z}_{t+1});\hspace*{40mm}
	\end{aligned}
	\end{equation}
	$\blacktriangleright$ \textbf{[Tracking $\mu_{t+1}$ and $\Sigma_{t+1}$ of (\ref{eq:sigmaideal1}) and (\ref{eq:sigmaideal2})]}
	\vspace*{-12mm}\\
	\begin{eqnarray} 
	\xi^{(0)}_{t+1} =& \xi^{(0)}_{t}+\beta_{t+1}\Delta \xi^{(0)}_{t+1}(\mathbf{z}_{t+1});\label{eqn:xi0}\hspace*{30mm}\\
	\xi^{(1)}_{t+1} =& \xi^{(1)}_{t}+\beta_{t+1}\Delta \xi^{(1)}_{t+1}(\mathbf{z}_{t+1});\hspace*{30mm}\label{eqn:xi1}
	\end{eqnarray}
	\vspace*{-10mm}\\
	\If{$\theta^{p}$ $\neq$ $NULL$}{
	\begin{equation}\label{eqn:algammaold}
	\left.
	\begin{aligned}
	\mathbf{z}^{p}_{t+1}  &\sim \widehat{f}_{\theta^{p}}(\cdot) \triangleq \lambda f_{\theta_{0}} + (1-\lambda)f_{\theta^{p}};\\
	\gamma^{p}_{t+1} &= \gamma^{p}_{t} -  \beta_{t+1} \Delta \gamma_{t+1}(\mathbf{z}^{p}_{t+1});
	\end{aligned}
	\hspace*{60mm}\right\}
	\end{equation}}
	\vspace*{2mm}
	$\blacktriangleright$ \textbf{[Threshold Comparison]}
	\vspace*{-8mm}\\
\begin{equation}\label{eq:Tt}
\hspace*{-18mm}T_{t+1} = T_{t} +  c\left(I_{\{\gamma_{t+1} > \gamma^{p}_{t+1}\}} - I_{\{\gamma_{t+1} \leq \gamma^{p}_{t+1}\}} - T_{t}\right);\hspace{50mm}
\end{equation}
	$\blacktriangleright$ \textbf{[Updating Model Parameter]}\\
	\vspace*{0mm}
	\eIf{$T_{t+1} > \epsilon_1$}{
	\vspace*{-4mm}
	\begin{equation*}
		\hspace*{-95mm}\theta^{p} = \theta_{t};
	\end{equation*}
	\begin{equation}\label{eq:thupd}
		\theta_{t+1} = \theta_{t} + \alpha_{t+1}\left((\xi^{(0)}_{t}, \xi^{(1)}_{t})^{\top} - \theta_{t}\right); \hspace{47mm}
	\end{equation}
	\begin{equation}\label{eq:gmstarupd}
			\hspace*{-58mm}\gamma^{p}_{t+1} = \gamma_{t};\hspace{4mm} T_{t} = 0; \hspace*{4mm} c = c_{t};
	\end{equation}
	}{
		\hspace*{30mm}$\gamma^{p}_{t+1} = \gamma^{p}_{t}$; \hspace{4mm} $\theta_{t+1} = \theta_{t}$;\\
	}
	$t := t+1$;	
	}	
	\caption{SCE-MSPBEM\label{algo:sce-mspbem}}
\end{algorithm}
\end{minipage}}

\begin{remark}In practice, different stopping criteria can be used. For instance, (a) $t$ reaches an \textit{a priori} fixed limit, (b) the computational resources are exhausted, or (c) the variable $T_{t}$ is unable to cross the $\epsilon_{1}$ threshold for an \textit{a priori} fixed number of iterations.
\end{remark}
\begin{figure}
{\includegraphics[scale=0.63]{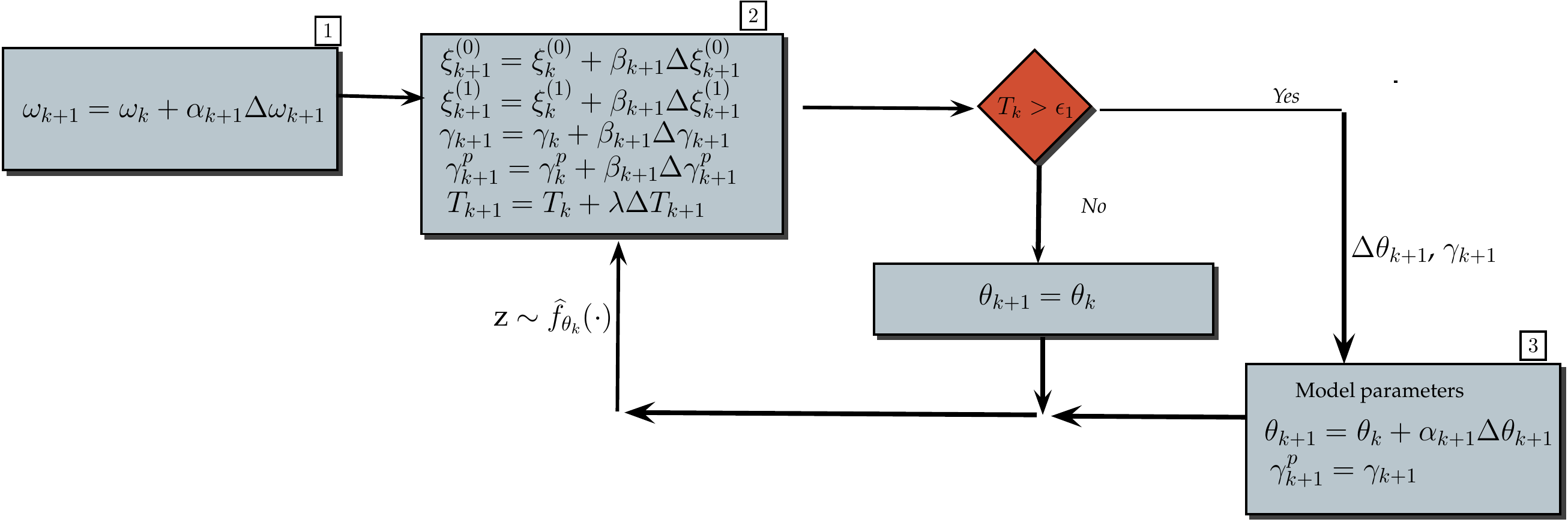}}
\caption{FlowChart representation of the algorithm SCE-MSPBEM.}\label{fig:algovisual}
\end{figure}
The pictorial depiction of the algorithm SCE-MSPBEM is shown in Figure \ref{fig:algovisual}. 

\begin{figure}
{\includegraphics[scale=0.85]{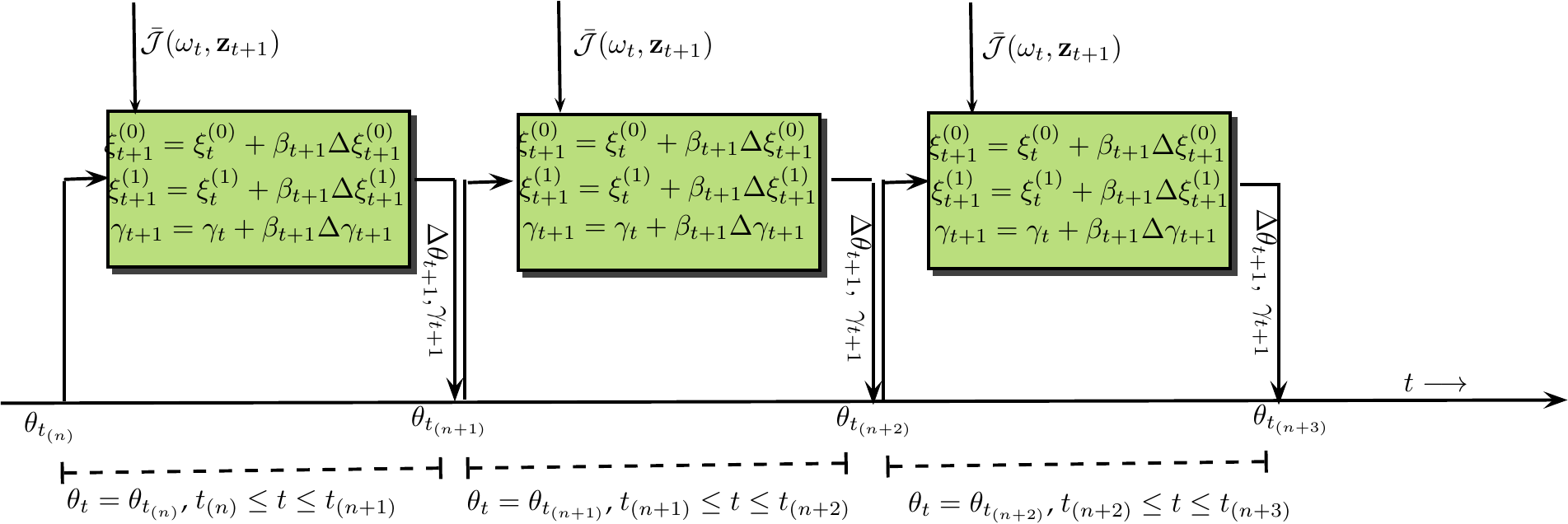}}
\caption{Timeline graph of the algorithm SCE-MSPBEM.}\label{fig:timepic}
\end{figure}

It is important to note that the model parameter $\theta_t$ is not updated at each $t$. Rather it is updated every time $T_t$ hits $\epsilon_1$ where $0 < \epsilon_1 < 1$. So the update of $\theta_t$ only happens along a subsequence $\{t_{(n)}\}_{n \in \mathbb{Z}_{+}}$ of $\{t\}_{t \in \mathbb{Z}_{+}}$. So between $t=t_{(n)}$ and $t=t_{(n+1)}$, the variable $\gamma_{t}$ estimates the quantity $\gamma_{\rho}(\bar{\mathcal{J}}_{\omega_{t}}, , \widehat{\theta}_{t_{(n)}})$. The threshold $\gamma^{p}_{t}$ is also updated during the $\epsilon_1$ crossover in (\ref{eq:gmstarupd}). Thus $\gamma^{p}_{t_{(n)}}$ is the estimate of $(1-\rho)$-quantile \emph{w.r.t.} $\widehat{f}_{\theta_{t_{(n-1)}}}$. Thus $T_t$ in recursion (\ref{eq:Tt}) is a elegant trick to ensure that the estimates $\gamma_t$ eventually become greater than the prior threshold $\gamma^{p}_{t_{(n)}}$, \emph{i.e.}, $\gamma_t > \gamma^{p}_{t_{(n)}}$ for all but finitely many $t$. A timeline map of the algorithm is shown in Figure \ref{fig:timepic}. 

It can be verified as follows that the random variable $T_t$ belongs to $(-1,1)$, $\forall t > 0$. We state it as a proposition here.
\begin{proposition}
For any $T_0 \in (0,1)$, $T_t$ in (\ref{eq:Tt}) belongs to $(-1,1)$, $\forall t > 0$.
\end{proposition}
\begin{Proof}
Assume $T_0 \in (0,1)$. Now the equation (\ref{eq:Tt}) can be rearranged as
\begin{gather*}
T_{t+1} = \left(1-c\right)T_{t} + c(I_{\{\gamma_{t+1} > \gamma^{p}_{t+1}\}} - I_{\{\gamma_{t+1} \leq \gamma^{p}_{t+1}\}}),
\end{gather*}
where $c \in (0,1)$. In the worst case, either $I_{\{\gamma_{t+1} > \gamma^{p}_{t+1}\}} = 1$, $\forall t$ or $I_{\{\gamma_{t+1} \leq \gamma^{p}_{t+1}\}} = 1$, $\forall t$.  Since the two events  $\{\gamma_{t+1} > \gamma^{p}_{t+1}\}$ and $\{\gamma_{t+1} \leq \gamma^{p}_{t+1}\}$ are mutually exclusive, we will only consider the former event $I_{\{\gamma_{t+1} > \gamma^{p}_{t+1}\}} = 1$, $\forall t$. In this case
\begin{equation*}
\begin{aligned}
\lim_{t \rightarrow \infty}T_{t} &= \lim_{t \rightarrow \infty}\left(c + c(1-c) + c(1-c)^{2} + \dots + c(1-c)^{t-1} + (1-c)^{t}T_0\right) \\
&= \lim_{t \rightarrow \infty}\frac{c(1-(1-c)^{t})}{c} + T_0(1-c)^{t} = \lim_{t \rightarrow \infty}(1-(1-c)^{t}) + T_0(1-c)^{t} = 1. \hspace*{5mm}(\because c \in (0,1))
\end{aligned}
\end{equation*}
Similarly for the latter event $I_{\{\gamma_{t+1} \leq \gamma^{p}_{t+1}\}} = 1$, $\forall t$, we can prove that $\lim_{t \rightarrow \infty} T_{t} = -1$.
\end{Proof}

\begin{remark}The recursion in equation (\ref{eqn:algammaold}) is not addressed in the discussion above. The update of $\gamma^{p}_{t}$ in equation (\ref{eq:gmstarupd}) happens along a subsequence $\{t_{(n)}\}_{n \geq 0}$. So $\gamma^{p}_{t_{(n)}}$ is the estimate of $\gamma_{\rho}(\bar{\mathcal{J}}_{\omega_{t_{(n)}}}, \theta_{t_{(n-1)}})$, where $\bar{\mathcal{J}}_{\omega_{t_{(n)}}}(\cdot) = \bar{\mathcal{J}}(\omega_{t_{(n)}}, \cdot)$. But at time $t_{(n)} < t \leq t_{(n+1)}$, $\gamma^{p}_{t}$ is compared with $\gamma_{t}$ in equation (\ref{eq:Tt}). But $\gamma_{t}$ is derived from a better estimate of $\bar{\mathcal{J}}(\omega_t, \cdot)$. Equation (\ref{eqn:algammaold}) ensures that  $\gamma^{p}_{t}$ is updated using the latest estimate of $\bar{\mathcal{J}}(\omega_{t}, \cdot)$. The variable $\theta^{p}$ holds the model parameter $\theta_{t_{(n-1)}}$ and the update of $\gamma^{p}_{t}$ in (\ref{eqn:algammaold}) is performed using the $\mathbf{z}^{p}_{t+1}$ sampled using $\widehat{f}_{\theta^{p}}(\cdot)$.
\end{remark}

\section{Convergence Analysis}

For analyzing the asymptotic behaviour of the algorithm, we apply the  ODE based analysis from \cite{kushner2012stochastic, kubrusly1973stochastic, borkar2008stochastic} where an ODE whose asymptotic behaviour is eventually tracked by the stochastic system is identified. The long run behaviour of the equivalent ODE is studied and it is argued that the algorithm asymptotically converges almost surely to the set of stable fixed points of the ODE. We define the filtration $\{\mathcal{F}_{t}\}_{t \in \mathbb{Z}_{+}}$ where the $\sigma$-field $\mathcal{F}_t$ = $\sigma\left(\omega_i, \gamma_i, \gamma^{p}_i, \xi^{(0)}_i, \xi^{(1)}_i, \theta_i, 0 \leq i \leq t; \mathbf{z}_{i}, 1 \leq i \leq t; \mathbf{s}_{i}, \mathbf{r}_{i}, \mathbf{s}^{\prime}_{i}, 0 \leq i < t \right)$, $t \in \mathbb{Z}_{+}$. 

It is worth mentioning that the recursion (\ref{eqn:alomgupd}) is independent of other recursions and hence can be analysed independently. For the recursion (\ref{eqn:alomgupd}) we have the following result.
\begin{lemma}\label{lemma:lm1}
Let the step-size sequences $\alpha_t$ and $\beta_t$ , $t \in \mathbb{Z}_{+}$ satisfy (\ref{eqn:learnrt}). For the sample trajectory $\{(\mathbf{s}_{t}, \mathbf{r}_{t}, \mathbf{s}^{\prime}_{t})\}_{t=0}^{\infty}$, we let assumption (A3) hold and let $\nu$ be the sampling distribution. Then, for a given $z \in \mathcal{Z}$, the iterates $\omega_{t}$ in equation (\ref{eqn:alomgupd}) satisfy with probability one,
\begin{gather*}
\lim_{t \rightarrow \infty}(\omega^{(0)}_{t} + \omega^{(1)}_{t}z) = \omega^{(0)}_{*} + \omega^{(1)}_{*}z,\hspace{51mm} \\
\lim_{t \rightarrow \infty}\omega^{(2)}_{t} = \omega^{(2)}_{*} \hspace{2mm} \mathrm{and} \hspace{2mm} \lim_{t \rightarrow \infty} \bar{\mathcal{J}}(\omega_{t}, z) = \mathcal{J}(z), \hspace{48mm}
\end{gather*}
where $\bar{\mathcal{J}}_{t}(z)$ is defined in equation (\ref{eqn:jval}), $\mathcal{J}(z)$ is defined in equation (\ref{eqn:mspbeobj}), $\Phi$ is defined in equation (\ref{eqn:phieq}) and $D^{\nu}$ is defined in equation (\ref{eq:norm}) respectively.
\end{lemma}
\begin{Proof}
By rearranging equations in (\ref{eqn:alomgupd}), for $t \in \mathbb{Z}_{+}$, we get
\begin{equation}\label{eqn:omgrec1}
\omega^{(0)}_{t+1} =  \omega^{(0)}_{t} + \alpha_{t+1}\big(\mathbb{M}^{(0,0)}_{t+1}+h^{(0,0)}(\omega^{(0)}_{t})\big), \hspace{75mm}
\end{equation}
where $\mathbb{M}^{(0,0)}_{t+1} = \mathbf{r}_{t}\phi_{t} - \mathbb{E}\left[\mathbf{r}_{t}\phi_{t}\right] \hspace{1mm} \mathrm{and}  \hspace{1mm} h^{(0,0)}(x)=\mathbb{E}\left[\mathbf{r}_{t}\phi_{t} \right]-x$.\\
Similarly,
\begin{equation}\label{eqn:omgrec2}
\omega^{(1)}_{t+1} =  \omega^{(1)}_{t} + \alpha_{t+1}\big(\mathbb{M}^{(0,1)}_{t+1}+h^{(0,1)}(\omega^{(1)}_{t})\big), \hspace*{75mm}
\end{equation}
where $\mathbb{M}^{(0,1)}_{t+1} = \phi_{t}(\gamma\phi^{\prime}_{t}-\phi_{t})^{\top} - \mathbb{E}\left[\phi_{t}(\gamma\phi^{\prime}_{t}-\phi_{t})^{\top}\right]$ and 
$h^{(0,1)}(x)=\mathbb{E}\left[\phi_{t}(\gamma\phi^{\prime}_{t}-\phi_{t})^{\top}\right]-x$.\\
Finally,
\begin{equation}\label{eqn:omgrec3}
\hspace*{-75mm}\omega^{(2)}_{t+1} =  \omega^{(2)}_{t} + \alpha_{t+1}\big(\mathbb{M}^{(0,2)}_{t+1}+h^{(0,2)}(\omega^{(2)}_{t})\big),
\end{equation}
where $\mathbb{M}^{(0,2)}_{t+1} = \mathbb{E}\left[\phi_{t}\phi_{t}^{\top}\omega^{(2)}_{t} \right] - \phi_{t}\phi_{t}^{\top}\omega^{(2)}_{t}  \textrm{ and }
h^{(0,2)}(x) = \mathbb{I}_{k \times k} -\mathbb{E}\left[\phi_{t}\phi_{t}^{\top}x\right]$.\\
It is easy to verify that $h^{(0,i)}, 0 \leq i \leq 2$ are Lipschitz continuous and  $\{\mathbb{M}^{(0,i)}_{t+1}\}_{t \in \mathbb{Z}_{+}}$, $0 \leq i \leq 2$ are martingale difference noise terms, \emph{i.e.}, for each $i$, $\mathbb{M}^{(0,i)}_{t}$ is $\mathcal{F}_{t}$-measurable, integrable and $\mathbb{E}\left[\mathbb{M}^{(0,i)}_{t+1} \vert \mathcal{F}_t\right] = 0$, $t \in \mathbb{Z}_{+}$, $0 \leq i \leq 2$.

Since $\phi_t$ and $\mathbf{r}_{t}$ have uniformly bounded second moments, the noise terms $\{\mathbb{M}^{(0,0)}_{t+1}\}_{t \in \mathbb{Z}_{+}}$ have uniformly bounded second moments as well and hence $\exists K_{0,0} > 0$ \emph{s.t.}
\[ \mathbb{E}\left[\Vert \mathbb{M}^{(0,0)}_{t+1} \Vert^{2} \vert \mathcal{F}_{t}\right] \leq K_{0,0}(1+\Vert \omega^{(0)}_{t} \Vert^{2}), t \geq 0. \]

Also $h^{(0,0)}_{c}(x) \triangleq \frac{h^{(0,0)}(cx)}{c} = \frac{\mathbb{E}\left[\mathbf{r}_{t}\phi_{t} \vert \mathcal{F}_{t}\right]-cx}{c} = \frac{\mathbb{E}\left[\mathbf{r}_{t}\phi_{t} \vert \mathcal{F}_{t}\right]}{c} - x$. So $h^{(0,0)}_{\infty}(x) = \lim_{t \rightarrow \infty}h^{(0,0)}_{c}(x) = -x$. Since the ODE $\dot{x}(t) = h^{(0,0)}_{\infty}(x)$ is globally asymptotically stable to the origin, we obtain that the iterates $\{\omega^{(0)}_{t}\}_{t \in \mathbb{Z}_{+}}$ are almost surely stable, \emph{i.e.}, $\sup_{t}{\Vert \omega^{(0)}_{t} \Vert} < \infty $ \emph{a.s.}, from Theorem 7, Chapter 3 of \cite{borkar2008stochastic}. 
Similarly we can show that $\sup_{t}{\Vert \omega^{(1)}_{t} \Vert} < \infty$ \emph{a.s.}

Since $\phi_{t}$ and $\phi^{\prime}_{t}$ have uniformly bounded second moments, the second moments of  $\{\mathbb{M}^{(0,1)}_{t+1}\}_{t \in \mathbb{Z}_{+}}$ are uniformly bounded  and therefore $\exists K_{0,1} > 0$ \emph{s.t.}
\[ \mathbb{E}\left[\Vert \mathbb{M}^{(0,1)}_{t+1} \Vert^{2} \vert \mathcal{F}_{t}\right] \leq K_{0,1}(1+\Vert \omega^{(1)}_{t} \Vert^{2}), t \geq 0.\]

Now define 
\begin{equation*}
\begin{aligned}
h^{(0,2)}_{c}(x) \triangleq \frac{h^{(0,2)}(cx)}{c} = \frac{\mathbb{I}_{k \times k}-\mathbb{E}\left[\phi_{t}\phi_{t}^{\top}cx \vert \mathcal{F}_{t}\right]}{c} = \frac{\mathbb{I}_{k \times k}}{c} - x\mathbb{E}\left[\phi_{t}\phi_{t}^{\top} \right].
\end{aligned}
\end{equation*} 
Hence $h^{(0,2)}_{\infty}(x) = \lim_{t \rightarrow \infty}h^{(0,2)}_{c}(x) = - x\mathbb{E}\left[\phi_{t}\phi_{t}^{\top} \right]$. The $\infty$-system ODE given by $\dot{x}(t) = h^{(0,2)}_{\infty}(x)$ is also globally asymptotically stable to the origin since $\mathbb{E}[\phi_t \phi_t^{\top}]$ is positive definite (as it is non-singular and positive semi-definite). So  $\sup_{t}{\Vert \omega^{(2)}_{t} \Vert} < \infty $ \emph{a.s.} from Theorem 7, Chapter 3 of \cite{borkar2008stochastic}. 

Since $\phi_{t}$ has uniformly bounded second moments, $\exists K_{0,2} > 0$ \emph{s.t.}
\[\mathbb{E}\left[\Vert \mathbb{M}^{(0,2)}_{t+1} \Vert^{2} \vert \mathcal{F}_{t}\right] \leq K_{0,2}(1+\Vert \omega^{(2)}_{t} \Vert^{2}), t \geq 0.\]

Now consider the following system of ODEs associated with (\ref{eqn:omgrec1})-(\ref{eqn:omgrec3}):
\begin{eqnarray}\label{eqn:omgodes}
\frac{d}{dt}\omega^{(0)}(t) = \mathbb{E}\left[\mathbf{r}_{t}\phi_{t}\right]-\omega^{(0)}(t), \hspace*{6mm} t \in \mathbb{R}_{+}, \label{eqn:omgode1} \\
\frac{d}{dt}\omega^{(1)}(t) = \mathbb{E}\left[\phi_{t}(\gamma\phi^{\prime}_{t}-\phi_{t})^{\top}\right]-\omega^{(1)}(t)), \hspace*{6mm} t \in \mathbb{R}_{+}, \label{eqn:omgode2} \\
\frac{d}{dt}\omega^{(2)}(t) = \mathbb{I}_{k \times k} -\mathbb{E}\left[\phi_{t}\phi_{t}^{\top}\right]\omega^{(2)}(t), \hspace*{6mm} t \in \mathbb{R}_{+}. \label{eqn:omgode3}
\end{eqnarray}
For the ODE (\ref{eqn:omgode1}), the point $\mathbb{E}\left[\mathbf{r}_{t}\phi_{t}\right]$ is a globally asymptotically stable equilibrium. Similarly for the ODE (\ref{eqn:omgode2}), the point $\mathbb{E}\left[\phi_{t}(\gamma\phi^{\prime}_{t}-\phi_{t})^{\top}\right]$ is a globally asymptotically stable equilibrium. For the ODE (\ref{eqn:omgode3}), since $\mathbb{E}\left[\phi_{t}\phi_{t}^{\top}\right]$ is non-negative definite and non-singular (from the assumptions of the lemma), the ODE (\ref{eqn:omgode3}) is globally asymptotically stable to the point $\mathbb{E}\left[\phi_{t}\phi_{t}^{\top}\right]^{-1}$.

It can now be shown from Theorem 2, Chapter 2 of \cite{borkar2008stochastic} that the asymptotic properties of the recursions (\ref{eqn:omgrec1}), (\ref{eqn:omgrec2}), (\ref{eqn:omgrec3}) and their associated ODEs (\ref{eqn:omgode1}), (\ref{eqn:omgode2}), (\ref{eqn:omgode3}) are similar and hence $\lim_{t \rightarrow \infty}\omega^{(0)}_{t} = \mathbb{E}\left[\mathbf{r}_{t}\phi_{t}\right]$ \emph{a.s.}, $\lim_{t \rightarrow \infty}\omega^{(1)}_{t} = \mathbb{E}\left[\phi_t(\gamma\phi^{\prime}_{t} -\phi_t)^{\top}\right]$ \emph{a.s.} and $\lim_{t \rightarrow \infty}\omega^{(2)}_{t+1} = \mathbb{E}\left[\phi_{t}\phi_{t}^{\top}\right]^{-1}$ \emph{a.s.}. So for any $z \in \mathbb{R}^{k}$, using  (\ref{eqn:comp1}), we have $\lim_{t \rightarrow \infty}(\omega^{(0)}_{t} + \omega^{(1)}_{t}z) = \Phi^{\top} D^{\nu} (T_{\pi} V_{z}-V_{z})$ \emph{a.s.} Also, from  (\ref{eqn:comp2}), we have $\lim_{t \rightarrow \infty}\omega^{(2)}_{t+1} = (\Phi^{\top}D^{\nu}\Phi)^{-1}$ \emph{a.s.}

Putting all the above together we get
$\lim_{t \rightarrow \infty} \bar{\mathcal{J}}(\omega_t, z) = \bar{\mathcal{J}}(\omega_{*}, z)$  = $\mathcal{J}(z)$ \emph{a.s.}
\end{Proof}

As mentioned before, the update of $\theta_t$ only happens along a subsequence $\{t_{(n)}\}_{n \in \mathbb{Z}_{+}}$ of $\{t\}_{t \in \mathbb{Z}_{+}}$. So between $t=t_{(n)}$ and $t=t_{(n+1)}$, $\theta_{t}$ is constant. The lemma and the theorems that follow in this paper depend on the timescale difference in the step-size schedules $\{\alpha_{t}\}_{t \geq 0}$ and $\{\beta_{t}\}_{t \geq 0}$. The timescale differences allow the different recursions to learn at different rates. The step-size $\{\beta_t\}_{t \geq 0}$ decays to $0$ at a slower rate than $\{\alpha_t\}_{t \geq 0}$ and hence the increments in the recursions (\ref{eqn:algamma}), (\ref{eqn:xi0}) and (\ref{eqn:xi1})  which are controlled by $\beta_t$ are larger and hence converge faster than the recursions (\ref{eqn:alomgupd}),(\ref{eqn:jval}) and (\ref{eq:thupd}) which are controlled by $\alpha_t$. So the relative evolution of  the variables from the slower timescale $\alpha_{t}$, \emph{i.e.}, $\omega_{t}$ is indeed slow and in fact can be considered constant when viewed from the faster timescale $\beta_t$, see Chapter 6, \cite{borkar2008stochastic} for a succinct description on multi-timescale stochastic approximation algorithms.
\begin{itemize}
\item[$\circledast$]\textbf{Assumption (A4):} The iterate sequence $\gamma_t$ in equation (\ref{eqn:algamma}) satisfies $\sup_{t}{\vert \gamma_{t} \vert} < \infty$ \emph{a.s.}.
\end{itemize}
\begin{remark}The assumption (A4) is a technical requirement to prove convergence. In practice, one may replace (\ref{eqn:algamma}) by its `projected version' whereby the iterates are projected back to an \textit{a priori} chosen compact convex set if they stray outside of this set.
\end{remark}
\textbf{Notation: } We denote by $\mathbb{E}_{\widehat{\theta}}[\cdot]$ the expectation \emph{w.r.t.} the mixture \emph{pdf} and $\mathbb{P}_{\widehat{\theta}}$ denotes its induced probability measure. Also $\gamma_{\rho}(\cdot, \widehat{\theta})$ represents the $(1-\rho)$-quantile \emph{w.r.t.} the mixture \emph{pdf} $\widehat{f}_{\theta}$.\\

The recursion (\ref{eqn:algamma}) moves on a faster timescale as compared to the recursion (\ref{eqn:alomgupd}) of $\omega_{t}$ and the recursion (\ref{eq:thupd}) of $\theta_{t}$. Hence, on the timescale of the recursion (\ref{eqn:algamma}), one may consider $\omega_{t}$ and $\theta_{t}$ to be fixed. For recursion (\ref{eqn:algamma}) we have the following result:
\begin{lemma}\label{lmn:gammacov}
Let $\omega_{t} \equiv \omega$, $\theta_{t} \equiv \theta$. Let $\bar{\mathcal{J}}_{\omega}(\cdot) \triangleq \bar{\mathcal{J}}(\omega, \cdot)$. Then $\gamma_t, t \in \mathbb{Z}_{+}$ in equation (\ref{eqn:algamma}) satisfy $\gamma_t \rightarrow \gamma_{\rho}({\bar{\mathcal{J}}}_{\omega}, \widehat{\theta})$ as $t \rightarrow \infty$ with probability one.
\end{lemma}
\begin{Proof}
Here, for easy reference we rewrite the recursion (\ref{eqn:algamma}),
\begin{equation}\label{eqn:preq1}
\gamma_{t+1} = \gamma_{t} - \beta_{t+1}\Delta \gamma_{t+1}(\mathbf{z}_{t+1}),\hspace*{60mm}
\end{equation}
Substituting the expression for $\Delta \gamma_{t+1}$ in (\ref{eqn:preq1}) with $\omega_t = \omega$ and $\theta_t = \theta$, we get
\begin{equation}\label{eqn:preq2}
\begin{aligned}
\gamma_{t+1} = \gamma_{t} - \beta_{t+1} \Big(-(1-\rho)I_{\{\bar{\mathcal{J}}(\omega, \mathbf{z}_{t+1}) \geq \gamma_{t}\}} +  \rho I_{\{\bar{\mathcal{J}}(\omega, \mathbf{z}_{t+1}) \leq \gamma_{t} \}}\Big), \hspace*{3mm} \textrm{where } \mathbf{z}_{t+1} \sim \widehat{f}_{\theta}(\cdot)\hspace*{8mm}\\
\end{aligned}
\end{equation}
The above equation can be apparently viewed as,
\begin{equation*}
\begin{aligned}
\gamma_{t+1} - \gamma_{t} \in - \beta_{t+1} \nabla_{y}\psi(\bar{\mathcal{J}}_{\omega}(\mathbf{z}_{t+1}), \gamma_{t}),
\end{aligned}
\end{equation*}
where $\nabla_{y}\psi$ is the sub-differential of $\psi(x,y)$ \emph{w.r.t.} $y$ (where $\psi$  is defined in Lemma \ref{lma:qnopt}). $\nabla_{y}\psi$ is a set function and is defined as follows:
\begin{equation}
\nabla_{y}\psi(\bar{\mathcal{J}}_{\omega}(z), y) = \left\{
                \begin{array}{ll}
                  \{-(1-\rho)I_{\{\bar{\mathcal{J}}(\omega, z) \geq y\}} +  \rho I_{\{\bar{\mathcal{J}}(\omega, z) \leq y \}}\}, \hspace{2mm} \textrm{for } \hspace{2mm} y \neq \bar{\mathcal{J}}(\omega, z)\\
                  \left[  \rho_1, \rho_2\ \right], \hspace{2mm} \textrm{for } \hspace{2mm} y = \bar{\mathcal{J}}(\omega, z),
                \end{array}
              \right.
\end{equation}
where $\rho_1 = \min{\{1-\rho, \rho\}}$ and $\rho_2 = \max{\{1-\rho, \rho\}}$.\\
Rearranging the terms in equation (\ref{eqn:preq2}) we get,
\begin{equation}
\gamma_{t+1} = \gamma_{t} + \beta_{t+1}\left(\mathbb{M}^{(1,0)}_{t+1} - \mathbb{E}_{\widehat{\theta}}\left[\Delta \gamma_{t+1}(\mathbf{z}_{t+1}) \right]\right),\hspace*{60mm}
\end{equation}
\vspace*{-1mm}
where $\mathbb{M}^{(1,0)}_{t+1} =  \mathbb{E}_{\widehat{\theta}}\left[\Delta \gamma_{t+1}(\mathbf{z}_{t+1})\right] - \Delta \gamma_{t+1}(\mathbf{z}_{t+1})$ with $\mathbf{z}_{t+1} \sim \widehat{f}_{\theta}(\cdot)$.\\
It is easy to verify that $\mathbb{E}_{\widehat{\theta}}\left[\Delta \gamma_{t+1}(\mathbf{z}_{t+1}) \right] = \nabla_{y}\mathbb{E}_{\widehat{\theta}}\left[\psi(\bar{\mathcal{J}}(\mathbf{z}_{t+1}), y)\right]$. For brevity, define $h^{(1,0)}(\gamma) \triangleq -\nabla_{y}\mathbb{E}_{\widehat{\theta}}\left[\psi(\bar{\mathcal{J}}_{\omega}(\mathbf{z}_{t+1}), \gamma)\right]$.\\
The set function $h^{(1,0)}:\mathbb{R} \rightarrow \{$subsets of $\mathbb{R}\}$ satisfies the following properties:
\begin{enumerate}
\item For each $y \in \mathbb{R}$, $h^{(1,0)}(y)$ is convex and compact.
\item For each $y \in \mathbb{R}$, $\sup_{y^{\prime} \in h(y)} \vert h^{(1,0)}(y^{\prime}) \vert < K_{1,0}(1+\vert y \vert), \textrm{ for some } 0 < K_{1,0}  < \infty$.
\item $h^{(1,0)}$ is upper semi-continuous.
\end{enumerate}
The noise term $\mathbb{M}^{(1,0)}_{t}$ satisfies the following properties:
\begin{enumerate}
\item $\mathbb{M}^{(1,0)}_{t}$ is $\mathcal{F}_{t}$-measurable $\forall t$ and integrable, $\forall t > 0$.
\item $\mathbb{M}^{(1,0)}_{t}$, $t \geq 0$ is a martingale difference noise sequence, \emph{i.e.}, $\mathbb{E}[\mathbb{M}^{(1,0)}_{t+1} | \mathcal{F}_t] = 0$ \emph{a.s.}
\item
$\mathbb{E}\left[\Vert \mathbb{M}^{(1,0)}_{t+1} \Vert^{2}\big|\mathcal{F}_{t}\right] \leq K_{1,1}(1+\Vert \gamma_{t} \Vert^{2} + \Vert \omega_{t} \Vert^{2}), \textrm{ for some } 0 < K_{1,1}  < \infty.$
This follows directly from the fact that $\Delta \gamma_{t+1}(\mathbf{z}_{t+1})$ has finite first and second order moments.
\end{enumerate}

Therefore by the almost sure boundedness of the sequence $\{\gamma_{t}\}$ in assumption (A4) and by Lemma 1, Chapter 2 in \cite{borkar2008stochastic}, we can claim that the stochastic sequence $\{\gamma_{t}\}$ asymptotically tracks the differential inclusion \begin{equation} \label{eqn:dtode}
\frac{d}{dt}\gamma(t) \in -\mathbb{E}_{\widehat{\theta}}\left[\nabla_{y} \psi(\bar{\mathcal{J}}_{\omega}(\mathbf{z}), \gamma(t))\right] = - \nabla_{y}\mathbb{E}_{\widehat{\theta}}\left[\psi(\bar{\mathcal{J}}_{\omega}(\mathbf{z}), \gamma(t))\right] = h^{(1,0)}(\gamma(t)).
\end{equation}
The interchange of $\nabla_{\gamma}$ and $\mathbb{E}_{\widehat{\theta}}[\cdot]$ in the above equation is guaranteed by the dominated convergence theorem.

Now we prove the  stability of the above differential inclusion. Note that by Lemma 1 of \cite{homem2007study}, we know that $\gamma^{*} \triangleq \gamma_{\rho}({\bar{\mathcal{J}}}_{\omega}, \widehat{\theta})$ is a root of the function $h^{(1,0)}(\gamma)$ and hence it is a fixed point of the flow induced by the above differential inclusion. Now define $V(\gamma) \triangleq \mathbb{E}_{\widehat{\theta}}\left[\psi(\bar{\mathcal{J}}_{\omega}(\mathbf{z}), \gamma)\right] - \gamma^{*}$. It is easy to verify that $V$ is continuously differentiable. Also by Lemma 1 of \cite{homem2007study}, we have $\mathbb{E}_{\widehat{\theta}}\left[\psi(\bar{\mathcal{J}}_{\omega}(\mathbf{z}), \gamma)\right]$ to be a convex function and $\gamma^{*}$ to be its global minimum. Hence $V(\gamma) > 0$, $\forall \gamma \in \mathbb{R}^{d} \backslash \{\gamma^{*}\}$. Further $V(\gamma^{*}) = 0$ and $V(\gamma) \rightarrow \infty$ as $\Vert \gamma \Vert \rightarrow \infty$. So $V(\cdot)$ is a Lyapunov function. Also note that $<\nabla V(\gamma), h(\gamma)> \leq 0$. So $\widehat{\gamma}$ is the global attractor of the differential inclusion defined in (\ref{eqn:dtode}). Thus by Theorem 2 of chapter 2 in \cite{borkar2008stochastic}, the iterates $\gamma_{t}$ converge almost surely to $\gamma^{*} = \gamma_{\rho}(\omega, \widehat{\theta})$.
\end{Proof}

The recursions (\ref{eqn:xi0})  and (\ref{eqn:xi1}) move on a faster timescale as
compared to the recursion (\ref{eqn:alomgupd}) of $\omega_{t}$ and the recursion (\ref{eq:thupd}) of $\bar{\theta}_{t}$. Hence, viewed from the timescale of the recursions (\ref{eqn:xi0}) and (\ref{eqn:xi1}), one may consider $\omega_{t}$ and $\theta_{t}$ to be fixed. For the recursions (\ref{eqn:xi0}) and (\ref{eqn:xi1}), we have the following result:

\begin{lemma}\label{lmn:xiconv}
Assume $\omega_{t} \equiv \omega$, $\theta_{t} \equiv \theta$. Let $\bar{\mathcal{J}}_{\omega}(\cdot) \triangleq \bar{\mathcal{J}}(\omega, \cdot)$. Then almost surely, 
\begin{equation*}	
\begin{aligned}
(&i)\lim_{t \rightarrow \infty} \xi^{(0)}_{t} = \xi^{(0)}_{*} = \frac{\mathbb{E}_{\widehat{\theta}}\left[ \mathbf{g}_{1}(\mathcal{J}_{\omega}(\mathbf{z}), \mathbf{z}, \gamma_{\rho}(\mathcal{J}, \widehat{\theta})\right]}{\mathbb{E}_{\widehat{\theta}}\left[\mathbf{g}_{0}(\mathcal{J}_{\omega}(\mathbf{z}), \gamma_{\rho}(\mathcal{J}, \widehat{\theta})\right]},\hspace*{20mm}\\
(&ii)\lim_{t \rightarrow \infty} \xi^{(1)}_{t} = \xi^{(1)}_{*} =  \frac{\mathbb{E}_{\widehat{\theta}}\left[\mathbf{g}_{2}\left(\bar{\mathcal{J}}_{\omega}(\mathbf{z}), \mathbf{z}, \gamma_{\rho}(\bar{\mathcal{J}}_{\omega}, \widehat{\theta}), \xi^{(0)}_{*}\right)\right]}{\mathbb{E}_{\widehat{\theta}}\left[\mathbf{g}_{0}\left(\bar{\mathcal{J}}_{\omega}(\mathbf{z}), \gamma_{\rho}(\bar{\mathcal{J}}_{\omega}, \widehat{\theta})\right)\right]}.
\end{aligned}
\end{equation*}
where $\mathbb{E}_{\widehat{\theta}}[\cdot]$ is the expectation w.r.t. the pdf $\widehat{f}_{\theta}(\cdot)$ and $\mathbf{z} \sim \widehat{f}_{\theta}(\cdot)$.\\
$(iii)$ If $\gamma_{\rho}(\bar{\mathcal{J}}_{\omega}, \widehat{\theta}) > \gamma_{\rho}(\bar{\mathcal{J}}_{\omega}, \widehat{\theta^{p}})$, then $T_{t}$, $t \in \mathbb{Z}_{+}$ in equation (\ref{eq:Tt}) satisfy $\lim_{t \rightarrow \infty} T_{t} = 1$ a.s.
\end{lemma}
\begin{Proof}
$(i)$ First, we recall equation (\ref{eqn:xi0}) below
\begin{equation}\label{eq:c0e1}
\begin{aligned}
\xi^{(0)}_{t+1} = \xi^{(0)}_{t} +  \beta_{t+1} \left(\mathbf{g}_{1}(\bar{\mathcal{J}}(\omega_{t}, \mathbf{z}_{t+1}), \mathbf{z}_{t+1}, \gamma_{t}) - \xi^{(0)}_{t}\mathbf{g}_{0}\left(\bar{\mathcal{J}}(\omega_{t}, \mathbf{z}_{t+1}), \gamma_{t}\right)\right). \hspace{1mm}
\end{aligned}
\end{equation}
Note that the above recursion of $\xi^{(0)}_{t}$ depends on $\gamma_{t}$, but not the other way. This implies that we can replace $\gamma_{t}$ by its limit point $\gamma_{\rho}(\bar{\mathcal{J}}_{\omega}, \widehat{\theta})$ and a bias term which goes to zero as $t \rightarrow \infty$. We denote the decaying bias term using the notation $o(1)$.
Further, using the assumption that $\omega_{t} = \omega$, $\theta_{t} = \theta$ and from the equation (\ref{eq:c0e1}), we get,
\begin{equation} \label{eq:c0}
\xi^{(0)}_{t+1} = \xi^{(0)}_{t} + \beta_{t+1} \left(h^{(2,0)}(\xi^{(0)}_{t}) + \mathbb{M}^{(2,0)}_{t+1} + o(1)\right),
\end{equation}
\begin{equation} \label{eq:hm0}
\begin{aligned}
\mathrm{where} \hspace{1mm}
h^{(2,0)}(x) \triangleq -\mathbb{E}\left[x \mathbf{g}_{0}\left(\bar{\mathcal{J}}_{\omega}(\mathbf{z}_{t+1}), \gamma_{\rho}(\bar{\mathcal{J}}_{\omega}, \widehat{\theta})\right) \Big| \mathcal{F}_{t}\right] + \mathbb{E}\left[ \mathbf{g}_{1}\left(\bar{\mathcal{J}}_{\omega}(\mathbf{z}_{t+1}), \mathbf{z}_{t+1}, \gamma_{\rho}(\bar{\mathcal{J}}_{\omega}, \widehat{\theta})\right) \Big|  \mathcal{F}_{t}\right],
\end{aligned}
\end{equation}
\begin{equation*}
\begin{aligned}
\mathbb{M}^{(2,0)}_{t+1} \triangleq \mathbf{g}_{1}\left(\bar{\mathcal{J}}_{\omega}(\mathbf{z}_{t+1}), \mathbf{z}_{t+1}, \gamma_{\rho}(\bar{\mathcal{J}}_{\omega}, \widehat{\theta})\right) - \mathbb{E}\left[\mathbf{g}_{1}\left(\bar{\mathcal{J}}_{\omega}(\mathbf{z}_{t+1}), \mathbf{z}_{t+1}, \gamma_{\rho}(\bar{\mathcal{J}}_{\omega}, \widehat{\theta})\right) \Big|  \mathcal{F}_{t}\right] - \hspace*{15mm}\\ \xi^{(0)}_{t}\mathbf{g}_{0}\left(\bar{\mathcal{J}}_{\omega}( \mathbf{z}_{t+1}), \gamma_{\rho}(\bar{\mathcal{J}}_{\omega}, \widehat{\theta})\right) +  \mathbb{E}\left[\xi^{(0)}_{t} \mathbf{g}_{0}\left(\bar{\mathcal{J}}_{\omega}(\mathbf{z}_{t+1}), \gamma_{\rho}(\bar{\mathcal{J}}_{\omega}, \widehat{\theta})\right) \Big| \mathcal{F}_{t}\right] \textrm{ and } \mathbf{z}_{t+1} \sim \widehat{f}_{\theta}(\cdot).
\end{aligned}
\end{equation*}
Since $\mathbf{z}_{t+1}$ is independent of the $\sigma$-field $\mathcal{F}_{t}$,
the function $h^{(2,0)}(\cdot)$ in equation (\ref{eq:hm0}) can be rewritten as
\begin{equation*}
\begin{aligned}
h^{(2,0)}(x) = -\mathbb{E}_{\widehat{\theta}}\left[x \mathbf{g}_{0}\left(\bar{\mathcal{J}}_{\omega}(\mathbf{z}), \gamma_{\rho}(\bar{\mathcal{J}}_{\omega}(\mathbf{z}), \widehat{\theta})\right)\right] + \mathbb{E}_{\widehat{\theta}}\left[\mathbf{g}_{1}\left(\bar{\mathcal{J}}_{\omega}(\mathbf{z}), \mathbf{z}, \gamma_{\rho}(\bar{\mathcal{J}}_{\omega}, \widehat{\theta})\right)\right], \textrm{ where } \mathbf{z} \sim \widehat{f}_{\theta}(\cdot).
\end{aligned}
\end{equation*}
It is easy to verify that $\mathbb{M}^{(2,0)}_{t}$, $t \in \mathbb{Z}_{+}$ is a martingale difference sequence, \emph{i.e.}, $\mathbb{M}^{(2,0)}_{t}$ is $\mathcal{F}_{t}$-measurable, integrable and $\mathbb{E}[\mathbb{M}^{(2,0)}_{t+1} | \mathcal{F}_t] = 0$ \emph{a.s.}, $\forall t \in \mathbb{Z}_{+}$. It is also easy to verify that $h^{(2,0)}(\cdot)$ is Lipschitz continuous. Also since $S(\cdot)$ is bounded above and $\widehat{f}_{\theta}(\cdot)$ has finite first and second moments we have almost surely,
\[\mathbb{E}\left[\Vert \mathbb{M}^{(2,0)}_{t+1} \Vert^{2} \vert \mathcal{F}_{t}\right] \leq K_{2,0}(1+\Vert \xi^{(0)}_t \Vert^{2}), \forall t \geq 0, \textrm{ for some } 0 < K_{2,0} < \infty. \]
Now consider the ODE
\begin{equation}\label{eq:ode2}
\frac{d}{dt}\xi^{(0)}(t) = h^{(2,0)}(\xi^{(0)}(t)).
\end{equation} 
We may rewrite the above ODE as,
\[\frac{d}{dt}\xi^{(0)}(t) = A\xi^{(0)}(t) + b^{(0)},\]
where $A$ is a  diagonal matrix with $A_{ii} = -\mathbb{E}_{\widehat{\theta}}\left[\mathbf{g}_{0}(\bar{\mathcal{J}}_{\omega}(\mathbf{z}), \gamma_{\rho}(\bar{\mathcal{J}}_{\omega}, \widehat{\theta}))\right]$, $0 \leq i < k$ and \\ $b^{(0)} = \mathbb{E}_{\widehat{\theta}}\left[\mathbf{g}_{1}(\bar{\mathcal{J}}_{\omega}(\mathbf{z}), \mathbf{z}, \gamma_{\rho}(\bar{\mathcal{J}}_{\omega}, \widehat{\theta}))\right]$. Now consider the ODE in the $\infty$-system $\frac{d}{dt}\xi^{(0)}(t)$ = $\lim_{\eta \rightarrow \infty}\frac{h^{(2,0)}(\eta\xi^{(0)}(t))}{\eta}$ = $A \xi^{(0)}(t)$. Since the matrix $A$ has the same value for all the diagonal elements, $A$ has only one eigenvalue: $\lambda(A) = -\mathbb{E}_{\widehat{\theta}}\left[\mathbf{g}_{0}(\bar{\mathcal{J}}_{\omega}(\mathbf{z}), \gamma_{\rho}(\bar{\mathcal{J}}_{\omega}, \widehat{\theta}))\right]$ with multiplicity $k$. Also observe that $\lambda(A) < 0$. Hence the ODE (\ref{eq:ode2}) is  globally asymptotically stable to the origin. Using Theorem 7, Chapter 3 of \cite{borkar2008stochastic}, the iterates $\{\xi^{(0)}_{t}\}_{t \in \mathbb{Z}_{+}}$ are stable \emph{a.s.}, \emph{i.e.}, $\sup_{t \in \mathbb{Z}_{+}}{\Vert \xi^{(0)}_{t} \Vert} < \infty$ \emph{a.s.}   

Again, by using the earlier argument that the eigenvalues $\lambda(A)$ of $A$ are negative and identical, the point $-A^{-1}b^{(0)}$ can be seen to be a globally asymptotically stable equilibrium of the ODE (\ref{eq:ode2}). By using Corollary 4, Chapter 2 of \cite{borkar2008stochastic}, we can conclude that
\begin{equation*}
\lim_{t \rightarrow \infty} \xi^{(0)}_{t} = -A^{-1}b^{(0)} \emph{a.s.} = \frac{E_{\widehat{\theta}}\left[\mathbf{g}_{1}(\bar{\mathcal{J}}_{\omega}(\mathbf{z}), \mathbf{z}, \gamma_{\rho}(\bar{\mathcal{J}}_{\omega}, \widehat{\theta}))\right]}{E_{\widehat{\theta}}\left[\mathbf{g}_{0}(\bar{\mathcal{J}}_{\omega}(\mathbf{z}), \gamma_{\rho}(\bar{\mathcal{J}}_{\omega}, \widehat{\theta}))\right]} \emph{a.s.}
\end{equation*}
$(ii)$ We recall first the matrix recursion (\ref{eqn:xi1}) below:
\begin{gather}\label{eq:c0e2}
\begin{aligned}
\xi^{(1)}_{t+1} = \xi^{(1)}_{t} + \beta_{t+1}\left(\mathbf{g}_{2}(\bar{\mathcal{J}}(\omega_{t}, \mathbf{z}_{t+1}), \mathbf{z}_{t+1}, \gamma_{t}, \xi^{(0)}_t) - \xi^{(1)}_{t}\mathbf{g}_{0}\left(\bar{\mathcal{J}}(\omega_{t}, \mathbf{z}_{t+1}), \gamma_{t}\right)\right).\hspace*{1mm}
\end{aligned}
\end{gather}
As in the earlier proof, we also assume $\omega_{t} = \omega$ and $\theta_{t} = \theta$. Also note that $\xi^{(1)}_{t}$, $\xi^{(0)}_{t}$ and $\gamma_{t}$ are on the same timescale. However, the recursion of $\gamma_{t}$ proceeds independently and in particular does not depend on $\xi^{(0)}_{t}$ and $\xi^{(1)}_{t}$. Also, there is a unilateral coupling of $\xi^{(1)}_{t}$ on $\xi^{(0)}_{t}$ and $\gamma_{t}$, but not the other way. Hence, while analyzing (\ref{eq:c0e1}), one may replace $\gamma_t$ and $\xi^{(0)}_{t}$ in equation (\ref{eq:c0e1}) with their limit points $\gamma_{\rho}(\bar{\mathcal{J}}_{\omega}, \theta)$  and $\xi^{(0)}_{*}$ respectively and a decaying bias term $o(1)$. Now, by considering all the above observations, we rewrite the equation (\ref{eq:c0e2}) as, 
\begin{equation} \label{eqn:c1}
\begin{aligned}
\xi^{(1)}_{t+1} = \xi^{(1)}_{t} + \beta_{t+1} \left(h^{(2,1)}(\xi^{(1)}_{t}) + \mathbb{M}^{(2,1)}_{t+1} + o(1)\right),
\end{aligned}
\end{equation}
\begin{equation}\label{eqn:hmc1}
\begin{aligned}
\hspace*{-6mm}\textrm{ where }  h^{(2,1)}(x) \triangleq
\mathbb{E}\left[\mathbf{g}_{2}\left(\bar{\mathcal{J}}_{\omega}(\mathbf{z}_{t+1}), \mathbf{z}_{t+1}, \gamma_{\rho}(\bar{\mathcal{J}}_{\omega}, \widehat{\theta}), \xi^{(0)}_{*}\right) \Big| \mathcal{F}_{t}\right] - \mathbb{E}\left[ x \mathbf{g}_{0}\left(\bar{\mathcal{J}}_{\omega}(\mathbf{z}_{t+1}), \gamma_{\rho}(\bar{\mathcal{J}}_{\omega},\widehat{\theta})\right) \Big|  \mathcal{F}_{t}\right]
\end{aligned}
\end{equation}
\begin{equation}
\begin{aligned}
\textrm{ and } \mathbb{M}^{(2,1)}_{t+1} \triangleq \mathbb{E}\left[\xi^{(1)}_{t}\mathbf{g}_{0}\left(\bar{\mathcal{J}}_{\omega}(\mathbf{z}_{t+1}), \gamma_{\rho}(\bar{\mathcal{J}}_{\omega}(\widehat{\theta})\right) \Big\vert \mathcal{F}_t\right] - \xi^{(1)}_{t}\mathbf{g}_{0}\left(\bar{\mathcal{J}}_{\omega}(\mathbf{z}_{t+1}), \gamma_{\rho}(\bar{\mathcal{J}}_{\omega}(\widehat{\theta})\right)   - \hspace*{10mm}\\  \mathbb{E}\left[\mathbf{g}_{2}\left(\bar{\mathcal{J}}_{\omega}(\mathbf{z}_{t+1}), \mathbf{z}_{t+1}, \gamma_{\rho}(\bar{\mathcal{J}}_{\omega}, \widehat{\theta}), \xi^{(0)}_{*}\right) \Big| \mathcal{F}_{t}\right] +  \mathbf{g}_{2}\left(\bar{\mathcal{J}}_{\omega}(\mathbf{z}_{t+1}), \mathbf{z}_{t+1}, \gamma_{\rho}(\bar{\mathcal{J}}_{\omega}, \widehat{\theta}), \xi^{(0)}_{*}\right),\\\hspace*{2mm} \textrm{ where } \mathbf{z}_{t+1} \sim \widehat{f}_{\theta}(\cdot).\hspace*{1mm}
\end{aligned}
\end{equation}
Since $\mathbf{z}_{t+1}$ is independent of the $\sigma$-field $\mathcal{F}_{t}$,
the function $h^{(2,1)}(\cdot)$ in equation (\ref{eqn:hmc1}) can be rewritten as
\begin{equation}
\begin{aligned}
h^{(2,1)}(x) = 
\mathbb{E}_{\widehat{\theta}}\left[\mathbf{g}_{2}\left(\bar{\mathcal{J}}_{\omega}(\mathbf{z}), \mathbf{z}, \gamma_{\rho}(\bar{\mathcal{J}}_{\omega}, \widehat{\theta}), \xi^{(0)}_{*}\right)\right] - \mathbb{E}_{\widehat{\theta}}\left[x \mathbf{g}_{0}\left(\bar{\mathcal{J}}_{\omega}(\mathbf{z}), \gamma_{\rho}(\bar{\mathcal{J}}_{\omega},\widehat{\theta})\right)\right], \textrm{ where } \mathbf{z} \sim \widehat{f}_{\theta}(\cdot).
\end{aligned}
\end{equation}
It is not difficult to verify that $\mathbb{M}^{(2,1)}_{t+1}$, $t \in \mathbb{Z}_{+}$ is a martingale difference noise sequence and $h^{(2,1)}(\cdot)$ is Lipschitz continuous. Also since $S(\cdot)$ is bounded and $\widehat{f}_{\theta}(\cdot)$ has finite first and second moments we get,
\[\mathbb{E}\left[\Vert \mathbb{M}^{(2,1)}_{t+1} \Vert^{2} \vert \mathcal{F}_{t}\right] \leq K_{2,1}(1+\Vert \xi^{(1)}_t \Vert^{2}), \forall t \in \mathbb{Z}_{+}, \textrm{ for some } 0 < K_{2,1} < \infty. \]
Now consider the ODE given by
\begin{equation}\label{eq:ode3}
\begin{aligned}
\frac{d}{dt}\xi^{(1)}(t) = h^{(2,1)}(\xi^{(1)}(t)), \hspace*{5mm} t \in \mathbb{R}_{+}.
\end{aligned}
\end{equation} 
By rewriting the above equation we get,
\[\frac{d}{dt}\xi^{(1)}(t) = A\xi^{(1)}(t) + b^{(1)}, \hspace*{5mm} t \in \mathbb{R}_{+},\]
where $A$ is a diagonal matrix as before, \emph{i.e.},  $A_{ii} = -\mathbb{E}_{\widehat{\theta}}\left[\mathbf{g}_{0}(\bar{\mathcal{J}}_{\omega}(\mathbf{z}), \gamma_{\rho}(\bar{\mathcal{J}}_{\omega}, \widehat{\theta})\right]$, $\forall i, 0 \leq i < k$ and $b^{(1)} = \mathbb{E}_{\widehat{\theta}}\left[\mathbf{g}_{2}(\bar{\mathcal{J}}_{\omega}(\mathbf{z}), \mathbf{z}, \gamma_{\rho}(\bar{\mathcal{J}}_{\omega}, \widehat{\theta}), \xi^{(0)}_{*})\right]$. Now consider the ODE in the $\infty$-system $\frac{d}{dt}{\xi}^{(1)}(t) = \lim_{\eta \rightarrow \infty}\frac{1}{\eta}h^{(2,1)}(\eta{\xi}^{(1)}(t)) = A{\xi}^{(1)}(t)$. Again, the eigenvalue $\lambda(A)$ =  $-\mathbb{E}_{\widehat{\theta}}\left[\mathbf{g}_{0}(\bar{\mathcal{J}}_{\omega}(\mathbf{z}), \gamma_{\rho}(\bar{\mathcal{J}}_{\omega}, \widehat{\theta})\right]$ of $A$ is negative and is of multiplicity $k$ and hence origin is the unique globally asymptotically stable equilibrium of the $\infty$-system. Therefore it follows that the iterates $\{{\xi}^{(1)}_{t}\}_{t \in \mathbb{Z}_{+}}$ are almost surely stable, \emph{i.e.}, $\sup_{t \in \mathbb{Z}_{+}}{\Vert {\xi}^{(0)}_{t} \Vert} < \infty$ \emph{a.s.}, see Theorem 7, Chapter 3 of \cite{borkar2008stochastic}.

Again, by using the earlier argument that the eigenvalues $\lambda(A)$ of $A$ are negative and identical, the point $-A^{-1}b^{(1)}$ can be seen to be a globally asymptotically stable equilibrium of the ODE (\ref{eq:ode3}). By Corollary 4, Chapter 2 of \cite{borkar2008stochastic},  it follows that 
\begin{equation*}
\lim_{t \rightarrow \infty}{\xi}^{(1)}_t = -A^{-1}b^{(1)}\hspace*{2mm}a.s. = \frac{\mathbb{E}_{\widehat{\theta}}\left[ \mathbf{g}_{2}(\bar{\mathcal{J}}_{\omega}(\mathbf{z}), \mathbf{z}, \gamma_{\rho}(\bar{\mathcal{J}}_{\omega}, \widehat{\theta}), \xi^{(0)}_{*})\right]}{\mathbb{E}_{\widehat{\theta}}\left[\mathbf{g}_{0}(\bar{\mathcal{J}}_{\omega}(\mathbf{z}), \gamma_{\rho}(\bar{\mathcal{J}}_{\omega}, \widehat{\theta})\right]}\hspace*{2mm} a.s.\\
\end{equation*}
$(iii)$ Here also we assume $\omega_t = \omega$. Then $\gamma_{t}$ in recursion (\ref{eqn:algamma}) and $\gamma^{p}_{t}$ in recursion (\ref{eqn:algammaold}) converge to $\gamma_{\rho}(\bar{\mathcal{J}}_{\omega}, \widehat{\theta})$  and  $\gamma_{\rho}(\bar{\mathcal{J}}_{\omega}, \widehat{\theta^{p}})$ respectively. So if $\gamma_{\rho}(\bar{\mathcal{J}}_{\omega}, \widehat{\theta}) > \gamma_{\rho}(\bar{\mathcal{J}}_{\omega}, \widehat{\theta^{p}})$, then $\gamma_{t} > \gamma^{p}_{t}$ eventually, \emph{i.e.}, $\gamma_{t} > \gamma^{p}_{t}$ for all but finitely many $t$. So almost surely $T_t$ in equation (\ref{eq:Tt}) will converge to $\mathbb{E}\left[I_{\{\gamma_{t} > \gamma^{p}_{t}\}} - I_{\{\gamma_{t} \leq \gamma^{p}_{t}\}}\right]$ = $\mathbb{P}\{\gamma_{t} > \gamma^{p}_{t}\} - \mathbb{P}\{\gamma_{t} \leq \gamma^{p}_{t}\} = 1-0 = 1$.
\end{Proof}\vspace*{4mm}\\
\textbf{Notation: }For the subsequence $\{t_{(n)}\}_{n > 0}$ of $\{t\}_{t \geq 0}$, we denote $t^{-}_{(n)} \triangleq t_{(n)}-1$ for $n > 0$.\\
As mentioned earlier, $\bar{\theta}_t$ is updated only along a subsequence $\{t_{(n)}\}_{n \geq 0}$ of $\{t\}_{t \geq 0}$ with $t_0 = 0$ as follows:
\begin{equation}\label{eqn:thetareal}
\bar{\theta}_{t_{(n+1)}} = \bar{\theta}_{t_{(n)}} + \alpha_{t_{(n+1)}}\left(({\xi}^{(0)}_{t^{-}_{(n+1)}}, {\xi}^{(1)}_{t^{-}_{(n+1)}})^{\top} - \bar{\theta}_{t_{(n)}}\right).
\end{equation} 
Now define $\Psi(\omega, \theta) = (\Psi_1(\omega, \theta), \Psi_2(\omega, \theta))^{\top}$, where 
\begin{gather}
\Psi_1(\omega, \theta) \triangleq \frac{\mathbb{E}_{\widehat{\theta}}\left[ \mathbf{g}_{1}(\bar{\mathcal{J}}_{\omega}(\mathbf{z}), \mathbf{z}, \gamma_{\rho}(\bar{\mathcal{J}}_{\omega}, \widehat{\theta})\right]}{\mathbb{E}_{\widehat{\theta}}\left[\mathbf{g}_{0}(\bar{\mathcal{J}}_{\omega}(\mathbf{z}), \gamma_{\rho}(\bar{\mathcal{J}}_{\omega}, \widehat{\theta})\right]},\hspace*{20mm}\\
\Psi_2(\omega, \theta) \triangleq  \frac{\mathbb{E}_{\widehat{\theta}}\left[\mathbf{g}_{2}\left(\bar{\mathcal{J}}_{\omega}(\mathbf{z}), \mathbf{z}, \gamma_{\rho}(\bar{\mathcal{J}}_{\omega}, \widehat{\theta}), \Psi_1(\omega, \theta)\right)\right]}{\mathbb{E}_{\widehat{\theta}}\left[\mathbf{g}_{0}\left(\bar{\mathcal{J}}_{\omega}(\mathbf{z}), \gamma_{\rho}(\bar{\mathcal{J}}_{\omega}, \widehat{\theta})\right)\right]}.
\end{gather}

We now state our main theorem. The theorem states that the model sequence $\{\theta_{t}\}$ generated by Algorithm \ref{algo:sce-mspbem} converges to $\theta^{*} = (z^{*}, 0_{k \times k})^{\top}$, which is the degenerate distribution concentrated at $z^{*}$.
\begin{theorem}\label{thm:main}
Let $S(z) = exp(rz)$, $r \in \mathbb{R}_{+}$.  Let $\rho \in (0,1)$, $\lambda \in (0,1)$ and $\lambda > \rho$. Let $\theta_0 = (\mu_0, qI_{k \times k})^{\top}$, where $q \in \mathbb{R}_{+}$. Let the step-size sequences $\alpha_t$, $\beta_t$, $t \in \mathbb{Z}_{+}$ satisfy (\ref{eqn:learnrt}). Also let $c_t \rightarrow 0$. Suppose $\{\theta_t = (\mu_t, \Sigma_t)^{\top}\}_{t \in \mathbb{Z}_{+}}$ is the sequence generated by Algorithm \ref{algo:sce-mspbem} and assume $\theta_{t} \in int(\Theta)$, $\forall t \in \mathbb{Z}_{+}$. Also, let the assumptions (A1), (A2), (A3) and (A4) hold. Further, we assume that there exists a continuously differentiable function $V:\Theta \rightarrow \mathbb{R}_{+}$ s.t. $\nabla V^{\top}(\theta)\Psi(\omega_{*}, \theta) < 0$, $\forall \theta \in \Theta\smallsetminus\{\theta^{*}\}$ and $\nabla V^{\top}(\theta^{*})\Psi(\omega_{*}, \theta^{*}) = 0$. Then, there exists $q^{*} \in \mathbb{R}_{+}$, $r^{*} \in \mathbb{R}_{+}$ and $\rho^{*} \in (0,1)$ s.t. $\forall q > q^{*}$, $\forall r > r^{*}$ and $\forall \rho < \rho^{*}$,
\begin{gather*}
\hspace{1mm} \lim_{t \rightarrow \infty} \bar{\mathcal{J}}(\omega_t, \mu_{t}) = \mathcal{J}^{*}  \hspace{3mm} and \hspace{2mm} \lim_{t \rightarrow \infty}\theta_{t} = \theta^{*} = (z^{*}, 0_{k \times k})^{\top} \textrm{ almost surely},
\end{gather*}
where $\mathcal{J}^{*}$ and $z^{*}$ are defined in (\ref{eqn:mspbeobj}). Further, since $\mathcal{J} = -\mathrm{MSPBE}$, the algorithm SCE-MSPBEM converges to the global minimum of MSPBE a.s.
\end{theorem}
\begin{Proof}
Rewriting the equation (\ref{eq:thupd}) along the subsequence $\{t_{(n)}\}_{n \in \mathbb{Z}_{+}}$, we have for $n \in \mathbb{Z}_{+}$,
\begin{equation}\label{eqn:thetap1}
\theta_{t_{(n+1)}} = \theta_{t_{(n)}} + \alpha_{t_{(n+1)}}\left(({\xi}^{(0)}_{t^{-}_{(n+1)}}, {\xi}^{(1)}_{t^{-}_{(n+1)}})^{\top} - \theta_{t_{(n)}}\right).
\end{equation} 
The iterates $\theta_{t_{(n)}}$ are stable, \emph{i.e.}, $\sup_{n}{\Vert \theta_{t_{(n)}}\Vert} < \infty$ \emph{a.s.} It is directly implied from the assumptions that $\theta_{t_{(n)}} \in int(\Theta)$ and $\Theta$ is a compact set.\\
Rearranging the equation (\ref{eqn:thetap1}) we get, for $n \in \mathbb{Z}_{+}$,
\begin{equation}
\theta_{t_{(n+1)}} = \theta_{t_{(n)}} + \alpha_{t_{(n+1)}}\left(\Psi(\omega_{*}, \theta_{t_{(n)}}) + \mathit{o}(1)\right).
\end{equation}
This easily follows from the fact that, for $t_{(n)} < t \leq t_{(n+1)}$, the random variables ${\xi}^{(0)}_t$ and ${\xi}^{(1)}_t$ estimates the quantities $\Psi_1(\omega_{t_{(n)}}, \theta_{t_{(n)}})$ and $\Psi_2(\omega_{t_{(n)}}, \theta_{t_{(n)}})$ respectively. Since $c_t \rightarrow 0$, the estimation error decays to $0$. Hence the term $\mathit{o}(1)$.\\
The limit points of the above recursion are the roots of $\Psi$. Hence by equating $\Psi_1(\omega_{*}, \theta)$ to $0_{k \times 1}$, we get,
\begin{eqnarray}\label{eqn:musol1}
\mu = \frac{\mathbb{E}_{\widehat{\theta}}\left[\mathbf{g_{1}}\bm{\big{(}}\mathcal{J}(\mathbf{z}), \mathbf{z},  \gamma_{\rho}(\mathcal{J}, \widehat{\theta})\bm{\big{)}}\right]}{\mathbb{E}_{\widehat{\theta}}\left[\mathbf{g_{0}}\bm{\big{(}}\mathcal{J}(\mathbf{z}), \gamma_{\rho}(\mathcal{J}, \widehat{\theta})\bm{\big{)}}\right]}.
\end{eqnarray}
Equating $\Psi_2(\omega_{*}, \theta)$ to $\mathbb{O}$ $(= 0_{k \times k})$, we get,
\begin{eqnarray}\label{eqn:sgmsol2}
\frac{\mathbb{E}_{\widehat{\theta}}\left[\mathbf{g_{2}}\bm{\big{(}}\mathcal{J}(\mathbf{z}), \mathbf{z}, \gamma_{\rho}(\mathcal{J},\widehat{\theta}), \mu\bm{\big{)}}\right]}{\mathbb{E}_{\widehat{\theta}}\left[\mathbf{g_{0}}\bm{\big{(}}\mathcal{J}(\mathbf{z}), \gamma_{\rho}(\mathcal{J}, \widehat{\theta})\bm{\big{)}}\right]} -  \Sigma = \mathbb{O}.
\end{eqnarray}
For brevity, we define 
\begin{equation}\label{eq:quanstar}
\gamma^{*}_{\rho}(\theta) \triangleq \gamma_{\rho}(\mathcal{J}, \theta), \hspace*{5mm}\mathbf{\hat{g}_{0}}(z, \theta) \triangleq \mathbf{g_{0}}\bm{\big{(}}\mathcal{J}(z), \gamma^{*}_{\rho}(\theta)\bm{\big{)}} \textrm{ and } L(\theta) \triangleq \mathbb{E}_{\theta}\left[\mathbf{\hat{g}_{0}}(\mathbf{z}, \theta)\right].
\end{equation}
Substituting the expression for $\mu$ from (\ref{eqn:musol1}) in (\ref{eqn:sgmsol2}) and after further simplification we get,
\begin{gather*}
(1/L(\widehat{\theta}))\mathbb{E}_{\widehat{\theta}}\left[\mathbf{\hat{g}_{0}}(\mathbf{z},\widehat{\theta})\mathbf{z} \mathbf{z}^{\top}\right]-\mu\mu^{\top}\hspace*{-2mm}-\Sigma = \mathbb{O}.\hspace*{57mm}
\end{gather*}
Since $\Sigma = \mathbb{E}_{\theta}\left[\mathbf{z} \mathbf{z}^{\top}\right]-\mu\mu^{\top}$, the above equation implies
\begin{gather*}
\begin{aligned}
&(1/L(\widehat{\theta}))\mathbb{E}_{\widehat{\theta}}\left[\mathbf{\hat{g}_{0}}(\mathbf{z}, \widehat{\theta})\mathbf{z} \mathbf{z}^{\top}\right] - \mathbb{E}_{\theta}\left[\mathbf{z} \mathbf{z}^{\top}\right]  = \mathbb{O} \hspace*{4mm}
\\&\Longrightarrow_1 \hspace*{4mm}\mathbb{E}_{\widehat{\theta}}\left[\mathbf{\hat{g}_{0}}(\mathbf{z}, \widehat{\theta})\mathbf{z}\mathbf{z}^{\top}\right] - L(\widehat{\theta})\mathbb{E}_{\theta}\left[\mathbf{z} \mathbf{z}^{\top}\right] = \mathbb{O},
\\&\Longrightarrow_2 \hspace*{4mm}(1-\lambda)\mathbb{E}_{\theta}\left[\mathbf{\hat{g}_{0}}(\mathbf{z}, \widehat{\theta})\mathbf{z}\mathbf{z}^{\top}\right] + \lambda\mathbb{E}_{\theta_0}\left[\mathbf{\hat{g}_{0}}(\mathbf{z}, \widehat{\theta})\mathbf{z}\mathbf{z}^{\top}\right]- L(\widehat{\theta})\mathbb{E}_{\theta}\left[\mathbf{z} \mathbf{z}^{\top}\right] = \mathbb{O}
\\&\Longrightarrow_3 \hspace*{4mm}(1-\lambda)\mathbb{E}_{\theta}\left[\mathbf{\hat{g}_{0}}(\mathbf{z}, \widehat{\theta})\mathbf{z}\mathbf{z}^{\top}\right] + \lambda\mathbb{E}_{\theta_0}\left[\mathbf{\hat{g}_{0}}(\mathbf{z}, \widehat{\theta})\mathbf{z}\mathbf{z}^{\top}\right] -  (1-\lambda)\mathbb{E}_{\theta}\left[\mathbf{\hat{g}_{0}}(\mathbf{z}, \widehat{\theta})\right]\mathbb{E}_{\theta}\left[\mathbf{z} \mathbf{z}^{\top}\right]  - \\&\hspace*{90mm} \lambda\mathbb{E}_{\theta_0}\left[\mathbf{\hat{g}_{0}}(\mathbf{z}, \widehat{\theta})\right]\mathbb{E}_{\theta}\left[\mathbf{z} \mathbf{z}^{\top}\right] = \mathbb{O}
\\&\Longrightarrow_4 \hspace*{4mm}(1-\lambda)\mathbb{E}_{\theta}\left[\left(\mathbf{\hat{g}_{0}}(\mathbf{z}, \widehat{\theta})-\mathbb{E}_{\theta}\left[\mathbf{\hat{g}_{0}}(\mathbf{z}, \widehat{\theta})\right]\right)\mathbf{z}\mathbf{z}^{\top}\right] + \lambda\mathbb{E}_{\theta_0}\left[\left(\mathbf{\hat{g}_{0}}(\mathbf{z}, \widehat{\theta})-\mathbb{E}_{\theta_0}\left[\mathbf{\hat{g}_{0}}(\mathbf{z}, \widehat{\theta})\right]\right)\mathbf{z}\mathbf{z}^{\top}\right] + \\&\hspace*{60mm}   - \lambda\left(\mathbb{E}_{\theta}\left[\mathbf{z} \mathbf{z}^{\top}\right] - \mathbb{E}_{\theta_0}\left[\mathbf{z} \mathbf{z}^{\top}\right]\right)\mathbb{E}_{\theta_0}\left[\mathbf{\hat{g}_{0}}(\mathbf{z}, \widehat{\theta})\right] = \mathbb{O} \\
&\Longrightarrow_5 \hspace*{4mm}(1-\lambda)\Sigma^{2}\mathbb{E}_{\theta}\left[\nabla_{z}^{2}\mathbf{\hat{g}_{0}}(\mathbf{z}, \widehat{\theta})\right] + \lambda\Sigma_{0}^{2}\mathbb{E}_{\theta_{0}}\left[\nabla_{z}^{2}\mathbf{g_{0}}(\mathbf{z}, \widehat{\theta})\right] + \\&\hspace*{50mm}\lambda\left(\mathbb{E}_{\theta}\left[\mathbf{z} \mathbf{z}^{\top}\right] - \mathbb{E}_{\theta_0}\left[\mathbf{z} \mathbf{z}^{\top}\right]\right)\mathbb{E}_{\theta_0}\left[\mathbf{\hat{g}_{0}}(\mathbf{z}, \widehat{\theta})\right] = \mathbb{O}
\\&\Longrightarrow_6 \hspace*{4mm}(1-\lambda)\Sigma^{2}\mathbb{E}_{\theta}\left[\nabla_{z}^{2}\mathbf{\hat{g}_{0}}(\mathbf{z}, \widehat{\theta})\right] + \lambda q^{2}\mathbb{E}_{\theta_{0}}\left[\nabla_{z}^{2}\mathbf{\hat{g}_{0}}(\mathbf{z}, \widehat{\theta})\right] + \\&\hspace*{50mm}\lambda\left(\mathbb{E}_{\theta}\left[\mathbf{z} \mathbf{z}^{\top}\right] - \mathbb{E}_{\theta_0}\left[\mathbf{z} \mathbf{z}^{\top}\right]\right)\mathbb{E}_{\theta_0}\left[\mathbf{\hat{g}_{0}}(\mathbf{z}, \widehat{\theta})\right] = \mathbb{O}
%\\&\Longrightarrow_7 \hspace*{4mm}(1-\lambda)\Sigma^{2}\mathbb{E}_{\theta}\left[\nabla_{z}^{2}\mathbf{\hat{g}_{0}}(\mathbf{z}, %%\widehat{\theta})\right] + \lambda q^{2}\mathbb{E}_{\theta_{0}}\left[\nabla_{z}^{2}\mathbf{\hat{g}_{0}}(\mathbf{z}, \widehat{\theta})%%%%\right] + \lambda\rho\left(S(B_1) - S(B_2)\right)\mathbb{E}_{\theta_0}\left[\mathbf{z} \mathbf{z}^{\top}\right] > \mathbb{O}
\end{aligned}
\end{gather*}
\begin{equation}\label{eqn:zeq}
\begin{aligned}
\hspace*{0mm}&\Longrightarrow_8 \hspace*{4mm}(1-\lambda)\Sigma^{2} \mathbb{E}_{\theta}\left[S(\mathcal{J}(\mathbf{z}))G^{r}(\mathbf{z})I_{\{\mathcal{J}(\mathbf{z}) \geq \gamma^{*}_{\rho}(\widehat{\theta})\}}\right] + \lambda q^{2} \mathbb{E}_{\theta_0}\left[S(\mathcal{J}(\mathbf{z}))G^{r}(\mathbf{z})I_{\{\mathcal{J}(\mathbf{z}) \geq \gamma^{*}_{\rho}(\widehat{\theta})\}}\right] + \\&\hspace*{50mm}\lambda\left(\mathbb{E}_{\theta}\left[\mathbf{z} \mathbf{z}^{\top}\right] - \mathbb{E}_{\theta_0}\left[\mathbf{z} \mathbf{z}^{\top}\right]\right)\mathbb{E}_{\theta_0}\left[\mathbf{\hat{g}_{0}}(\mathbf{z}, \widehat{\theta})\right] = \mathbb{O},
\end{aligned}
\end{equation}
where $G^{r}(z) \triangleq r^{2}\nabla\mathcal{J}(z)\nabla\mathcal{J}(z)^{\top} + r\nabla^{2}\mathcal{J}(z)$. Note that $\Longrightarrow_5$ follows from ``integration by parts" rule for multivariate Gaussian and $\Longrightarrow_8$ follows from the assumption $S(z) = exp(rz)$. Note that for each $z \in \mathcal{Z}$, $G^{r}(z) \in \mathbb{R}^{k \times k}$. Hence we denote $G^{r}(z)$ as $\left[G^{r}_{ij}(z)\right]_{i=1,j=1}^{i=k,j=k}$. For brevity, we also define
\begin{gather}
F^{r, \rho}(z,\theta) \triangleq S(\mathcal{J}(z))G^{r}(z)I_{\{\mathcal{J}(z) \geq \gamma^{*}_{\rho}(\theta)\}},
\end{gather}
where $F^{r, \rho}(z,\theta) \in \mathbb{R}^{k \times k}$ which is also denoted as $\left[F^{r, \rho}_{ij}(z)\right]_{i=1,j=1}^{i=k,j=k}$.\\
Hence equation (\ref{eqn:zeq}) becomes,
\begin{equation}\label{eqn:zeq2}
\begin{aligned}
(1-\lambda)\Sigma^{2} \mathbb{E}_{\theta}\left[F^{r, \rho}(\mathbf{z}, \widehat{\theta})\right] + \lambda q^{2} \mathbb{E}_{\theta_0}\left[F^{r, \rho}(\mathbf{z}, \widehat{\theta})\right] + \lambda\left(\mathbb{E}_{\theta}\left[\mathbf{z} \mathbf{z}^{\top}\right] - \mathbb{E}_{\theta_0}\left[\mathbf{z} \mathbf{z}^{\top}\right]\right)\mathbb{E}_{\theta_0}\left[\mathbf{\hat{g}_{0}}(\mathbf{z}, \widehat{\theta})\right] = \mathbb{O}.
\end{aligned}
\end{equation}
Note that $(\nabla_{i}\mathcal{J})^{2} \geq 0$. Hence we can find a $r^{*} \in \mathbb{R}_{+}$ \emph{s.t.} $G^{r}_{ii}(z) > 0$, $\forall r > r^{*}$, $1 \leq i \leq k$, $\forall z \in \mathcal{Z}$. This further implies that $\mathbb{E}_{\theta}[F^{r, \rho}_{ii}(\mathbf{z}, \widehat{\theta}] > 0$, $\forall \theta \in \Theta$. Also since $\mathcal{Z}$ is compact and $\mathcal{J}$ is continuous, we have $J(z) > B_1 > -\infty$, $\forall z \in \mathcal{Z}$.  Hence we obtain the following bound:
\begin{equation}\label{eqn:zeq3}
\left(\mathbb{E}_{\theta}\left[\mathbf{z}^{2}_{i}\right] - \mathbb{E}_{\theta_0}\left[\mathbf{z}^{2}_{i}\right]\right)\mathbb{E}_{\theta_0}\left[\mathbf{\hat{g}_{0}}(\mathbf{z}, \widehat{\theta})\right] > \rho K_3 S(B_1), \textrm{ where } 0 < K_3 < \infty.
\end{equation}
Now from (\ref{eqn:zeq2}) and (\ref{eqn:zeq3}), we can find a $\rho^{*} \in (0,1)$ and $q^{*} \in \mathbb{R}$ \emph{s.t.} $\forall \rho < \rho^{*}$, $\forall q > q^{*}$, $\forall r > r^{*}$, we have, 
\begin{equation}
(1-\lambda)\Sigma^{2} \mathbb{E}_{\theta}\left[F^{r, \rho}_{ii}(\mathbf{z}, \widehat{\theta})\right] + \lambda q^{2} \mathbb{E}_{\theta_0}\left[F^{r, \rho}_{ii}(\mathbf{z}, \widehat{\theta})\right] + \lambda\left(\mathbb{E}_{\theta}\left[\mathbf{z}^{2}_{i}\right] - \mathbb{E}_{\theta_0}\left[\mathbf{z}^{2}_{i}\right]\right)\mathbb{E}_{\theta_0}\left[\mathbf{\hat{g}_{0}}(\mathbf{z}, \widehat{\theta})\right] > 0.
\end{equation}
This contradicts equation (\ref{eqn:zeq2}) for such a choice of $\rho$, $q$ and $r$. This implies that each of the terms in equation (\ref{eqn:zeq2}) is $0$, \emph{i.e.},
\begin{gather}
\Sigma^{2} \mathbb{E}_{\theta}\left[F^{r, \rho}(\mathbf{z}, \widehat{\theta})\right] = \mathbb{O},\label{eq:zeq3}\\ 
q^{2} \mathbb{E}_{\theta_0}\left[F^{r, \rho}(\mathbf{z}, \widehat{\theta})\right]  = \mathbb{O} \hspace*{10mm}\textrm{ and }\label{eq:zeq4} \\
\left(\mathbb{E}_{\theta}\left[\mathbf{z} \mathbf{z}^{\top}\right] - \mathbb{E}_{\theta_0}\left[\mathbf{z} \mathbf{z}^{\top}\right]\right)\mathbb{E}_{\theta_0}\left[\mathbf{\hat{g}_{0}}(\mathbf{z}, \widehat{\theta})\right] = \mathbb{O}.\label{eq:zeq5}
\end{gather}  
It is easy to verify that $\Sigma = \mathbb{O}$ simultaneously satisfies (\ref{eq:zeq3}), (\ref{eq:zeq4}) and (\ref{eq:zeq5}). Besides $\Sigma = \mathbb{O}$ is the only solution for $r > r^{*}$. This is because we have already established earlier that $\forall r > r^{*}$, $1 \leq i \leq k$, $\forall z \in \mathcal{Z}$ $\mathbb{E}_{\theta}[F^{r, \rho}_{ii}(\mathbf{z}, \widehat{\theta}] > 0$, $\forall \theta \in \Theta$. This proves that for any $z \in \mathcal{Z}$, the degenerate distribution concentrated on $z$ given by $\theta_z = (z, 0_{k \times k})^{\top}$ is a potential limit point of the recursion. From (\ref{eq:zeq4}), we have  
\begin{equation*}
\mathbb{E}_{\theta_0}\left[F^{r, \rho}(\mathbf{z}, \widehat{\theta})\right]  = \mathbb{O} \hspace*{10mm} \Longrightarrow \hspace*{10mm}\gamma^{*}_{\rho}(\widehat{\theta}) = \mathcal{J}(z^{*}).
\end{equation*}
\textbf{Claim A:} The only degenerate distribution which satisfies the above condition is $\theta^{*} = (z^{*}, 0_{k \times k})^{\top}$.\\
The above claim can be verified as follows: if there exists $z^{\prime} (\in \mathcal{Z}) \neq z^{*}$ \emph{s.t.} $\gamma^{*}_{\rho}(\widehat{\theta}_{z^{\prime}}) = \mathcal{J}(z^{*})$ is satisfied (where $\widehat{\theta}_{z^{\prime}}$ represents the mixture distribution $\widehat{f}_{\theta_{z^{\prime}}}$) , then from the definition of $\gamma^{*}_{\rho}(\cdot)$ in (\ref{eq:quantile}) and (\ref{eq:quanstar}), we can find an increasing sequence $\{l_i\}$, where $l_i > J(z^{\prime})$ \emph{s.t.} the following property is satisfied:
\begin{equation}\label{eq:sqeq}
\lim_{i \rightarrow \infty}l_i = \mathcal{J}(z^{*}) \textrm{ and } \mathbb{P}_{\widehat{\theta}}(\mathcal{J}(z) \geq l_i) \geq \rho.
\end{equation}
But $\mathbb{P}_{\widehat{\theta_{z^{\prime}}}}(\mathcal{J}(z) \geq l_i) = (1-\lambda)\mathbb{P}_{\theta_{z^{\prime}}}(\mathcal{J}(z) \geq l_i) + \lambda \mathbb{P}_{\theta_{0}}(\mathcal{J}(z) \geq l_i)$ and $\mathbb{P}_{\theta_{z^{\prime}}}(\mathcal{J}(z) \geq l_i) = 0$, $\forall i$. Therefore from (\ref{eq:sqeq}), we get,
\begin{gather*}
\begin{aligned}
&\mathbb{P}_{\widehat{\theta_{z^{\prime}}}}(\mathcal{J}(z) \geq l_i) \geq \rho\\
&\Rightarrow (1-\lambda)\mathbb{P}_{\theta_{z^{\prime}}}(\mathcal{J}(z) \geq l_i) + \lambda \mathbb{P}_{\theta_{0}}(\mathcal{J}(z) \geq l_i) \geq \rho\\
&\Rightarrow \lambda \mathbb{P}_{\theta_{0}}(\mathcal{J}(z) \geq l_i) \geq \rho\\
&\Rightarrow \mathbb{P}_{\theta_{0}}(\mathcal{J}(z) \geq l_i) \geq \frac{\rho}{\lambda} < 1.
\end{aligned}
\end{gather*}
Recollect that $l_i \rightarrow \mathcal{J}(z^{*})$. Thus by the continuity of probability measures we get 
\begin{equation*}
0 = \mathbb{P}_{\theta_{0}}(\mathcal{J}(z) \geq J(z^{*})) = \lim_{i \rightarrow \infty}\mathbb{P}_{\theta_{0}}(\mathcal{J}(z) \geq l_i) \geq \frac{\rho}{\lambda},
\end{equation*}
which is a contradiction. This proves the Claim A. Now the only remaining task is to prove $\theta^{*}$ is a stable attractor. This easily follows from the assumption regarding the existence of the Lyapunov function $V$ in the statement of the theorem.
\end{Proof}

\subsection{Computational Complexity}
\textit{The computational load of this algorithm is $\Theta(k^{2})$ per iteration} which comes from (\ref{eqn:alomgupd}). Least squares algorithms like LSTD and LSPE also require $\Theta(k^{2})$ per iteration. However, LSTD requires an extra operation of inverting the $k \times k$ matrix $A_{T}$ which requires an extra computational effort of $\Theta(k^{3})$. (Note that LSPE also requires a $k \times k$ matrix inversion). This makes the \textit{overall complexity of LSTD and LSPE to be $\Theta(k^{3})$}. Further in some cases the matrix $A_{T}$ may not be invertible. In that case, the pseudo inverse of $A_{T}$ needs to be obtained in LSTD, LSPE which is computationally even more expensive. Our algorithm does not require such an inversion procedure. Also \textit{even though the complexity of the first order temporal difference algorithms such as TD($\lambda$) and GTD2 is $\Theta(k)$, the approximations they produced in the experiments we conducted turned out to be inferior to ours and also showed a slower rate of convergence than our algorithm.} Another noteworthy characteristic exhibited by our algorithm is \textit{stability}. Recall that the convergence of TD($0$) is guaranteed by the requirements that the Markov Chain of $\mathrm{P}^{\pi}$ should be ergodic and the sampling distribution $\nu$ to be its stationary distribution. The classic example of Baird's 7-star \cite{baird1995residual} violates those restrictions and hence TD(0) is seen to diverge. However, our algorithm does not impose such restrictions and shows stable behaviour even in non-ergodic off policy cases such as the Baird's example.

\section{Experimental Results}
We present here a numerical comparison of SCE-MSPBEM with various state-of-the-art algorithms in the literature on some benchmark Reinforcement Learning problems. In each of the experiments, a random trajectory $\{(s_t, r_{t}, s^{\prime}_{t})\}_{t=0}^{\infty}$ is chosen and all the algorithms are updated using it. Each $s_t$ in $\{(s_t, r_{t}, s^{\prime}_{t}), t \geq 0\}$ is sampled using an arbitrary distribution $\nu$ over $\mathbb{S}$. The algorithms are run on multiple trajectories and the average of the results obtained are plotted. The $x$-axis in the plots is $t/1000$, where $t$ is the iteration number. In each case, the learning rates $\alpha_{t}, \beta_{t}$ are chosen so that the condition (\ref{eqn:learnrt}) is satisfied. The function $S(\cdot)$ is chosen as $S(x) = \exp{(rx)}$, where $r \in \mathbb{R}$ is chosen appropriately.

SCE-MSPBEM was tested on the following benchmark problems:
\begin{enumerate}
\item
Linearized Cart-Pole Balancing \cite{dann2014policy}
\item
5-Link Actuated Pendulum Balancing \cite{dann2014policy}
\item
Baird's 7-Star MDP \cite{baird1995residual}
\item
10-state Ring MDP \cite{kveton2006solving}
\item
Large state space and action space with Radial Basis Functions
\item
Large state space and action space with Fourier Basis Functions \cite{konidaris2011value}
\end{enumerate}
\subsection{Experiment 1: Linearized Cart-Pole Balancing \cite{dann2014policy}}
\textbf{Setup: } A pole with mass $m$ and length $l$ is connected to a cart of mass $M$ . It can rotate $360^{\circ}$ and the cart is free to move in either direction within the bounds of a linear track. \\
\textbf{Goal: } To balance the pole upright and the cart at the centre of the track. \\ 
\textbf{State space: }The 4-tuple $[x, \dot{x}, \psi, \dot{\psi}]$ where $\psi$ is  the angle of the pendulum \emph{w.r.t.} the vertical axis, $\dot{\psi}$ is the angular velocity, $x$ the relative cart position from the centre of the track and $\dot{x}$ is its velocity.\\
\textbf{Control space: }The controller applies a horizontal force $a$ on the cart parallel to the track. The stochastic policy used in this setting corresponds to $\pi(a|s) = \mathcal{N}(a | \beta_{1}^{\top}s, \sigma_1^{2})$.\\
\textbf{System dynamics: }
The dynamical equations of the system are given by
\begin{equation}
\ddot{\psi} = \frac{-3ml\dot{\psi}^{2}\sin{\psi}\cos{\psi}+(6M+m)g\sin{\psi}-6(a-b\dot{\psi})\cos{\psi}}{4l(M+m)-3ml\cos{\psi}},
\end{equation}
\begin{equation}
\ddot{x} = \frac{-2ml\dot{\psi}^{2}\sin{\psi}+3mg\sin{\psi}\cos{\psi}+4a-4b\dot{\psi}}{4(M+m)-3m\cos{\psi}}.
\end{equation}
By making further assumptions on the initial conditions, the system dynamics can be approximated accurately by the linear system 
\begin{equation}
\begin{bmatrix} 
x_{t+1}\\
\dot{x}_{t+1}\\
\psi_{t+1}\\
\dot{\psi}_{t+1}
\end{bmatrix} = \begin{bmatrix} 
x_{t}\\
\dot{x}_{t}\\
\psi_{t}\\
\dot{\psi}_{t}
\end{bmatrix} + \Delta t \begin{bmatrix}
\dot{\psi}_{t} \\
\frac{3(M+m)\psi_t-3a+3b\dot{\psi_t}}{4Ml-ml} \\
\dot{x}_{t} \\
\frac{3mg\psi_t + 4a - 4b\dot{\psi_t}}{4M-m}
\end{bmatrix} + 
\begin{bmatrix}
0 \\
0 \\
0 \\
\mathbf{z}
\end{bmatrix},
\end{equation}
where $\Delta t$ is the integration time step, \emph{i.e.}, the time difference between two transitions and $\mathbf{z}$ is a Gaussian noise on the velocity of the cart with standard deviation $\sigma_{2}$. \\
%\end{center}
\textbf{Reward function: }
$\mathrm{R}(s, a) = \mathrm{R}(\psi, \dot{\psi}, x, \dot{x}, a) = -100\psi^2 - x^2 - \frac{1}{10}a^2$. \\
\textbf{Feature vectors: } $\phi(s \in \mathbb{R}^{4}) = (1, s_{1}^{2}, s_{2}^{2} \dots, s_{1}s_{2}, s_{1}s_{3}, \dots, s_{3}s_{4})^{\top} \in \mathbb{R}^{11}$.\\
\textbf{Evaluation policy: }The policy evaluated in the experiment is the optimal policy $\pi^{*}(a | s) = \mathcal{N}(a | {\beta_{1}^{*}}^{\top}s, {\sigma_{1}^{*}}^{2})$. The parameters $\beta_{1}^{*}$ and $\sigma_{1}^{*}$ are computed using dynamic programming. The feature set chosen above is a perfect feature set, \emph{i.e.}, $V^{\pi^{*}} \in \{\Phi z \vert z \in \mathbb{R}^{k}\}$.
\begin{figure}[!h]
	\centering
	\fbox{\includegraphics[height=40mm, width=80mm]{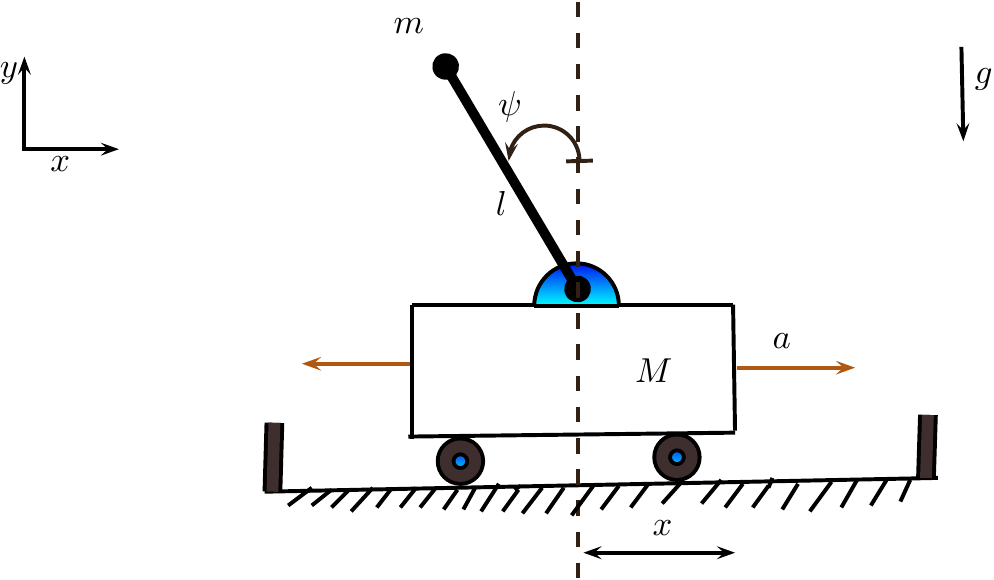}}
	\caption{The Cart-Pole System. The goal is to keep the pole in the  upright position and the cart at the center of the track by pushing the cart with a force $a$ either to the left or the right. The system is parameterized by the position $x$ of the cart, the angle of the pole $\psi$, the velocity $\dot{x}$ and the angular velocity $\dot{\psi}$.}
\end{figure}%
\begin{figure}[!h]
        \begin{subfigure}[h]{0.5\textwidth}
                \fbox{\includegraphics[height=65mm, width=75mm]{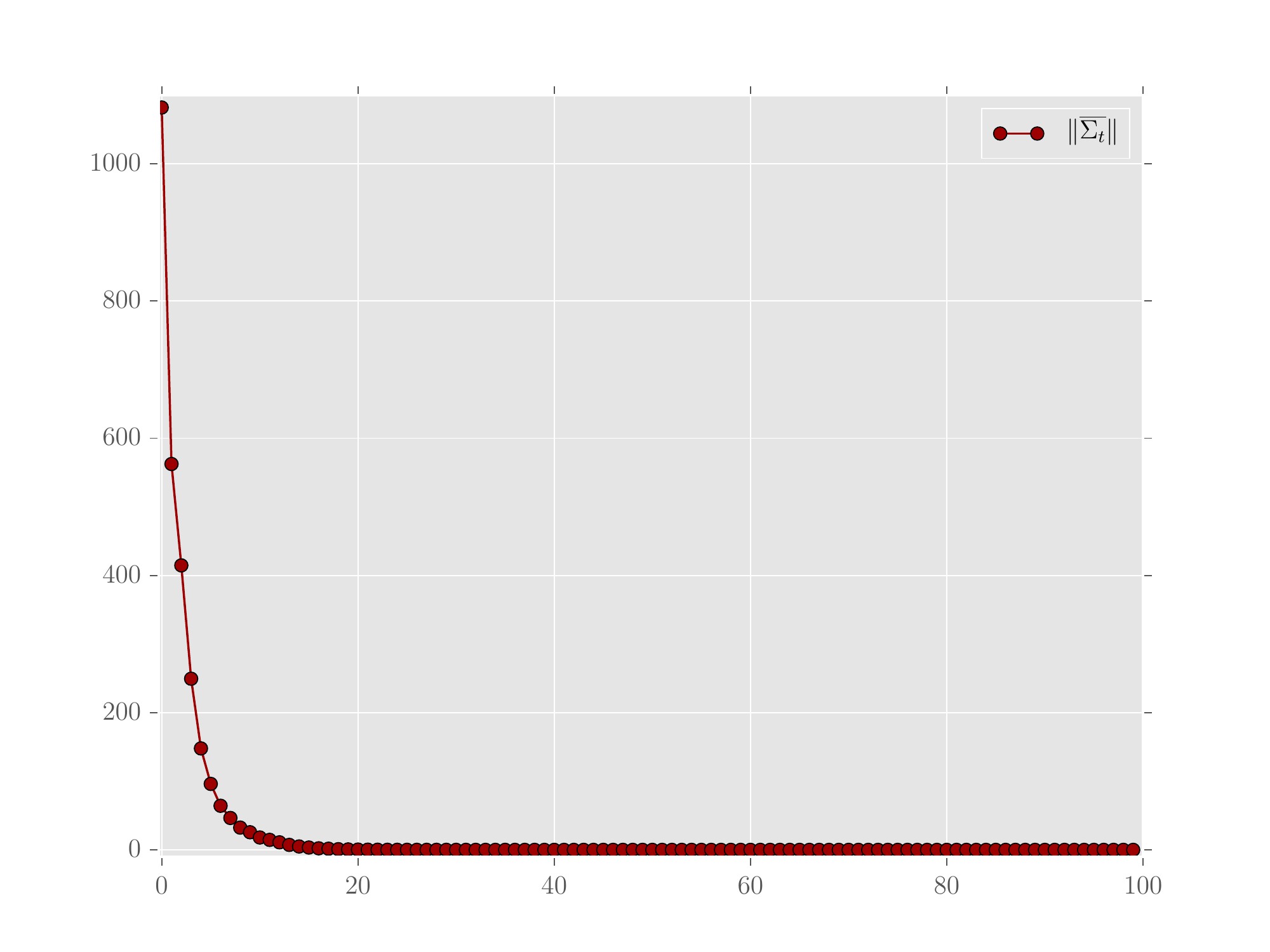}}
                	\subcaption{$\Vert \bar{\Sigma}_{t} \Vert_{F}$  (where $\Vert\cdot\Vert_{F}$ is the Frobenius norm)}
        \end{subfigure}%
        \begin{subfigure}[h]{0.5\textwidth}
                \fbox{\includegraphics[height=65mm, width=75mm]{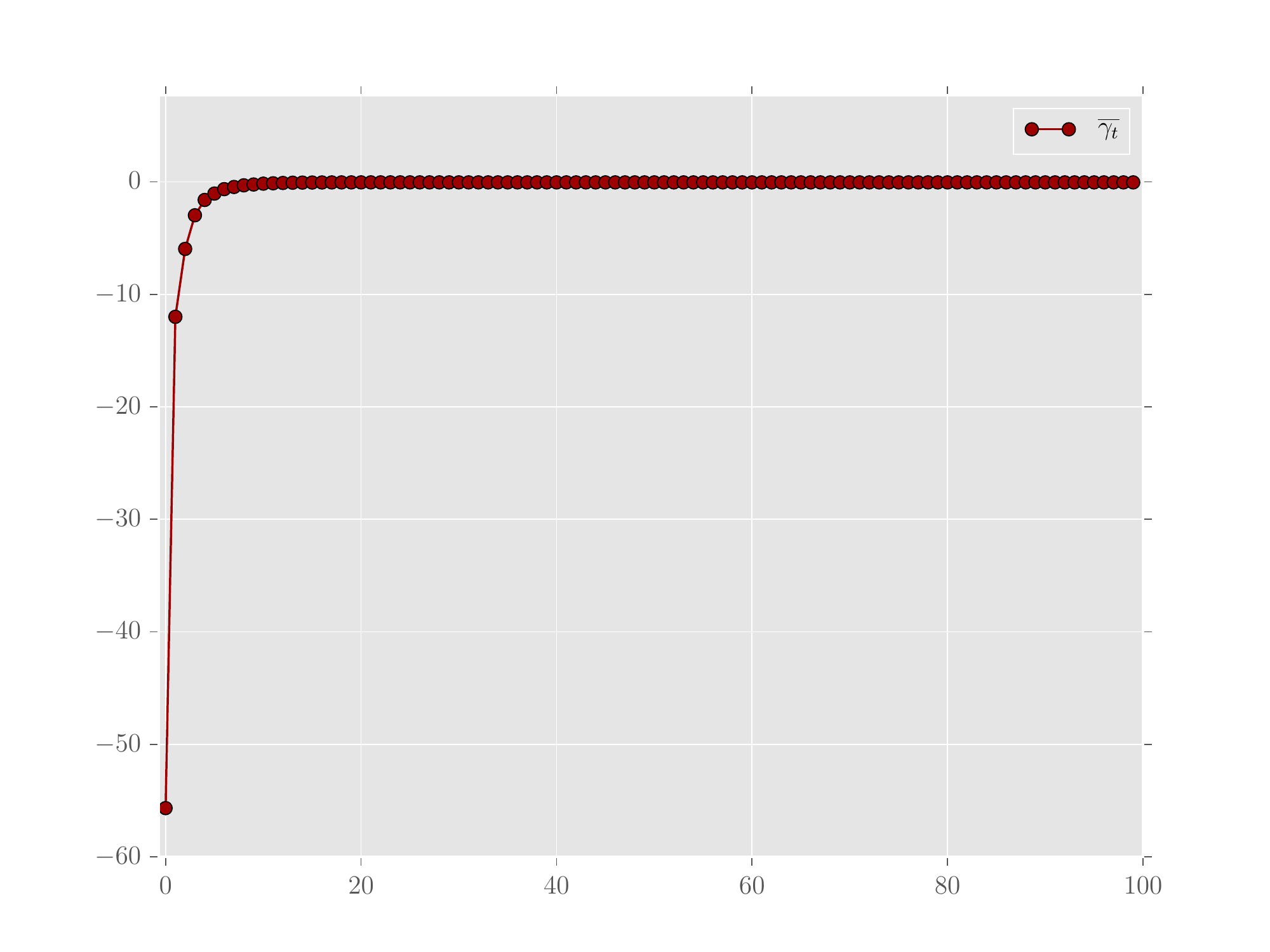}}
                \subcaption{$\bar{\gamma}^{*}_{t}$}
        \end{subfigure}%        
        
        \begin{subfigure}[h]{0.5\textwidth}
                \fbox{\includegraphics[height=65mm, width=75mm]{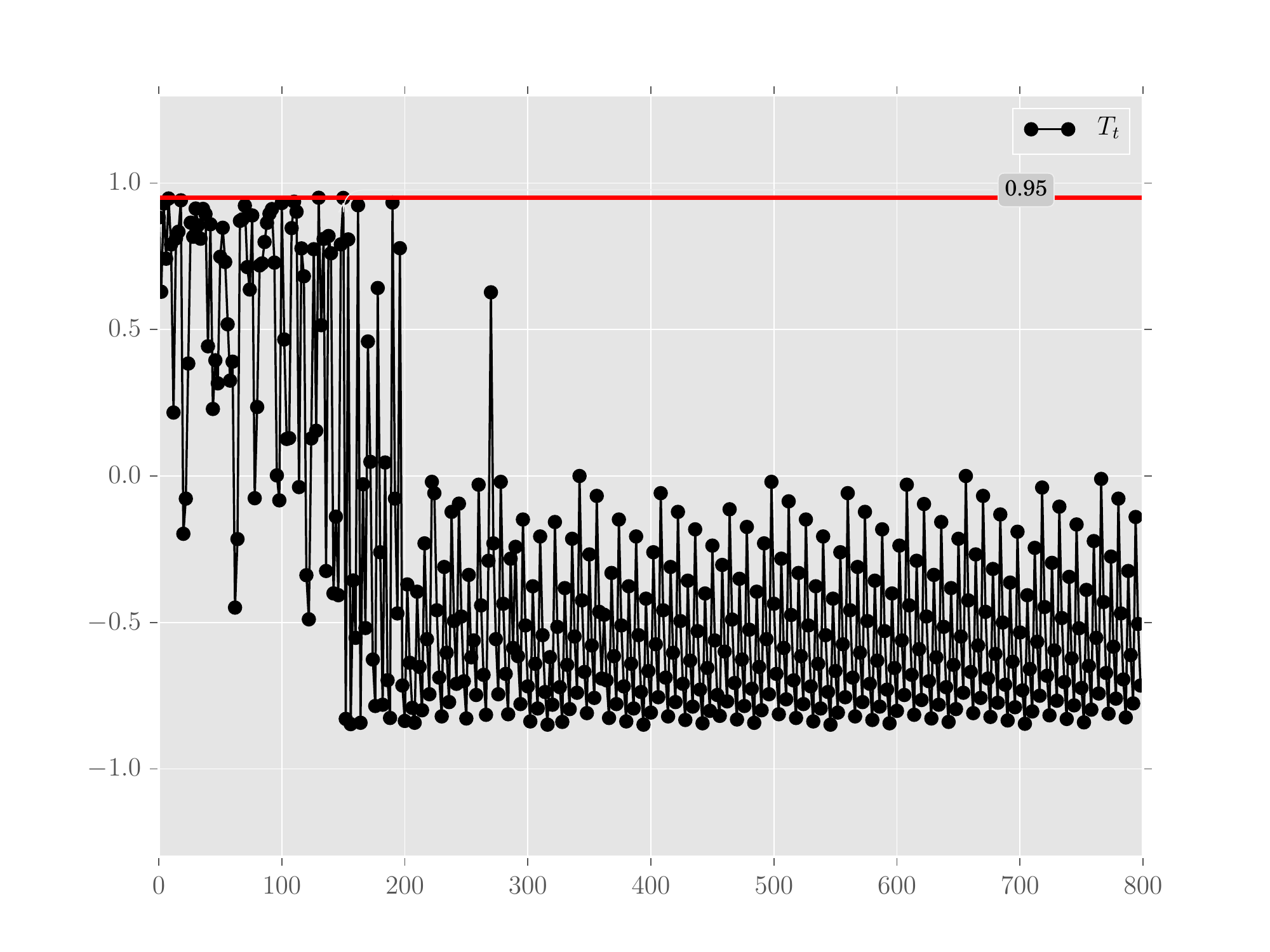}}
                \subcaption{$T_{t}$}
        \end{subfigure}%
        \begin{subfigure}[h]{0.5\textwidth}
                \fbox{\includegraphics[height=65mm, width=75mm]{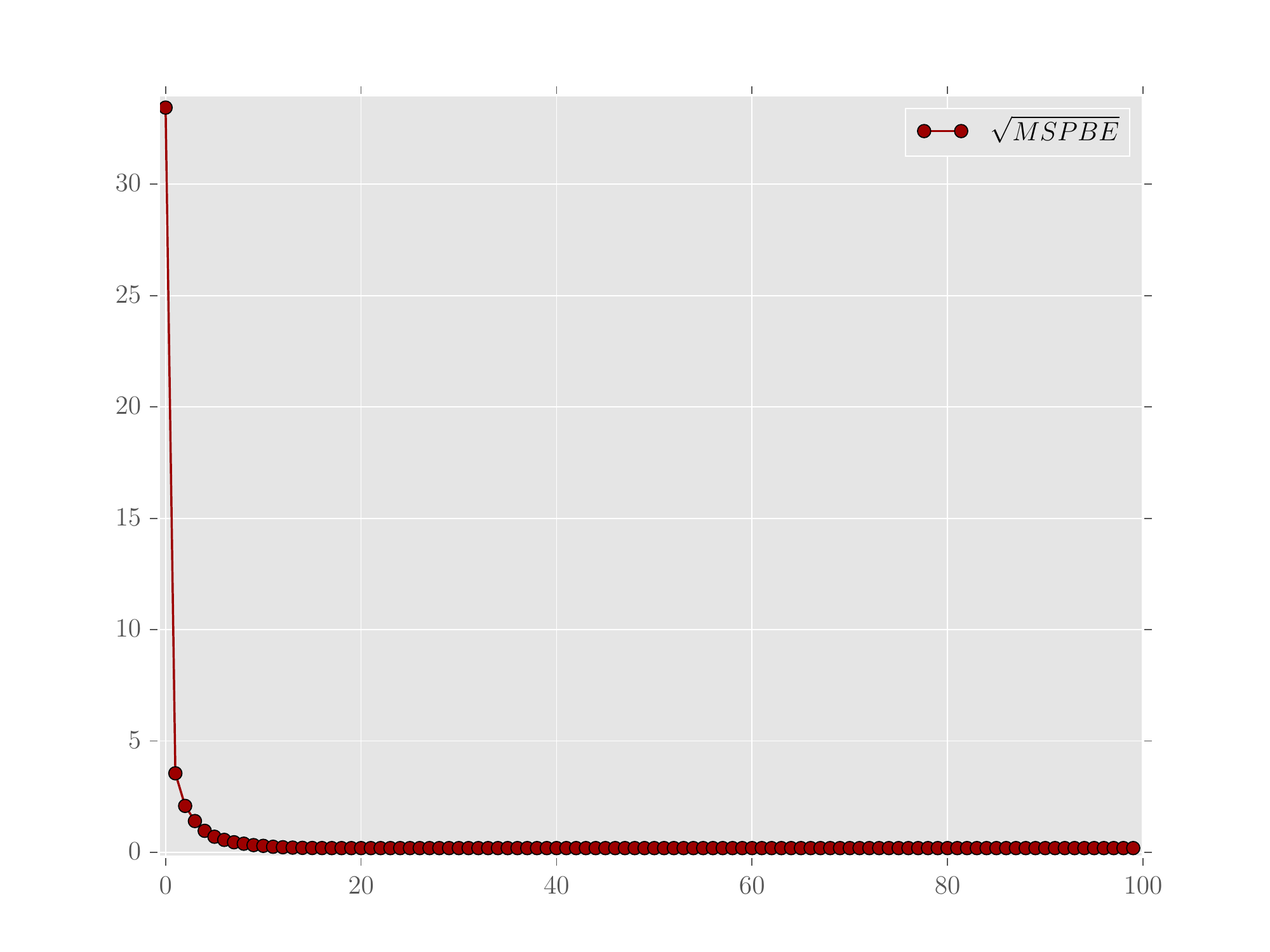}}
                \subcaption{$\sqrt{\mathrm{MSPBE}(\bar{\mu}_{t})}$}
        \end{subfigure}%
		\caption{The Cart-Pole setting. The evolutionary trajectory of the variables $\Vert \bar{\Sigma}_{t} \Vert_{F}$  (where $\Vert\cdot\Vert_{F}$ is the Frobenius norm), $\bar{\gamma}^{*}_{t}$, $T_{t}$ and $\sqrt{\mathrm{MSPBE}(\bar{\mu}_{t})}$. Note that both $\bar{\gamma}^{*}_{t}$ and $\sqrt{\mathrm{MSPBE}(\bar{\mu}_{t})}$  converge to $0$ as $t \rightarrow \infty$, while $\Vert \bar{\Sigma}_{t} \Vert_{F}$ also converges to $0$. This implies that the model $\bar{\theta}_{t} = (\bar{\mu}_t, \bar{\Sigma}_t)^{\top}$ converges to the degenerate distribution concentrated on $z^{*}$. The evolutionary track of $T_{t}$ shows that $T_{t}$ does not cross the $\epsilon_{1} = 0.95$ line after the model $\bar{\theta}_t = (\bar{\mu}_{t}, \bar{\Sigma}_{t})^{\top}$ reaches a close neighbourhood of its limit. }\label{fig:cartpoleres}
\end{figure}
The table of the various parameter values we used in our experiment is given below.\vspace*{4mm}\\
%\begin{center}
\small
\hspace*{2cm}\begin{tabular}{ | l | c |}
  \specialrule{.2em}{.02em}{.02em} 
    Gravitational acceleration ($g$) & $9.8\frac{m}{s^{2}}$ \\ \hline
    Mass of the pole ($m$) & $0.5kg$ \\ \hline
    Mass of the cart ($M$) & $0.5kg$ \\ \hline
    Length of the pole ($l$) & $0.6m$ \\ \hline
    Friction coefficient ($b$) & $0.1N(ms)^{-1}$ \\ \hline
    Integration time step ($\Delta t$) & $0.1s$\\ \hline
    Standard deviation of $z$ ($\sigma_{2}$) & $0.01$ \\ \hline
    Discount factor ($\gamma$) & $0.95$ \\ \hline
   \specialrule{.1em}{.02em}{.02em} 
  \end{tabular}
\hspace*{10mm} \begin{tabular}{ | c | c |}
  \specialrule{.2em}{.04em}{.04em} 
    $\alpha_{t}$\hspace*{25mm} & $t^{-1.0}$ \\ \hline
    $\beta_{t}$\hspace*{25mm} & $t^{-0.6}$ \\ \hline
    $c_t$\hspace*{25mm} & $0.01$ \\ \hline
    $\epsilon_1$\hspace*{25mm} & $0.95$ \\
   \specialrule{.1em}{.02em}{.02em} 
  \end{tabular}
\normalsize
\vspace*{4mm}\\
The results of the experiments are shown in Figure \ref{fig:cartpoleres}.

\subsection{Experiment 2: 5-Link Actuated Pendulum Balancing \cite{dann2014policy}}
\textbf{Setup: } $5$ independent poles each with mass $m$ and length $l$ with the top pole being a pendulum connected using $5$ rotational joints.\\
\textbf{Goal: } To keep all the poles in the upright position by applying independent torques at each joint.\\
\textbf{State space: }The state $s = (q, \dot{q})^{\top} \in \mathbb{R}^{10}$ where $q = (\psi_{1}, \psi_{2}, \psi_{3}, \psi_{4}, \psi_{5}) \in \mathbb{R}^{5}$ and $\dot{q} = (\dot{\psi}_{1}, \dot{\psi}_{2}, \dot{\psi}_{3}, \dot{\psi}_{4}, \dot{\psi}_{5})  \in \mathbb{R}^{5}$ where $\psi_{i}$ is  the angle of the pole $i$ \emph{w.r.t.} the vertical axis and $\dot{\psi}_{i}$ is the angular velocity.\\
\textbf{Control space: }The action $a = (a_{1}, a_{2}, \dots, a_{5})^{\top} \in \mathbb{R}^{5}$ where $a_{i}$ is the torque applied to the joint $i$. The stochastic policy  used in this setting corresponds to $\pi(a|s) = \mathcal{N}(a | \beta_{1}^{\top}s, \sigma_1^{2})$.\\
\textbf{System dynamics: } The approximate linear system dynamics is given by
\begin{equation} 
\begin{bmatrix} 
q_{t+1}\\
\dot{q}_{t+1}
\end{bmatrix} = 
\begin{bmatrix} 
I && \Delta t\hspace*{1mm} I\\
-\Delta t \hspace*{1mm}M^{-1}U && I
\end{bmatrix}\begin{bmatrix}q_{t}\\ \dot{q}_{t}\end{bmatrix} + \Delta t \begin{bmatrix}
0 \\
M^{-1}
\end{bmatrix}a + 
\mathbf{z}
\end{equation}
where $\Delta t$ is the integration time step, \emph{i.e.}, the time difference between two transitions, $M$ is the mass matrix in the upright position where $M_{ij} = l^{2}(6-max(i,j))m$ and $U$ is a diagonal matrix with $U_{ii} = -gl(6-i)m$.  Each component of $\mathbf{z}$ is a Gaussian noise.\\
\begin{figure}[!h]
	\centering
	\includegraphics[height=55mm, width=60mm]{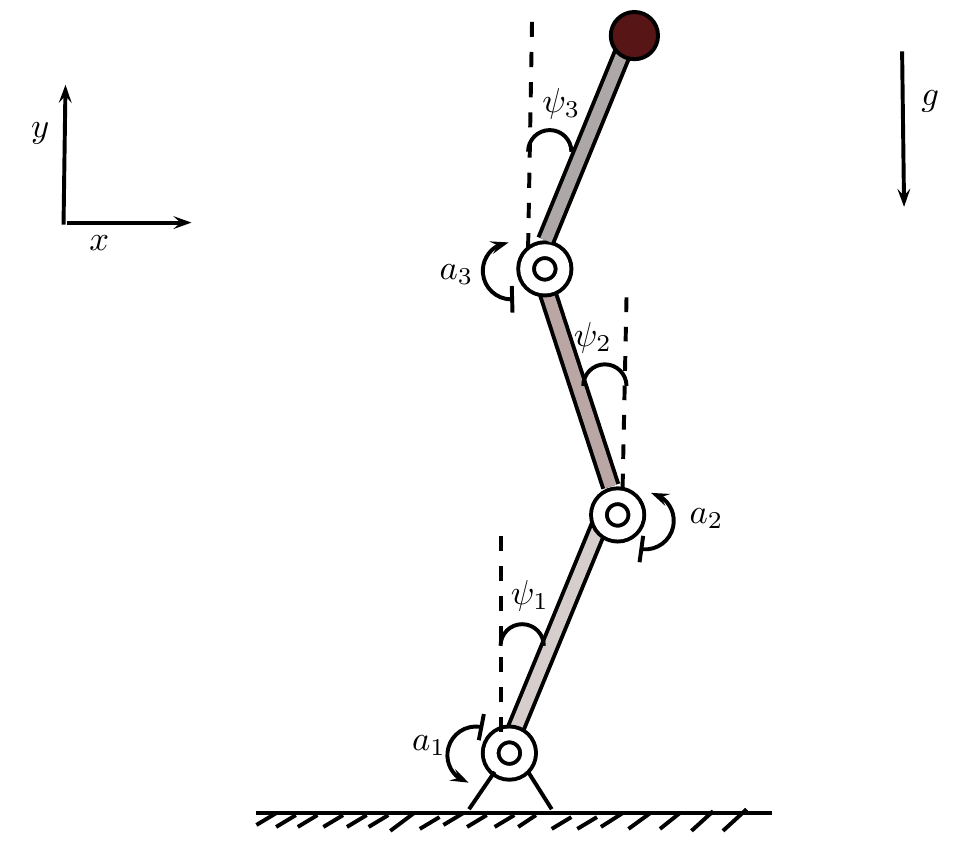}
	\caption{3-link actuated pendulum setting. Each rotational joint $i$, $1 \leq i \leq 3$ is actuated by a torque $a_{i}$ . The system is parameterized by the angle $\psi_{i}$ against the vertical direction and the angular velocity $\dot{\psi}_{i}$. The goal is to balance the pole in the upright direction, \emph{i.e.}, all $\psi_{i}$ should be as close to $0$ as possible. }
\end{figure}\\
\textbf{Reward function: } $\mathrm{R}(q, \dot{q}, a) = -q^{\top}q$.\\
\textbf{Feature vectors: } $\phi(s \in \mathbb{R}^{10}) = (1, s_{1}^{2}, s_{2}^{2} \dots, s_{1}s_{2}, s_{1}s_{3}, \dots, s_{9}s_{10})^{\top} \in \mathbb{R}^{46}$.\\
\textbf{Evaluation policy: }The policy evaluated in the experiment is the optimal policy $\pi^{*}(a | s) = \mathcal{N}(a | {\beta_{1}^{*}}^{\top}s, {\sigma_{1}^{*}}^{2})$. The parameters $\beta_{1}^{*}$ and $\sigma_{1}^{*}$ are computed using dynamic programming. The feature set chosen above is a perfect feature set, \emph{i.e.}, $V^{\pi^{*}} \in \{\Phi z \vert z \in \mathbb{R}^{k}\}$.

\begin{figure}[!h]
	\begin{subfigure}[h]{0.5\textwidth}
		\fbox{\includegraphics[height=55mm, width=80mm]{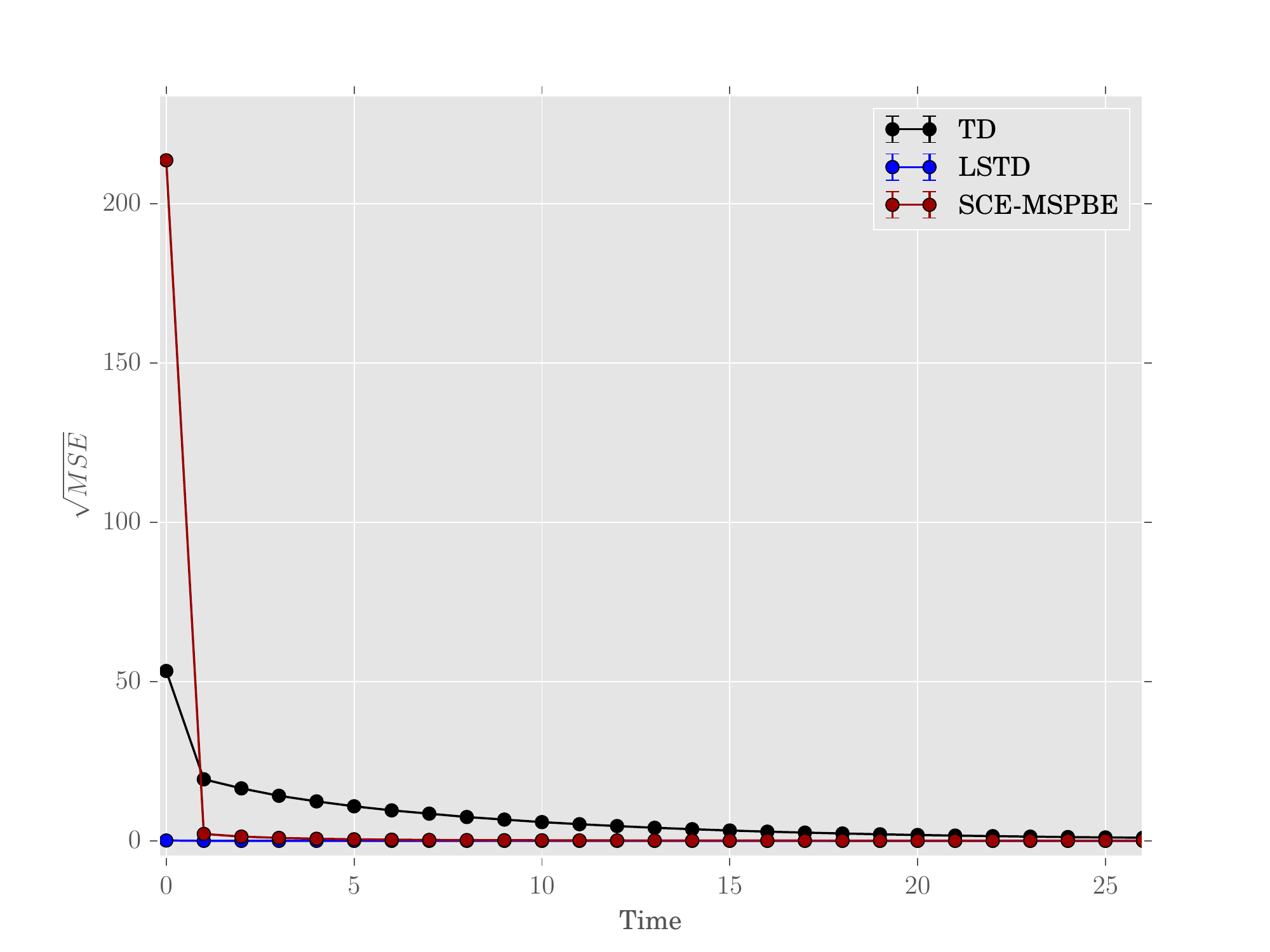}}
        \subcaption{$\sqrt{\mathrm{MSPBE}(\bar{\mu}_{t})}$}
    \end{subfigure}%
    \begin{subfigure}[h]{0.5\textwidth}
            \fbox{\includegraphics[width=80mm, height=55mm]{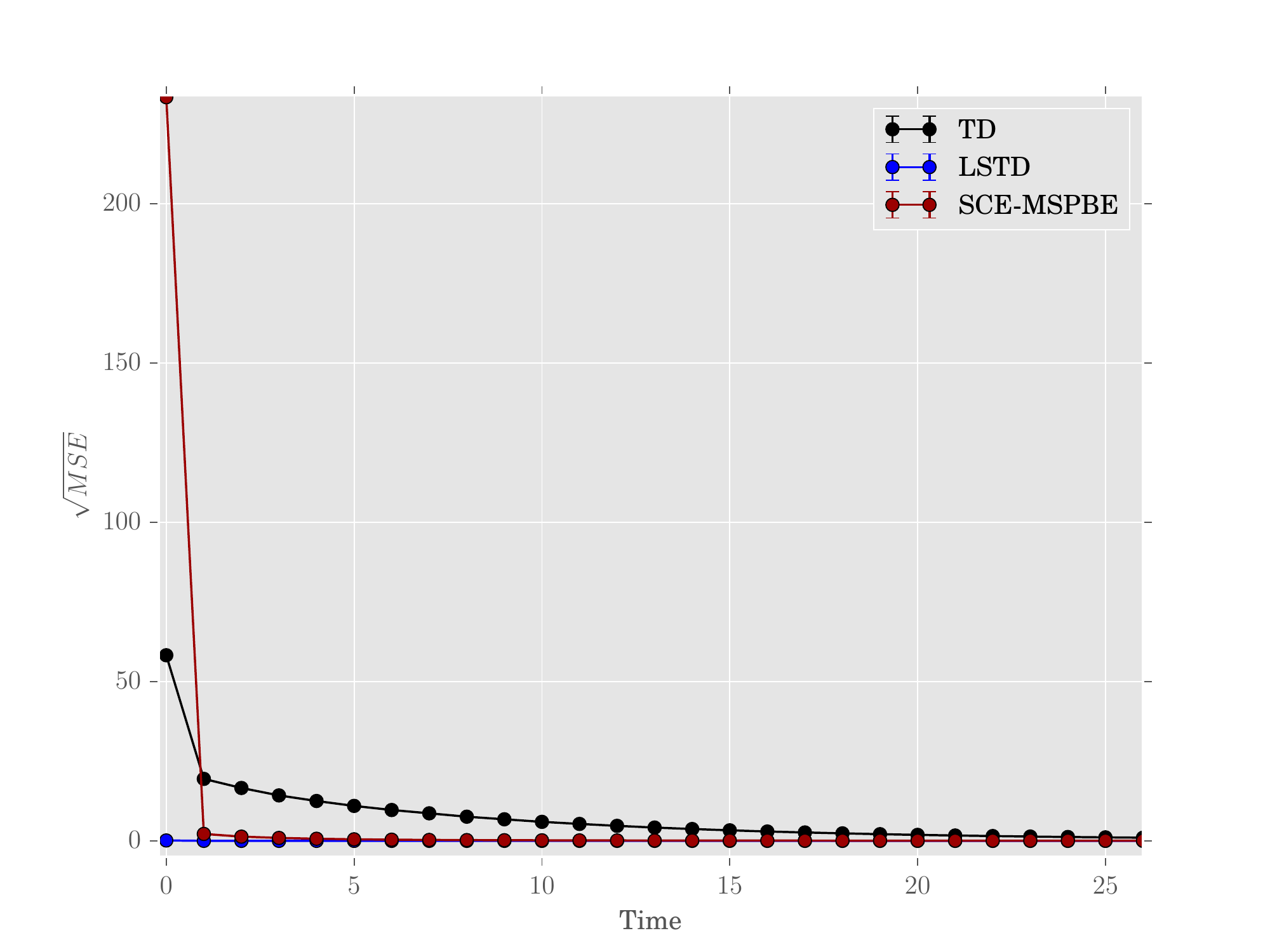}}
            \subcaption{$\sqrt{\mathrm{MSE}(\bar{\mu}_{t})}$}
    \end{subfigure}	
	\caption{5-link actuated pendulum setting. The respective trajectories of the $\sqrt{\mathrm{MSPBE}}$ and $\sqrt{\mathrm{MSE}}$ generated by TD(0), LSTD(0) and SCE-MSPBEM algorithms are plotted. The graph on the left is for $\sqrt{\mathrm{MSPBE}}$ , while on the right is that of $\sqrt{\mathrm{MSE}}$. Note that $\sqrt{\mathrm{MSE}}$ also converges to $0$ since the feature set is perfect.}\label{eqn:linkres}
\end{figure}

The table of the various parameter values we used in our experiment is given below. Note that we have used constant step-sizes in this experiment.\vspace*{4mm}\\
%\begin{center}
\small
\hspace*{2cm}\begin{tabular}{ | l | c |}
  \specialrule{.2em}{.02em}{.02em} 
    Gravitational acceleration ($g$) & $9.8\frac{m}{s^{2}}$ \\ \hline
    Mass of the pole ($m$) & $1.0kg$ \\ \hline
    Length of the pole ($l$) & $1.0m$ \\ \hline
    Integration time step ($\Delta t$) & $0.1s$\\ \hline
    Discount factor ($\gamma$) & $0.95$ \\ \hline
   \specialrule{.1em}{.02em}{.02em} 
  \end{tabular}
%\end{center}
\hspace*{10mm} \begin{tabular}{ | c | c |}
  \specialrule{.2em}{.04em}{.04em} 
    $\alpha_{t}$\hspace*{25mm} & $0.001$ \\ \hline
    $\beta_{t}$\hspace*{25mm} & $0.05$ \\ \hline
    $c_t$\hspace*{25mm} & $0.05$ \\ \hline
    $\epsilon_1$\hspace*{25mm} & $0.95$ \\
   \specialrule{.1em}{.02em}{.02em} 
  \end{tabular}
\normalsize
\vspace*{4mm}\\The results of the experiment are shown in Figure \ref{eqn:linkres}.
\subsection{Experiment 3: Baird's 7-Star MDP \cite{baird1995residual}}
\begin{figure}[!h]
        \begin{subfigure}[h]{0.5\textwidth}
            \fbox{\includegraphics[width=80mm, height=75mm]{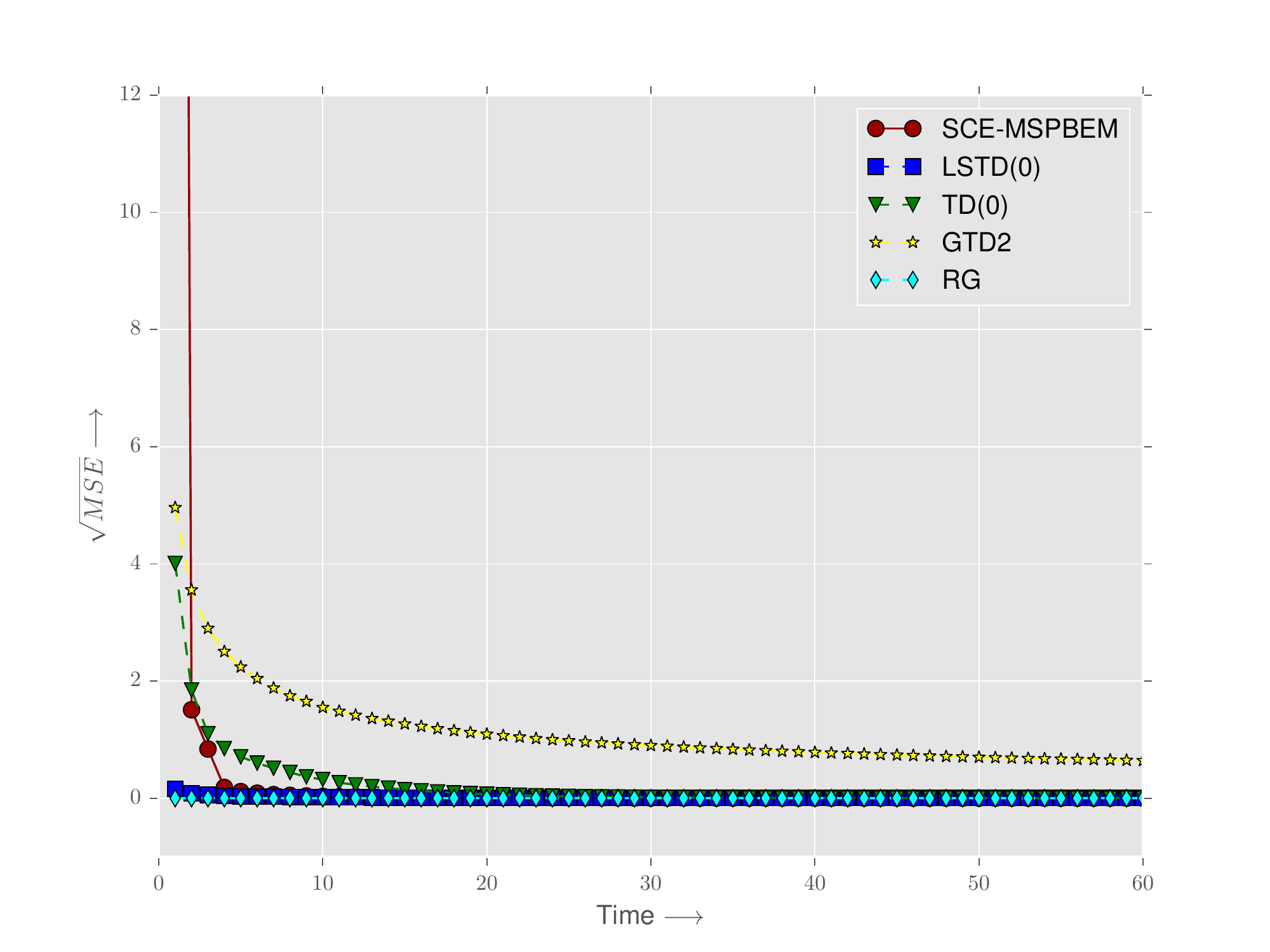}}
            	\subcaption{Discount factor $\gamma = 0.1$}
        \end{subfigure}%
        \begin{subfigure}[h]{0.5\textwidth}
            \fbox{\includegraphics[width=80mm, height=75mm]{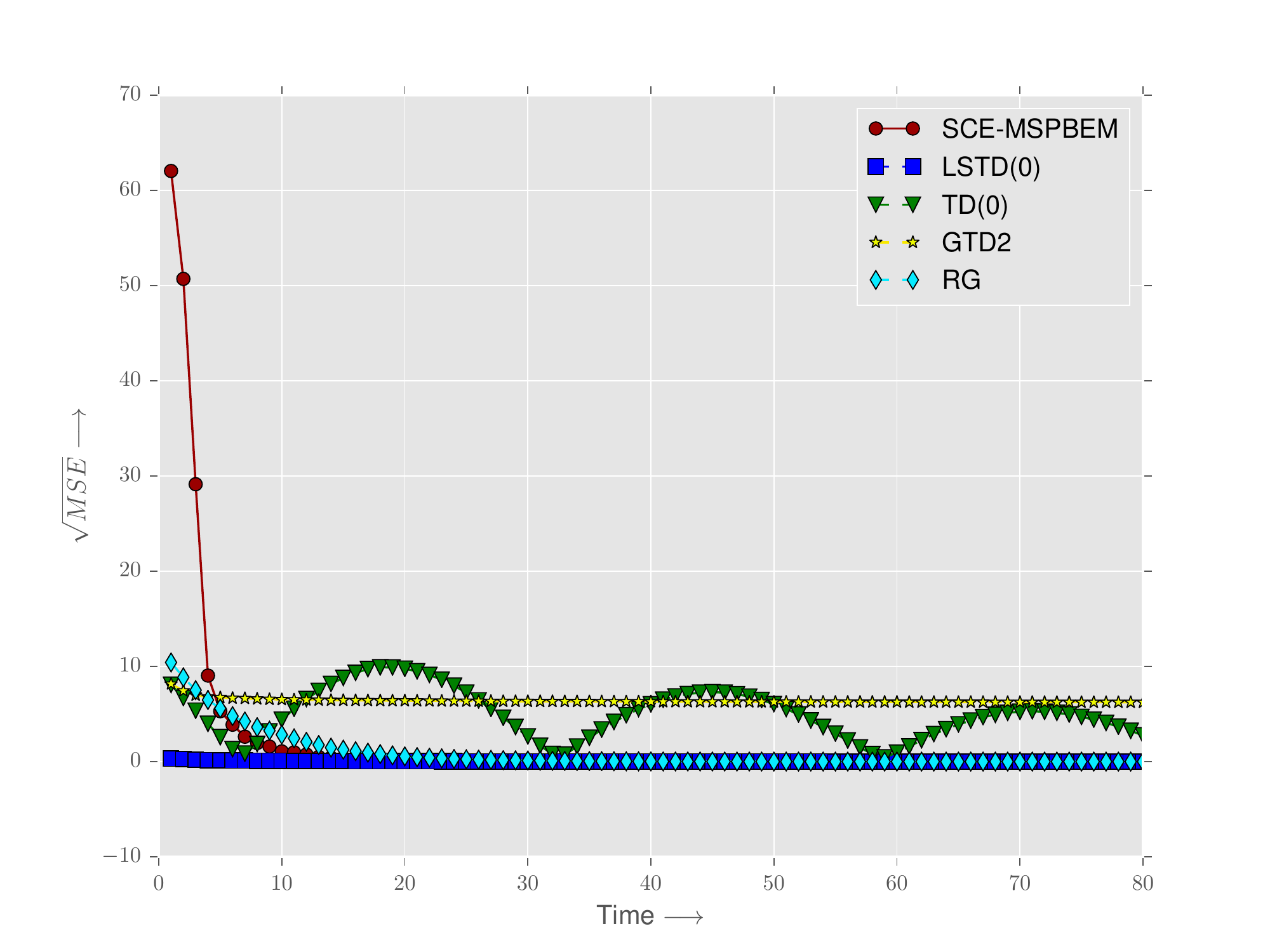}}
            \subcaption{Discount factor $\gamma = 0.9$}
        \end{subfigure}%
        	\caption{Baird's 7-Star MDP with perfect feature set. For $\gamma = 0.1$, all the algorithms show almost the same rate of convergence. The initial jump of SCE-MSPBEM is due to the fact that the initial value is far from the limit. For $\gamma = 0.9$, TD(0) does not converge and GTD2 is slower. However, SCE-MSPBEM exhibits good convergence behaviour. }\label{fig:starperf}
\end{figure}

Our algorithm was also tested on Baird's star problem \cite{baird1995residual} with $\vert \mathbb{S} \vert = 7$, $\vert \mathbb{A} \vert = 2$ and $k=8$. We let $\nu$ be the uniform distribution over $\mathbb{S}$ and the feature matrix $\Phi$ and the transition matrix $\mathrm{P}^{\pi}$ are given by\\
%\begin{equation}
\small
\hspace*{4cm}
\resizebox{0.2\linewidth}{!}{$\Phi = \begin{pmatrix}
 1 & 2 & 0 & 0 & 0 & 0 & 0 & 0\\
 1 & 0 & 2 & 0 & 0 & 0 & 0 & 0\\
 1 & 0 & 0 & 2 & 0 & 0 & 0 & 0\\
 1 & 0 & 0 & 0 & 2 & 0 & 0 & 0\\
 1 & 0 & 0 & 0 & 0 & 2 & 0 & 0\\
 1 & 0 & 0 & 0 & 0 & 0 & 2 & 0\\
 2 & 0 & 0 & 0 & 0 & 0 & 0 & 1\\
\end{pmatrix}$}
\hspace*{1cm}
%\end{equation}
%\begin{equation}5
\resizebox{0.2\linewidth}{!}{$\mathrm{P}^{\pi} = \begin{pmatrix}
 0 & 0 & 0 & 0 & 0 & 0 & 1 \\
 0 & 0 & 0 & 0 & 0 & 0 & 1 \\
 0 & 0 & 0 & 0 & 0 & 0 & 1 \\
 0 & 0 & 0 & 0 & 0 & 0 & 1 \\
 0 & 0 & 0 & 0 & 0 & 0 & 1 \\
 0 & 0 & 0 & 0 & 0 & 0 & 1 \\
 0 & 0 & 0 & 0 & 0 & 0 & 1 \\
\end{pmatrix}$}.\vspace*{2mm}\\
\normalsize
The reward function is given by $\mathrm{R}(s, s^{\prime}) = 0$, $\forall s, s^{\prime} \in \mathbb{S}$. The Markov Chain in this case is not ergodic and hence belongs to an off-policy setting. This is a classic example where TD(0) is seen to diverge \cite{baird1995residual}. The performance comparison of the algorithms GTD2, TD($0$) and LSTD($0$) with SCE-MSPBEM is shown in Figure \ref{fig:starperf}. The performance metric used here is the $\sqrt{MSE(\cdot)}$ of the prediction vector returned by the corresponding algorithm at time $t$. The algorithm parameters for the problem are given below:\\
\begin{wrapfigure}{r}{0.3\textwidth}
  \vspace{-60pt}
  \begin{center}
    \includegraphics[width=0.22\textwidth]{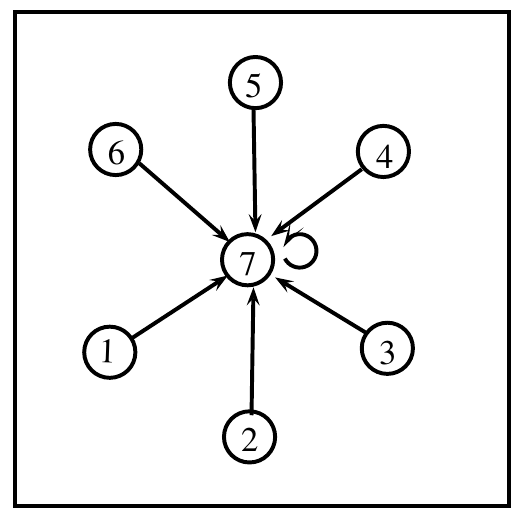}
  \end{center}
 \vspace{-15pt}
  \caption{Baird's 7-star MDP}
  \vspace{-100pt}
\end{wrapfigure}
\vspace*{15mm}
\small
\hspace*{15mm} \begin{tabular}{ | c | c |}
  \specialrule{.2em}{.04em}{.04em} 
    $\alpha_{t}$\hspace*{25mm} & $0.001$ \\ \hline
    $\beta_{t}$\hspace*{25mm} & $0.05$ \\ \hline
    $c_t$\hspace*{25mm} & $0.01$ \\ \hline
    $\epsilon_1$\hspace*{25mm} & $0.8$ \\
   \specialrule{.1em}{.02em}{.02em} 
  \end{tabular}
%\end{center}
\normalsize

A careful analysis in \cite{schoknecht2002convergent} has shown that when the discount factor $\gamma \leq 0.88$, with appropriate learning rate, TD($0$) converges. Nonetheless, it is also shown in the same paper that for discount factor $\gamma = 0.9$, TD($0$) will diverge for all values of the learning rate. This is explicitly demonstrated in Figure \ref{fig:starperf}. However our algorithm SCE-MSPBEM converges in both cases, which demonstrates the stable behaviour exhibited by our algorithm.

\begin{figure}[!h]
        \begin{subfigure}[h]{0.5\textwidth}
            \fbox{\includegraphics[width=85mm, height=70mm]{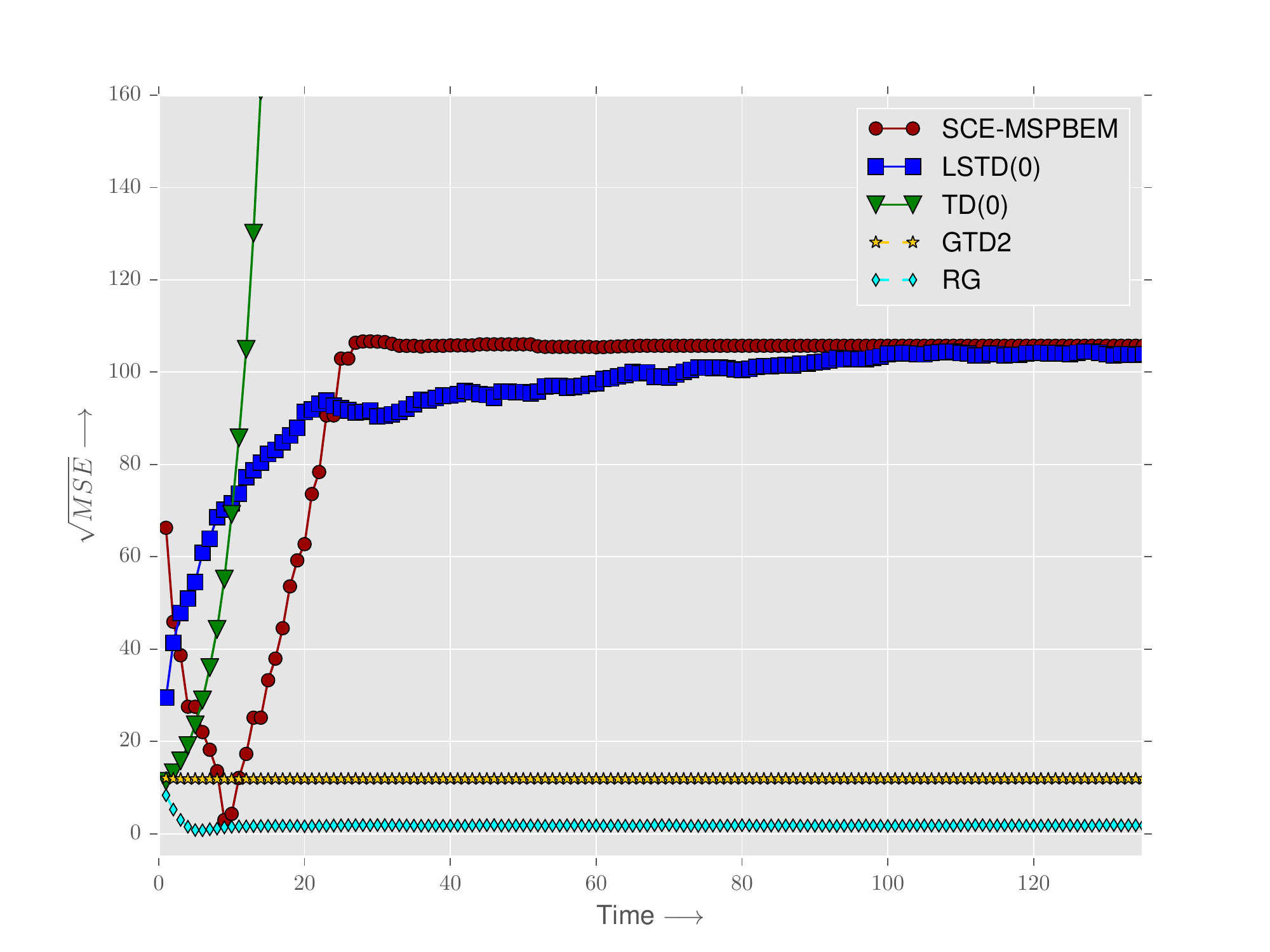}}
            	\subcaption{$\sqrt{\mathrm{MSE}}$.}
        \end{subfigure}%
        \begin{subfigure}[h]{0.5\textwidth}
            \fbox{\includegraphics[width=75mm, height=70mm]{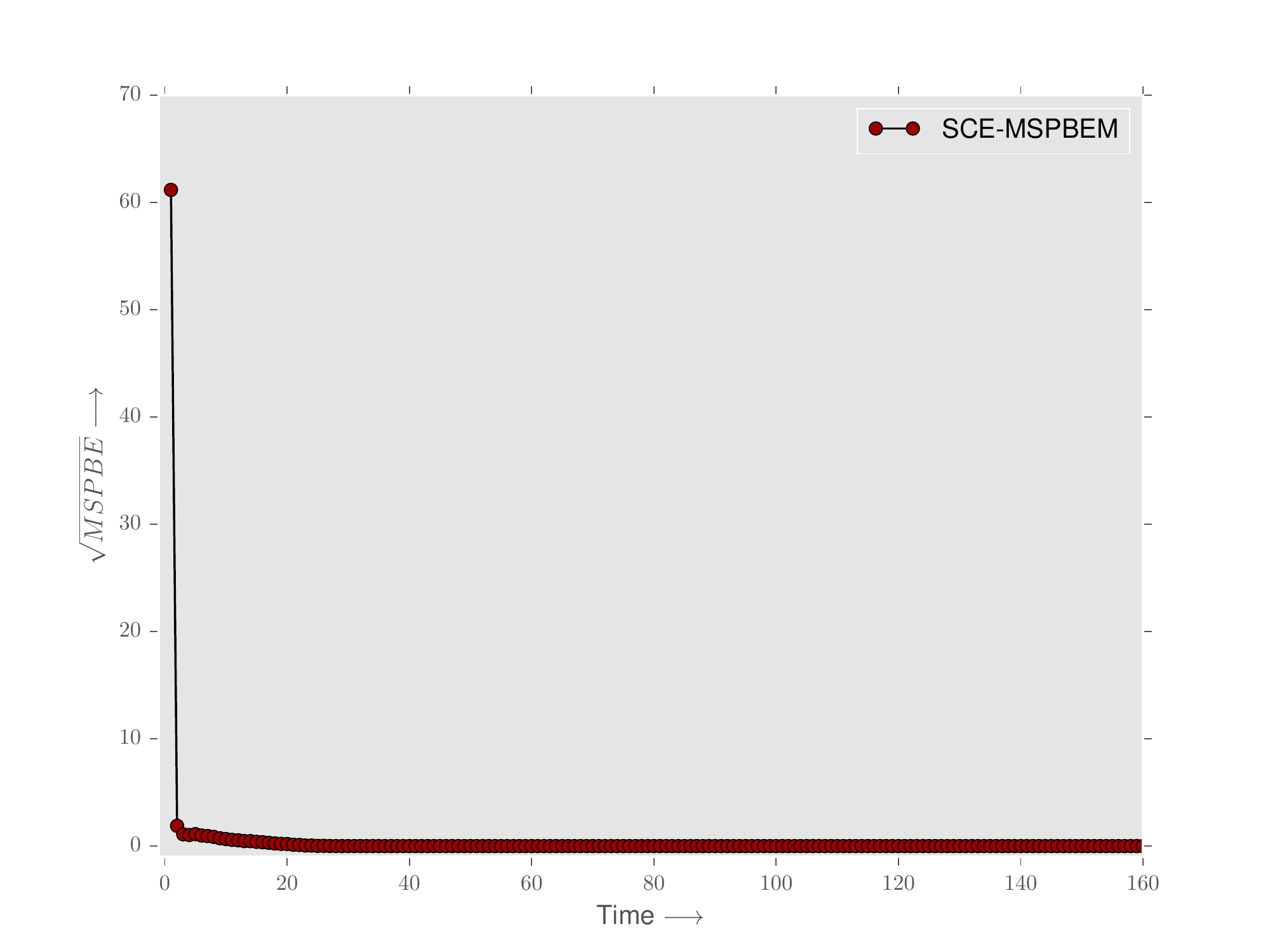}}
            \subcaption{$\sqrt{\mathrm{MSPBE}}$.}
        \end{subfigure}%
        	\caption{Baird's 7-Star MDP with imperfect feature set.	 Here the discount factor $\gamma = 0.99$. In this case, TD(0) method diverges. However, $\sqrt{\mathrm{MSE}}$ of SCE-MSPBEM and LSTD($0$) converge to the same limit point (=$103.0$), while RG converges to a different limit (= $1.6919$). This is because the feature set is imperfect and also due to the fact that RG minimizes MSBR, while SCE-MSPBEM and LSTD minimize MSPBE. To verify this fact, note that in (b), $\sqrt{\mathrm{MSPBE}(\bar{\mu}_{t})}$ of SCE-MSPBEM converges to $0$.}\label{fig:starperf2}
\end{figure}

The algorithms were also compared on the same Baird's 7-star, but with a different feature matrix $\Phi_{1}$.\vspace*{2mm}\\
\hspace*{50mm}\resizebox{0.2\linewidth}{!}{$\Phi_{1} = \begin{pmatrix}
 1 & 2 & 0 & 0 & 0 & 0 & 1 & 0\\
 1 & 0 & 2 & 0 & 0 & 0 & 0 & 0\\
 1 & 0 & 0 & 2 & 0 & 0 & 0 & 0\\
 1 & 0 & 0 & 0 & 2 & 0 & 0 & 0\\
 1 & 0 & 0 & 0 & 0 & 0 & 0 & 2\\
 1 & 0 & 0 & 0 & 0 & 0 & 0 & 3\\
 2 & 0 & 0 & 0 & 0 & 0 & 0 & 1\\
\end{pmatrix}$}.
\vspace*{2mm}\\
In this case, the reward function is given by $\mathrm{R}(s, s^{\prime})=2.0$, $\forall s, s^{\prime} \in \mathbb{S}$. Note that $\Phi_{1}$ gives an imperfect feature set. The algorithm parameter values used are same as earlier. The results are show in Figure \ref{fig:starperf2}. In this case also, TD($0$) diverges. However, SCE-MSPBEM is exhibiting good stable behaviour.
\subsection*{Experiment 4: 10-State Ring MDP \cite{kveton2006solving}}
\begin{figure}[!h]
        \begin{subfigure}[h]{0.5\textwidth}
            \fbox{\includegraphics[width=80mm, height=60mm]{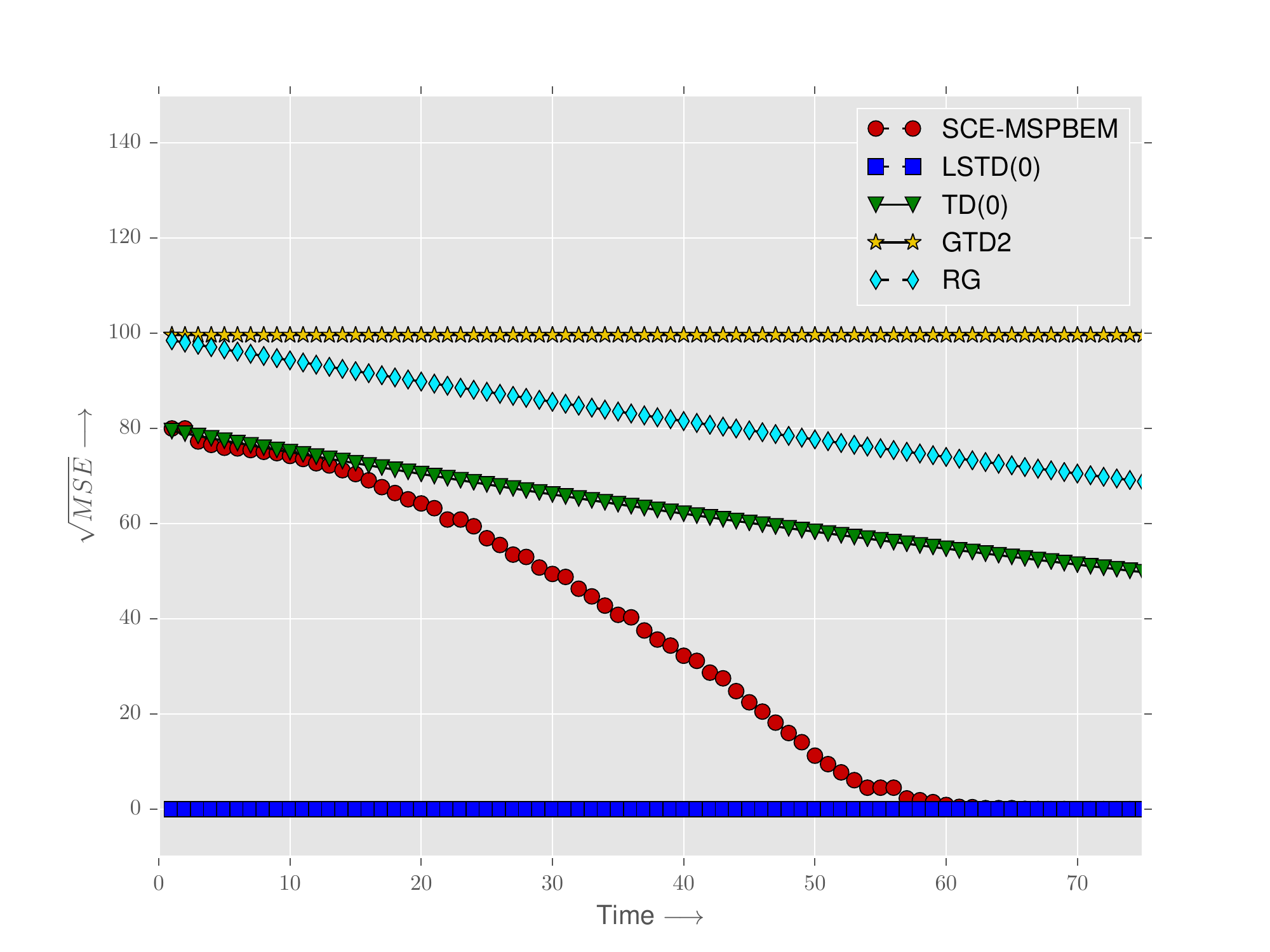}}
            \subcaption{Discount factor $\gamma = 0.99$}
        \end{subfigure}%
        \begin{subfigure}[h]{0.5\textwidth}
            \fbox{\includegraphics[width=80mm, height=60mm]{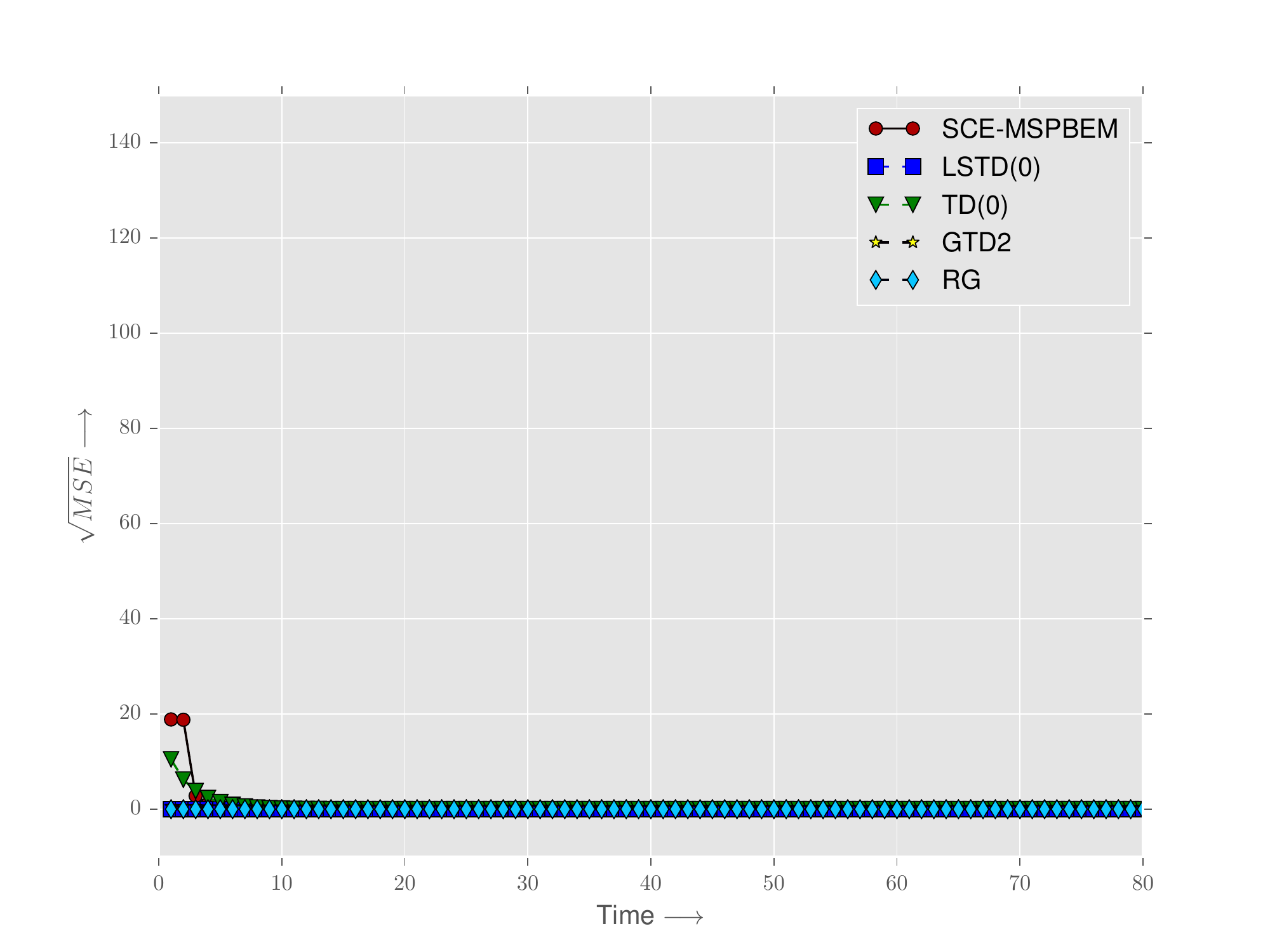}}
            \subcaption{Discount factor $\gamma = 0.1$}
        \end{subfigure}%
        	\caption{10-Ring MDP with perfect feature set: For $\gamma = 0.1$, all the algorithms exhibit almost the same rate of convergence. For $\gamma = 0.99$, SCE-MSPBEM converges faster than TD($0$), GTD2 and RG.}\label{fig:ringper}
\end{figure}

Next, we studied the performance comparisons of the algorithms on a $10$-ring MDP with $\vert\mathbb{S}\vert = 10$ and $k=8$. We let $\nu$ be the uniform distribution over $\mathbb{S}$. The transition matrix $\mathrm{P}^{\pi}$ and the feature matrix $\Phi$ are given by\\
\small
\hspace*{2cm}\resizebox{0.2\linewidth}{!}{$\mathrm{P}^{\pi} = \begin{pmatrix}
 0 & 1 & 0 & 0 & 0 & 0 & 0 & 0 & 0 & 0 \\
 0 & 0 & 1 & 0 & 0 & 0 & 0 & 0 & 0 & 0 \\
 0 & 0 & 0 & 1 & 0 & 0 & 0 & 0 & 0 & 0 \\
 0 & 0 & 0 & 0 & 1 & 0 & 0 & 0 & 0 & 0 \\
 0 & 0 & 0 & 0 & 0 & 1 & 0 & 0 & 0 & 0 \\
 0 & 0 & 0 & 0 & 0 & 0 & 1 & 0 & 0 & 0 \\
 0 & 0 & 0 & 0 & 0 & 0 & 0 & 1 & 0 & 0 \\
 0 & 0 & 0 & 0 & 0 & 0 & 0 & 0 & 1 & 0 \\
 0 & 0 & 0 & 0 & 0 & 0 & 0 & 0 & 0 & 1 \\
 1 & 0 & 0 & 0 & 0 & 0 & 0 & 0 & 0 & 0 
\end{pmatrix}$},
%\begin{gather*}
\hspace*{5mm}\resizebox{0.2\linewidth}{!}{\scriptsize{$\Phi$} = $\begin{pmatrix}
	1 & 0 & 0 & 0 & 0 & 0 & 0 & 0 \\
    0 & 1 & 0 & 0 & 0 & 0 & 0 & 0 \\
    0 & 0 & 1 & 0 & 0 & 0 & 0 & 0 \\
    0 & 0 & 0 & 1 & 0 & 0 & 0 & 0 \\
    0 & 0 & 0 & 0 & 1 & 0 & 0 & 0 \\
    0 & 0 & 0 & 0 & 0 & 1 & 0 & 0 \\
    0 & 0 & 0 & 0 & 0 & 0 & 1 & 0 \\
    0 & 0 & 0 & 0 & 0 & 0 & 0 & 1 \\
    0 & 0 & 0 & 0 & 0 & 0 & 0 & 1 \\
    0 & 0 & 0 & 0 & 0 & 1 & 0 & 0
\end{pmatrix}$}.\\
\normalsize
%\end{gather*}
\begin{wrapfigure}{r}{0.26\textwidth}
  \vspace{-54mm}
 \begin{center}
    \includegraphics[width=0.23\textwidth]{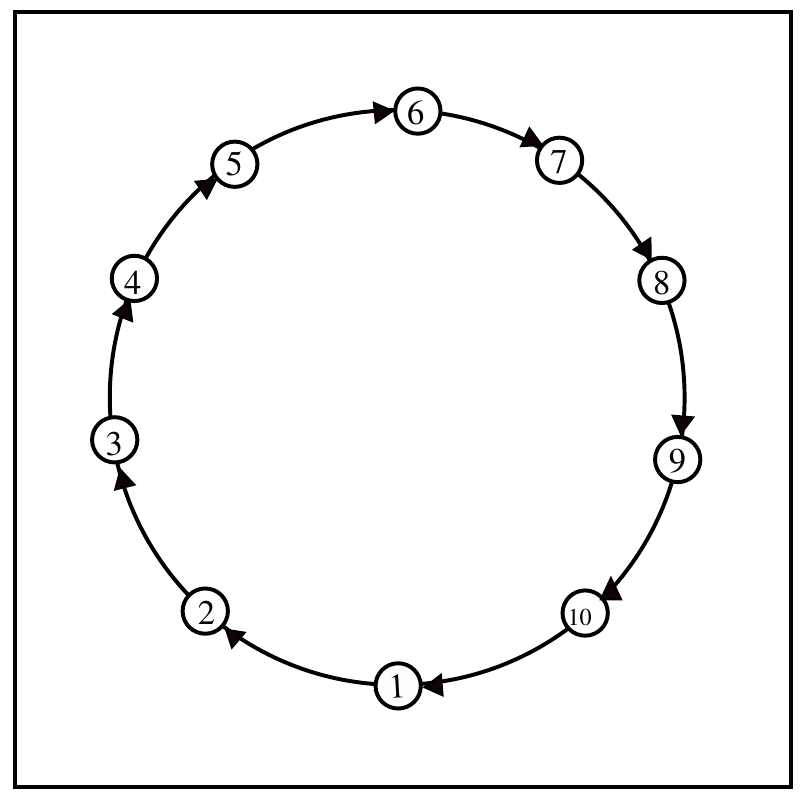}
  \end{center}
  \vspace{-20pt}
  \caption{10-Ring MDP}
  \vspace{-45pt}
\end{wrapfigure}
\normalsize
The reward function is $\mathrm{R}(s, s^{\prime})=1.0, \forall s, s^{\prime} \in \mathbb{S}$.
The performance comparisons of the algorithms GTD2, TD($0$) and LSTD($0$) with SCE-MSPBEM are shown in Figure \ref{fig:ringper}. The performance metric used here is the $\sqrt{MSE(\cdot)}$ of the prediction vector returned by the corresponding algorithm at time $t$.
\newpage
The algorithm parameters for the problem are as follows:
\small
\vspace*{-4mm}
\begin{center}
\hspace*{75mm} \begin{tabular}{ | c | c |}
  \specialrule{.2em}{.04em}{.04em} 
    $\alpha_{t}$\hspace*{25mm} & $0.001$ \\ \hline
    $\beta_{t}$\hspace*{25mm} & $0.05$ \\ \hline
    $c_t$\hspace*{25mm} & $0.075$ \\ \hline
    $\epsilon_1$\hspace*{25mm} & $0.85$ \\
   \specialrule{.1em}{.02em}{.02em} 
  \end{tabular}
\end{center}
\normalsize

\subsection{Experiment 5: Large State Space Random MDP with Radial Basis Functions and Fourier Basis}
These experiments were designed by us. Here, tests were performed by varying the feature set to prove that the algorithm is not dependent on any particular feature set. Two types of feature sets are used here: Fourier Basis Functions and Radial Basis Functions (RBF).

\begin{figure}[h]
	  \centering
      \fbox{\includegraphics[height=80mm, width=125mm]{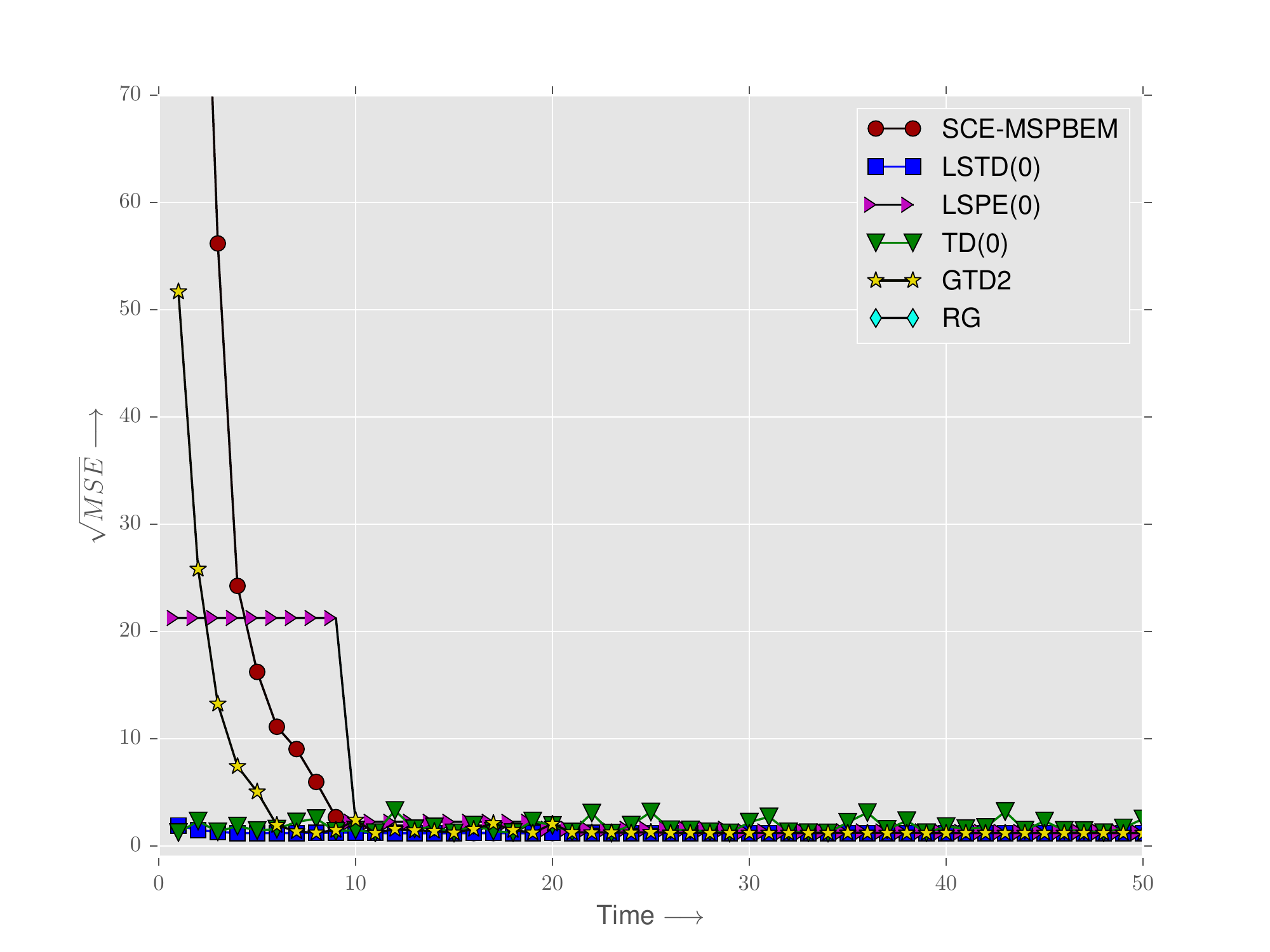}}
      \caption{Fourier Basis Function: Here, $\vert\mathbb{S}\vert = 1000$, $\vert\mathbb{A}\vert = 200$, $k = 50$ and $\gamma = 0.9$. In this case, SCE-MSPBEM shows good convergence behaviour.}\label{fig:fourier}
\end{figure}

\begin{figure}[h]
	\centering
    \fbox{\includegraphics[height=90mm, width=125mm]{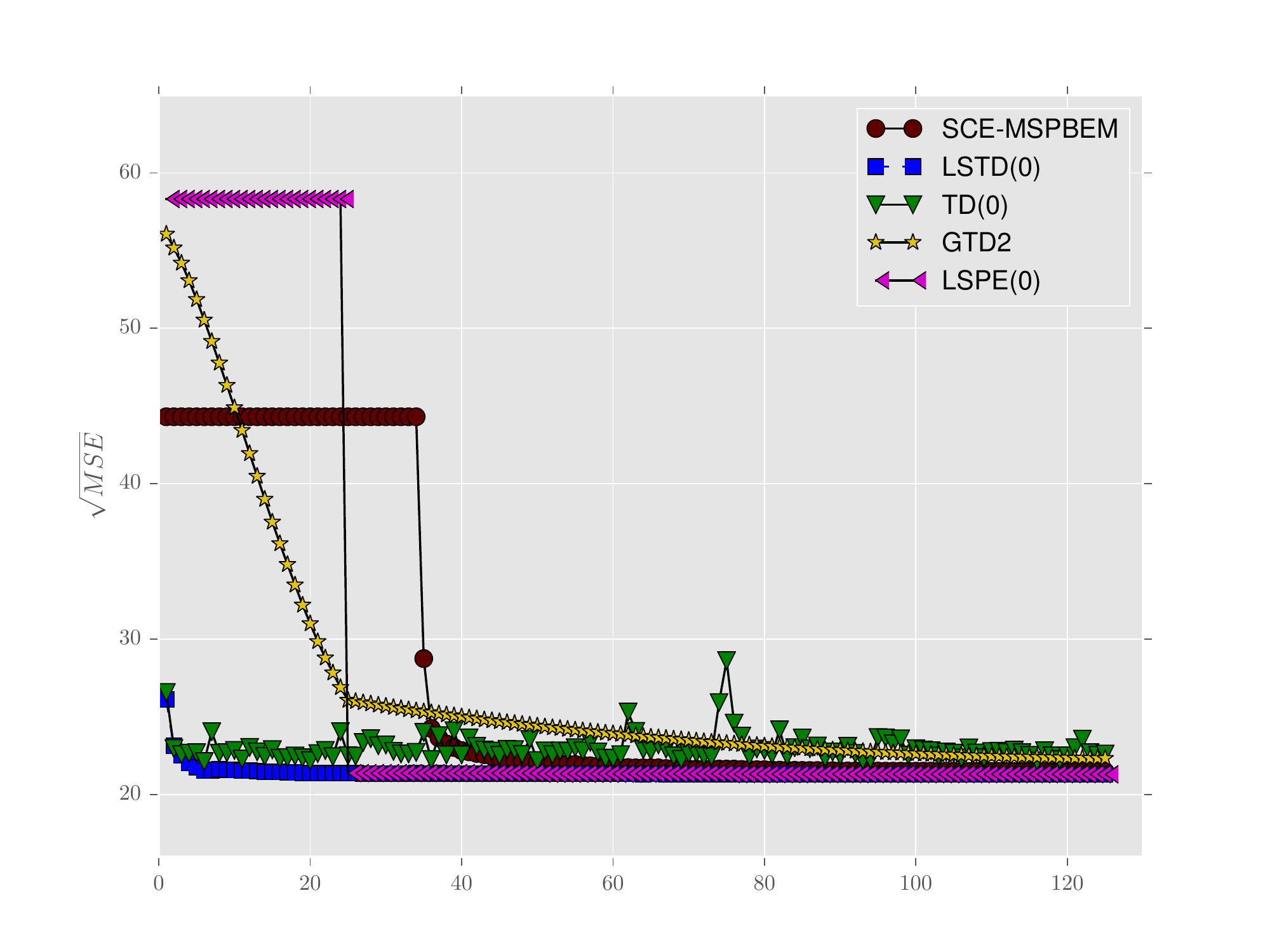}}
	\caption{Radial Basis Function. Here, $\vert\mathbb{S}\vert = 1000$, $\vert\mathbb{A}\vert = 200$, $k=50$ and $\gamma=0.01$. In this case, SCE-MSPBEM converges to the same limit point as other algorithms.}\label{fig:rbfres}
\end{figure}

Figure \ref{fig:fourier} shows the performance comparisons when Fourier Basis Functions are used for the features $\lbrace\phi_i\rbrace_{i=1}^{k}$, where
\begin{eqnarray}
\phi_i(s) = \begin{cases}  1  &\mbox{if } i = 1, \\
		  \cos{\frac{(i+1)\pi s}{2}} & \mbox{if } i \mbox{ is odd}, \\
		  \sin{\frac{i\pi s}{2}} & \mbox{if } i \mbox{ is even}.
		  \end{cases}
\end{eqnarray}
Figure \ref{fig:rbfres} shows the performance comparisons when RBF is used instead for the features $\lbrace\phi_i\rbrace_{i=1}^{k}$, where 
\begin{equation}
\phi_{i}(s) = e^{-\frac{(s-m_{i})^{2}}{2.0v_{i}^{2}}}
\end{equation}
with $m_i$ and $v_i$ fixed \emph{a priori}. 

In both the cases, the reward function is given by
\begin{equation}\label{eqn:rwdrnd}
\mathrm{R}(s, s^{\prime}) = G(s)G(s^{\prime})\left(\frac{1}{(1.0+s{\prime})^{0.25}}\right), \hspace*{1cm} \forall s, s^{\prime} \in \mathbb{S},
\end{equation}
where the vector $G \in (0,1)^{\vert \mathbb{S} \vert}$ is initialized for the algorithm with $G(s) \sim U(0,1), \forall s \in \mathbb{S}$.

Also in both the cases, the transition probability matrix $\mathrm{P}^{\pi}$ is  generated as follows \\
\begin{equation}\label{eqn:prnd}
\mathrm{P}^{\pi}(s, s^{\prime}) =  {\vert \mathbb{S} \vert \choose s^{\prime}}b(s)^{s^{\prime}}(1.0-b(s))^{\vert \mathbb{S} \vert - s^{\prime}}, \hspace*{1cm} \forall s, s^{\prime} \in \mathbb{S},
\end{equation}
where the vector $b \in (0,1)^{\vert \mathbb{S} \vert}$ is initialized for the algorithm with $b(s) \sim U(0,1), \forall s \in \mathbb{S}$. It is easy to verify that the Markov Chain defined by $\mathrm{P}^{\pi}$ is ergodic in nature.

In the case of RBF, we have set $\vert\mathbb{S}\vert = 1000$, $\vert\mathbb{A}\vert = 200$, $k=50$, $m_i = 10+20(i-1)$ and $v_i = 10$, while for Fourier Basis Functions, $\vert\mathbb{S}\vert = 1000$, $\vert\mathbb{A}\vert = 200$, $k=50$. In both the cases, the distribution $\nu$ is the stationary distribution of the Markov Chain. The simulation is run sufficiently long to ensure that the chain achieves its steady state behaviour, \emph{i.e.}, the states appear with the stationary distribution.\\
The algorithm parameters for the problem are as follows:
\small
\vspace*{-5mm}
\begin{center}
\hspace*{90mm} \begin{tabular}{ | c | c |}
  \specialrule{.2em}{.04em}{.04em} 
    \multicolumn{2}{|c|}{Both RBF \& Fourier Basis}  \\ \hline
    $\alpha_{t}$\hspace*{35mm} & $0.001$\hspace*{10mm} \\ \hline
    $\beta_{t}$\hspace*{35mm} & $0.05$\hspace*{10mm} \\ \hline
    $c_t$\hspace*{35mm} & $0.075$\hspace*{10mm} \\ \hline
    $\epsilon_1$\hspace*{35mm} & $0.85$\hspace*{10mm} \\
   \specialrule{.1em}{.02em}{.02em} 
  \end{tabular}
\end{center}
\vspace*{2mm}
\normalsize
Also note that when Fourier basis is used, the discount factor $\gamma=0.9$ and for RBFs, $\gamma=0.01$. SCE-MSPBEM exhibits good convergence behaviour in both cases, which shows the non-dependence of SCE-MSPBEM on the discount factor $\gamma$. This is important because in \cite{schoknecht2003td}, the performance of TD methods is shown to be dependent on the discount factor $\gamma$.

To get a measure of how well our algorithm performs on much larger problems, we applied it on a large MDP where $|\mathbb{S}| = 2^{15}, |\mathbb{A}| = 50$, $k = 100$ and $\gamma=0.9$. The reward function $\mathrm{R}$ and the transition probability matrix $\mathrm{P}^{\pi}$ are generated using the equations (\ref{eqn:rwdrnd}) and (\ref{eqn:prnd}) respectively. RBFs are used as the features in this case. Since the MDP is huge, the algorithms were run on Amazon cloud servers. The true value function $V^{\pi}$ was computed and the $\sqrt{MSE}$s of the prediction vectors generated by the different algorithms were compared. The performance results are shown in Table \ref{tab:comptable}.
\small
\begin{table}[!h]
\begin{center}
\begin{tabular}{|c | c | c | c | c | c | c|}
\specialrule{.2em}{.04em}{.04em} 
%\hline\noalign{\smallskip}
   \textbf{Ex\#} & \textbf{SCE-MSPBEM} & \textbf{LSTD(0)} & \textbf{TD(0)} & \textbf{RLSTD(0)} & \textbf{LSPE(0)} & \textbf{GTD2} \\
%\noalign{\smallskip}
\specialrule{.2em}{.04em}{.04em} 
%\hline
\noalign{\smallskip}
 	1 & 23.3393204 &  23.3393204 & 24.581849219 & 23.3393204 & 23.354929410 & 24.93208571 \\ 
 	2 & 23.1428622 &  23.1428622 & 24.372033722 & 23.1428622 & 23.178814326 & 24.75593565 \\ 
	3 & 23.3327844 &  23.3327848 & 24.537556372 & 23.3327848 & 23.446585398 & 24.88119648 \\ 
    4 & 22.9786909 &  22.9786909 & 24.194543862 & 22.9786909 & 22.987520761 & 24.53206023 \\  
	5 & 22.9502660 &  22.9502660 & 24.203561613 & 22.9502660 & 22.965571900 & 24.55473382 \\ 
	6 & 23.0609354 &  23.0609354 & 24.253239213 & 23.0609354 & 23.084399716 & 24.60783237 \\ 
    7 & 23.2280270 &  23.2280270 & 24.481937450 & 23.2280270 & 23.244345617 & 24.83529005 \\
\specialrule{.2em}{.04em}{.04em} 
\end{tabular}
\end{center}
\caption{Performance comparison of various algorithms with large state space. Here $|\mathbb{S}| = 2^{15}, |\mathbb{A}| = 50, k = 100,$ and $\gamma = 0.9$. RBF is used as the feature set. The feature set is imperfect. The entries in the table correspond to the $\sqrt{\mathrm{MSE}}$ values obtained from the respective algorithms on $7$ different random MDPs. While the entries of SCE-MSPBEM, LSTD($0$) and RLSTD($0$) appear to be similar, they actually differed in decimal digits that are not shown here for lack of space.}\label{tab:comptable}
\end{table}
\normalsize

\section{Conclusion and Future Work}
We proposed, for the first time, an application of the Cross Entropy (CE) method to the problem of prediction in Reinforcement Learning (RL) under the linear function approximation architecture. This task is accomplished by remodelling the original CE algorithm as a multi-timescale stochastic approximation algorithm and using it to minimize the Mean Squared Projected Bellman Error (MSPBE). The proof of convergence to the optimum value using the ODE method is also provided. The theoretical analysis is supplemented by extensive experimental evaluation which is shown to corroborate the claim. Experimental comparisons with the state-of-the-art algorithms show the superiority in the accuracy of our algorithm while being competitive enough with regard to computational efficiency and rate of convergence. 

The algorithm can be extended to non-linear approximation settings also. In \cite{bhatnagar2009convergent}, a variant of the TD($0$) algorithm is developed and applied in the non-linear function approximation setting, where the convergence to the local optima is proven. But we believe our approach can converge to the global optimum in the non-linear case because of the application of a CE-based approach, thus providing a better approximation to the value function. The algorithm can also be extended to the off-policy case \cite{sutton2009convergent, sutton2009fast}, where the sample trajectories are developed using a behaviour policy which is different from the target policy whose value function is approximated. This can be achieved by appropriately integrating a weighting ratio \cite{glynn1989importance} in the recursions. TD learning methods are shown to be divergent in the off-policy setting \cite{baird1995residual}. So it will be interesting to see how our algorithm behaves in such a setting. Another future work includes extending this optimization technique to the control problem to obtain an optimum policy. This can be achieved by parametrizing the policy space and using the optimization technique in SCE-MSPBEM to search in this parameter space.

% Appendix here
% Options are (1) APPENDIX (with or without general title) or
%             (2) APPENDICES (if it has more than one unrelated sections)
% Outcomment the appropriate case if necessary
%
% \begin{APPENDIX}{<Title of the Appendix>}
% \end{APPENDIX}
%
%   or
%
% \begin{APPENDICES}
% \section{<Title of Section A>}
% \section{<Title of Section B>}
% etc
% \end{APPENDICES}

%%
% Acknowledgments here

% References here (outcomment the appropriate case)

% CASE 1: BiBTeX used to constantly update the references
%   (while the paper is being written).
%\bibliographystyle{ormsv080} % outcomment this and next line in Case 1
%\bibliography{<your bib file(s)>} % if more than one, comma separated
\bibliographystyle{ieeetr}
% CASE 2: BiBTeX used to generate mypaper.bbl (to be further fine tuned)
%\input{Operations-Research-template.bbl} % outcomment this line in Case 2

%If you don't use BiBTex, you can manually itemize references as shown below.
\bibliography{Operations-Research-template.bib}

%%%%%%%%%%%%%%%%%
\end{document}